\documentclass[a4paper, 11pt]{article}
\usepackage{latexsym,amsmath,amsfonts,amssymb}
\usepackage{tikz}
\usetikzlibrary{decorations.pathmorphing,cd,decorations.markings,calc}
\usepackage{mathrsfs}
\usepackage[american]{babel}
\usepackage{graphicx}
\usepackage{bbm}
\usepackage{cite}
\usepackage{tcolorbox}

\usepackage[colorlinks=true, citecolor=blue!90!black, linkcolor=blue!90!black, linktocpage=true, urlcolor=red!70!black]{hyperref}

\renewcommand{\baselinestretch}{1.2}
\newcommand{\ket}[1]{\ensuremath{| #1 \rangle}}
\newcommand{\bra}[1]{\ensuremath{\langle #1 |}}

\numberwithin{equation}{section}

\newcommand{\be}{\begin{equation}} \newcommand{\ee}{\end{equation}}
\newcommand{\bea}{\begin{equation} \begin{aligned}} \newcommand{\eea}{\end{aligned} \end{equation}}

\newcommand{\cA}{\mathcal{A}}
\newcommand{\cB}{\mathcal{B}}
\newcommand{\cC}{\mathcal{C}}
\newcommand{\cD}{\mathcal{D}}

\newcommand{\cL}{\mathcal{L}}

\newcommand{\cM}{\mathcal{M}}

\newcommand{\cN}{\mathcal{N}}
\newcommand{\cO}{\mathcal{O}}

\newcommand{\cS}{\mathcal{S}}

\newcommand{\bC}{\mathbb{C}}

\newcommand{\bH}{\mathbb{H}}

\newcommand{\bN}{\mathbb{N}}

\newcommand{\bR}{\mathbb{R}}

\newcommand{\bZ}{\mathbb{Z}}

\newcommand{\unit}{\mathbbm{1}}

\def\su{\mathfrak{su}}

\def\repa{\raise4pt\hbox{$\square$}\mkern-14mu\raise-4pt\hbox{$\square$}}
\def\repab{\overline{\raise4pt\hbox{$\square$}\mkern-14mu\raise-4pt\hbox{$\square$}\mkern-1mu}}

\renewcommand{\thefootnote}{\fnsymbol{footnote}}

\begin{document}

\thispagestyle{empty}
\fontsize{12pt}{20pt}
\vspace{13mm}
\begin{center}
	{\huge Non-Invertible Symmetry in\\ Calabi-Yau Conformal Field Theories}
	\\[13mm]
    	{\large Clay Córdova$^{a}$\footnote{\href{clayc@uchicago.edu}{clayc@uchicago.edu}} and Giovanni Rizi$^{b,\, c}$\footnote{\href{grizi@sissa.it}{grizi@sissa.it}}}
	
	\bigskip
	{\it 
        $^a$ Kadanoff Center for Theoretical Physics \& Enrico Fermi Institute, University of Chicago \\
		$^b$ SISSA, Via Bonomea 265, 34136 Trieste, Italy \\[.0em]
		$^c$ INFN, Sezione di Trieste, Via Valerio 2, 34127 Trieste, Italy \\[.6em]
	}
\end{center}

\bigskip

\begin{abstract}

\noindent We construct examples of non-invertible global symmetries in two-dimensional superconformal field theories described by sigma models into Calabi-Yau target spaces.  Our construction provides some of the first examples of non-invertible symmetry in irrational conformal field theories.  Our approach begins at a Gepner point in the conformal manifold where the sigma model specializes to a rational conformal field theory and we can identify all supersymmetric topological Verlinde lines.  By deforming away from this special locus using exactly marginal operators, we then identify submanifolds in moduli space where some non-invertible symmetry persists. For instance, along ten-dimensional loci in the complex structure moduli space of quintic Calabi-Yau threefolds there is a symmetry characterized by a Fibonacci fusion category.  The symmetries we identify provide new constraints on spectra and correlation functions.  As an application we show how they constrain conformal perturbation theory, consistent with recent results about scaling dimensions in the K3 sigma model near its Gepner point.

\end{abstract}

\vfill

\begin{flushleft}
December 2023
\end{flushleft}

\newpage
\pagenumbering{arabic}
\setcounter{page}{1}
\setcounter{footnote}{0}
\renewcommand{\thefootnote}{\arabic{footnote}}

{\renewcommand{\baselinestretch}{.88} \parskip=0pt
\setcounter{tocdepth}{2}
\tableofcontents}

\newpage

\section{Introduction}

Symmetry plays a central role in quantum mechanical systems.  In particular, discrete internal symmetries can organize eigenstates into multiplets and constrain matrix elements by selection rules. Recently the concept of symmetry has been dramatically broadened, allowing the possibility that symmetries do not form a group but instead a richer algebraic structure of a (higher) fusion category (see e.g.\ \cite{Gaiotto:2014kfa, Cordova:2022ruw, Shao:2023gho, Schafer-Nameki:2023jdn} for reviews with references).  These novel non-invertible symmetries lead to new multiplet structures and selection rules \cite{Lin:2022dhv, Choi:2022jqy, Cordova:2022ieu, Bhardwaj:2023ayw, Bhardwaj:2023wzd, Bartsch:2023pzl, Bartsch:2023wvv} and hence can provide a new window into strongly-coupled physics \cite{Chang:2018iay,Komargodski:2020mxz,Apte:2022xtu, Zhang:2023wlu, Jacobsen:2023isq,Choi:2023xjw, Antinucci:2023ezl}. 

Despite their seemingly exotic nature, non-invertible symmetries exist in many familiar quantum field theories.  Concentrating on two spacetime dimensions, the focus of this paper,  non-invertible symmetries  have been particularly well investigated for topological systems \cite{Bhardwaj:2017xup,Thorngren:2019iar, Thorngren:2021yso, Huang:2021zvu, Bhardwaj:2023fca, Bhardwaj:2023idu} and rational conformal field theories (CFTs) \cite{Fuchs:2002cm,Fuchs:2003id,Fuchs:2004dz, Fuchs:2004xi, Frohlich:2004ef, Frohlich:2006ch, Chang:2018iay, Choi:2023vgk, Diatlyk:2023fwf, Perez-Lona:2023djo, Cordova:2023jip} which have the simplifying property of having a finite number of primary operators with respect to an infinite dimensional chiral algebra.  For instance, in diagonal rational CFTs, there are internal global symmetries in one-to-one correspondence with the primary local operators.  Like all global symmetries, these are represented by line operators $\mathcal{L}_{i}$, so-called Verlinde lines \cite{Verlinde:1988sn, Petkova:2000ip, Drukker:2010jp, Moore:1988qv}, which have the property of being topological, i.e.\ they commute with the energy-momentum tensor.  Their action on a primary operator $\mathcal{O}_{j}$ is determined by the modular $S$-matrix:
\begin{equation}
    \mathcal{L}_{i}\cdot \mathcal{O}_{j}=\frac{S_{ij}}{S_{0j}}\mathcal{O}_{j}.
\end{equation}
Meanwhile, the fusion rule of these topological lines is:
\begin{equation}
    \mathcal{L}_{i}\times \mathcal{L}_{i}=\sum_{k}N_{ij}^{k}\mathcal{L}_{k},
\end{equation}
with $N_{ij}^{k}$ the integral fusion coefficients.  

The Verlinde lines described above are a useful example to study non-invertible symmetry in part because the associated rational CFTs are essentially exactly solvable.  Looking beyond these examples, our goal here is instead to find \emph{irrational} CFTs that still possess non-invertible global symmetry, where the existence of new symmetries may yield new insight into the dynamics.  Below we will carry this out by constructing non-invertible global symmetries for supersymmetric CFTs described by non-linear sigma models with Calabi-Yau target spaces, specifically, K3 surfaces, and quintic threefolds \cite{Greene:1996cy, Hori:2003ic, Sevrin:2013oca}. These models are interesting in that they provide some of the first examples of irrational theories with non-invertible symmetries (excluding interesting symmetries of the compact boson and orbifolds at generic radii \cite{Thorngren:2021yso, Nagoya:2023zky,Damia:2024xju}.) Moreover, in string theory such CFTs can describe the string worldsheet and it is expected, though not proven, that any worldsheet global symmetry should give rise to a spacetime gauge symmetry.  If that is the case we expect string theory on the spacetimes we describe below to exhibit novel gauge structures, perhaps implying the existence of novel branes as in \cite{Dierigl:2022reg,Debray:2023yrs, Dierigl:2023jdp,Kaidi:2023tqo}, which we leave as a target for future investigation.

\subsection{Summary of Results}

Our starting point is the observation that Calabi-Yau CFTs come in moduli spaces known as conformal manifolds.  Along these manifolds the scaling dimensions and operator product coefficients of generic operators vary.  The typical point in the conformal manifold is an irrational CFT with a current algebra given by the appropriate (extended) superconformal algebra.  The moduli space is parameterized by exactly marginal local operators, which moreover preserve the supercharges.  It is locally factorized into the K\"{a}hler moduli, corresponding to size deformations of the target space, and complex structure moduli corresponding to deformations in the shape of the target space (or more pragmatically deformations in the coefficients of the equations defining the target space.)

At special loci in these conformal manifolds, some Calabi-Yau CFTs admit Gepner points \cite{Gepner:1986hr, Gepner:1987qi, Gepner:1987vz} where the models are described formally by Fermat Calabi-Yau's with a suitable discrete gauging (orbifolding).  At these Gepner points the CFT specializes and becomes rational.  In such theories, there are many non-invertible symmetries described by the Verlinde lines reviewed above.  Our first task, carried out in Sections \ref{LGCYsec} and \ref{sec:symgep} below is thus to construct explicitly the symmetries of these Gepner models by reviewing their operator content and $S$-matrices following \cite{Witten:1993jg, Witten:1993yc, Greene:1996cy, Fuchs:2000gv,Maldacena:2001ky, Hori:2003ic, Gray:2008je}.  We achieve this by using a Cardy condition and demanding that the twisted sector Hilbert spaces, describing operators at the end of the lines, has a well-defined positive integral dimension at each energy level and spin \cite{Petkova:2000ip, Cardy:2004hm, Thorngren:2021yso, Collier:2021ngi}.  We also provide an equivalent perspective using an associated three-dimensional topological field theory which encodes the symmetries of these models \cite{Gaiotto:2014kfa, Gaiotto:2020iye, Apruzzi:2021nmk, Apruzzi:2022rei, Freed:2022iao, Freed:2022qnc,Kaidi:2022cpf, Kaidi:2023maf}.

In Section \ref{sec:symloci} we derive our main results.  Specifically, we move away from the Gepner point by deforming the action by exactly marginal operators.  We then track which non-invertible symmetries are preserved by these deformations. We content ourselves to an investigation of the fusion rules, leaving a detailed study of the F-symbols to future work.  Remarkably, we find that many special loci in the moduli space preserve some of the non-invertible symmetry.  These theories are generically irrational, and correspondingly little is known about their non-BPS spectrum of scaling dimensions and OPE coefficients. While the exact non-invertible symmetry depends in detail on the special locus investigated, we highlight several special cases to get a feeling for our results:
\begin{itemize}
    \item We begin with a warm up of the torus SCFT.  The Gepner point of interest corresponds to a square torus of unit volume.  As we deform away from is point, we find that an interesting non-invertible symmetry is preserved at any modulus $\tau$ as long as the $B$ field is taken to always vanish and the volume (defined by the metric $G$) obeys:
    \begin{equation}
        \tau=x+iy, \hspace{.2in}\det(G)=\frac{1+x^{2}+y^{2}}{2y}.
    \end{equation}
    Specifically, along this locus in moduli space there are topological defect lines (TDLs) with fusion algebra of a $\bZ_2 \times \bZ_2$ Tambara-Yamagami category \cite{tambara1998tensor, tambara2000representations,Thorngren:2021yso}. This fusion category has invertible lines $\eta_{i}$ generating the group $\bZ_2 \times \bZ_2$ as well as a single non-invertible line $\mathcal{D}$ obeying:
    \begin{equation}\label{eq:fusiontorus}
    \cD \times  \cD^{\dagger} = \unit + \eta_1 + \eta_2 + \eta_1 \times \eta_2\, . 
    \end{equation}
    
    \item In the case of the K3 CFT we find a variety of subloci in moduli space preserving non-invertible symmetries.  For instance along a four-dimensional subspace preserving the full $\mathcal{N}=4$ superconformal algebra we find a fusion algebra containing of a Tambara-Yamagami $\bZ_{2}^{4}$ category
    \begin{equation}\label{eq:fusionK3}     \cD\times \cD^{\dagger} =  \unit+\sum_{i=1}^{4}\eta_{i}+\sum_{i<j}\eta_{i}\times \eta_{j}\sum_{i<j<k}\eta_{i}\times \eta_{j}\times \eta_k+\eta_{1}\times \eta_{2}\times_{3}\times \eta_4\, . 
    \end{equation}
As discussed in \cite{Thorngren:2021yso, Chang:2018iay, Choi:2021kmx, Choi:2022zal ,Choi:2023vgk,Diatlyk:2023fwf} Tambara-Yamagami fusion category symmetries arise when a model is self-dual under gauging invertible symmetries.  Thus, our identification of Tambara-Yamagami fusion category symmetry in the K3 sigma model implies that at these special loci there are new self-dualities of these CFTs.
    
    \item In the case of the quintic threefold, we again find many loci preserving non-invertible symmetry.  Notably in this case, these are all loci in the complex structure moduli space; the unique K\"{a}hler modulus is frozen to its Gepner value.  As a particular highlight, we mention that along various ten-dimensional loci in complex structure moduli space we find (at least) a fusion category symmetry characterized by Fibonacci line $W$ obeying the fusion rule:
\begin{equation}
    W\times W = \unit+W.
\end{equation}

The fact that non-invertible symmetry appears in the irrational CFTs discussed above implies new constraints on these models which may be useful, for example, in constraining correlation functions or in a bootstrap type analysis  of their spectral data \cite{Keller:2012mr,Friedan:2013cba,Lin:2015wcg, Lin:2016gcl,Collier:2021ngi}.  To this end in Section \ref{sec:select} we review results of \cite{Lin:2022dhv} which characterize these selection rules in representation theoretic terms. Finally we then apply these considerations in Section \ref{sec:perturb} to conformal perturbation theory of the K3 sigma model near its Gepner point as recently studied in \cite{Keller:2023ssv}.  In particular, we show that the non-invertible symmetry at the Gepner point implies that only certain powers of the coupling can appear with non-zero coefficients in calculations of the perturbed scaling dimensions. This is consistent with observations of similar phenomena in \cite{Keller:2023ssv}.
    
\end{itemize}

\section{LG/CY Correspondence and Supersymmetric Minimal Models}\label{LGCYsec}

In this section we briefly review the Landau-Ginzburg/Calabi-Yau correspondence to provide the necessary context, then we  prepare the stage to study the topological defect lines in Gepner models analyzing the case of a single minimal model.

\subsection{Landau-Ginzburg/Calabi-Yau Correspondence}
Gepner \cite{Gepner:1987vz, gepner1988yukawa} was the first one to provide evidence relating $\cN=2$ minimal models with Calabi-Yau sigma models. We won't attempt to give an historically accurate account of the subsequent developments, the interested reader can consult \cite{Greene:1996cy, Hori:2003ic}. Here we instead recall the arguments of \cite{Witten:1993yc}, providing evidence for this relation. We consider $\cN = (2,2)$ supersymmetric theories in two dimensions. The particular class of models we are going to look at are $U(1)$ gauge theories with $r+1$ chiral multiplets $P$ and $X_1, \ldots, X_r$ with a specific superpotential
\begin{equation}
    W = P \left(X_1 ^{k_1+2} + \cdots+ X_r^{k_r +2} \right)\, , \qquad k_i \in \bN_{0}, \,\, \forall \,\, i = 1,\ldots, r
\end{equation}
and a linear twisted superpotential 
\begin{equation}
    \widetilde{W} = t \Sigma
\end{equation}
where $\Sigma = D_+ D_- V$, $V$ being the vector multiplet, and $t = r-i\theta $ encodes the FI parameter $r$ and the theta angle. The $U(1)$ gauge group transformations are
\begin{equation}
P \rightarrow \mathrm{e}^{-i H \lambda} P, \quad X_i \rightarrow \mathrm{e}^{i w_i \lambda} X_i
\end{equation}
where
\begin{equation}
\begin{aligned}
H & :=\operatorname{lcm}\left\{k_i+2\right\}, \\
w_i & :=\frac{H}{k_i+2} \, .
\end{aligned}
\end{equation}
Before analysing the phases of this system let's look at the symmetries of the action, in particular to the $R$-symmetries. In superspace notation the F-terms involving the two superpotentials are of the form 
\begin{equation}
    \int \, d\theta_+\, d\theta_- W(P,X_i) + \text{h.c} \qquad \qquad  \int \, d\theta_+\, d\bar{\theta}_- \widetilde{W}(\Sigma) + \text{h.c} \, ,
\end{equation}
then, in order for the theory to enjoy both left and right moving R-symmetries we need to find charge assigments for the chiral multiplets such that $W$ has charges $(-1,-1)$ under $U(1)_R^+ \times U(1)_R^-$. For the twisted superpotential instead one can check that a consistent charge assigments for the vector multiplet are possible only if $\widetilde{W}$ is linear\cite{Witten:1993yc}. This ensures that the $R$-symmetries are classical symmetries of the action, but they may still be affected by ABJ-anomalies as they couple to fermions of a given chirality only. It is simple to check that the anomaly cancellation condition is
\begin{equation}\label{eq:CYcond}
   - 1+ \sum_{i=1}^{r}\frac{1}{k_i +2} = 0\, ,
\end{equation}
i.e. that the gauge charges of the chirals sum to zero. If this condition is verified the twisted superpotential is not renormalized and the FI parameter $r$ is a true parameter of the theory\cite{Witten:1993yc, Hori:2003ic}. We now look at the space of classical vacua to try to understand the IR phases of this system. The scalar potential is (in the following the lowercase letters are the expectation values of the scalar components of the superfield denoted by the corresponding uppercase symbol)
\begin{equation}\label{eq:scalarpot}
\begin{split}
    U(x_i, \sigma, p) & = \left|\sum_{i}x_i^{k_i+2} \right|^2 + |p|^2 \sum_{i}|(k_i+2)x_i^{k_i+1}|^2 + \frac{e^{2}}{2}\left(\sum_{i}\frac{H|x_i|^2}{k_i+2} - H |p|^2- r\right)\\& + 2 |\sigma|^2 \left(\sum_{i}\frac{H^2|x_i|^2}{(k_i+2)^2} + H^2 |p|^2\right)\, .
\end{split}    
\end{equation}
Notice that the second term vanishes if $|p|=0$ and/or $|x_i|= 0$ for all $i =1,\ldots r$\footnote{For a generic superpotential of the form $W = P G(X_1,\ldots,X_r)$, where $G$ is a quasi homogeneous function, one requires that the equations $\partial_{X_i}G = 0$ for $i =1,\ldots r$ have a unique common solution at $X_1 =X_2=\ldots=X_r= 0$. In our example this transversality condition is automatically built in. }. We can identify two regions with qualitatively different behaviour. For $r>0$ we see that the susy vacua necessarily have $|p|=|\sigma|= 0$, the leftover equations are
\begin{equation}
\begin{split}
    &  \sum_{i}\frac{H}{k_i+2}|x_i|^2 = r\\
    &  \sum_{i}x_i^{k_i+2} = 0\, . 
\end{split}
\end{equation}
The first one identifies a compact submanifold of $\bC^r$, for instance in the case $k_i = k$ this is simply a sphere. Modding out by the $U(1)$ gauge group we find that the first equation identifies a complex weighted projective space $\mathbb{CP}[w_1,\ldots,w_n]$ with weights given by the gauge charges (notice that if $k_i = k$ for all $i$ then $w_i = 1$). The second equation then identifies a complex hypersurface $M$ inside the weighted projective space, in particular $M$ is K\"ahler and  the vanishing of its first Chern class turns out to be equivalent to the anomaly cancellation condition for the $R$-symmetries \eqref{eq:CYcond} (see e.g. Chapter 14 of \cite{Blumenhagen:2013fgp}). The system at low energies reduces to a non-linear sigma model on $M$ (one checks that all fields except the components of the $x_i$ tangent to $M$ get a mass at tree level), thus an anomaly free $R$-symmetry in the IR sigma model is equivalent to the CY condition. At least classically we found a region of the parameter space in which the low energy theory is a Calabi-Yau sigma model, although classical these computations are qualitatively correct also taking into account quantum corrections \cite{Witten:1993yc}. The other regime of interest is instead $r<0$, in this case the susy vacua are 
\begin{equation}
    |\sigma| = |x_1|= \cdots = |x_r| = 0\, , \qquad |p| = \sqrt{\frac{-r}{H}}\, 
\end{equation}
around these vacua the only the $x_i$ are massless (unless some $k_i = 0$) with interactions determined by a superpotential obtained integrating out $P$. This can be done replacing $P$ with its expectation value, so that, modulo reabsorbing an overall constant in the $x_i$'s, the new superpotential is simply 
\begin{equation}\label{eq:superpotLG}
    X_{1}^{k_1+2} +\cdots + X_{r}^{k_r+2}\, .
\end{equation}
Notice that, since $P$ is charged under the gauge group, we have an Higgsing from $U(1)$ to $\bZ_H$ with the unbroken gauge group acting as 
\begin{equation} \label{eq:symmorb}
    X_i \mapsto e^{i w_i \lambda}X_i
\end{equation}
where $e^{iH\lambda}=1$. Thus in this regime the theory below the breaking scale is a theory of only chirals with superpotential \eqref{eq:superpotLG}, \textit{a.k.a} a Landau-Ginzburg model, orbifolded by the action \eqref{eq:symmorb}. The theory of a single chiral superfield with superpotential $W = X^{k+2}$ flows in the IR to an $\cN=2$ minimal model $M_k$ of central charge 
\begin{equation}
    c = \frac{3k}{k+2}\, ,
\end{equation}
in particular to the so called $A$-series minimal model corresponding to choosing the diagonal modular invariant partition function \cite{Witten:1993jg}. Then our initial model, for negative values of the FI parameter, flows in the IR to an orbifolded product of minimal models
\begin{equation}
    \prod_{i=1}^{r} M_{k_i} \Big/ \bZ_H
\end{equation}
this is the Gepner model.

From the perspective of the NLSM we discussed before this Gepner model should correspond to a particular value of the K\"ahler modulus, that is it should correspond to a particular point on the conformal manifold, the Gepner point. The path connecting a generic point and this special point is smooth and there are no singularities (at least as long as one keeps a generic value of the theta angle \cite{Witten:1993yc}). There is much more to this story, with many possible generalizations (for instance taking a gauge group $U(1)^n$) and further subtleties to be addressed, however this general picture is enough to provide context.

\subsection{Minimal Models: Representations and Fusion Rules }
Minimal models for the $\cN=2$ superconformal algebra (see Appendix \ref{app:conv} for our conventions) form a discrete series with central charge
\begin{equation}
    c = \frac{3k}{k+2}\, ,
\end{equation}
where $k \in \bN$ is the level. At these values of $c$ unitarity allows only a finite number of superconformal primaries, these are labelled by their weight and $U(1)_R$ charges
\begin{equation}
    \begin{split}
h_{l, m}^{(\lambda)} & =\frac{l(l+2)}{4 (k+2)}+\frac{\lambda^2}{2}-\frac{(m-2 \lambda)^2}{4 (k+2)} \\
q_{l, m}^{(\lambda)} & = \frac{m+k \lambda}{k+2} 
     \end{split}
\end{equation}
with $\lambda = 0$ in the NS sector and $\lambda = - 1/2$ in the R sector. The variables $l,m$ are valued in
\begin{equation}
    (l,m) \in P_k = \left\{ l=0,\ldots,k\, , \, |m| \le l\, , \, l+m=0 \text{ (mod 2)} \right\} \, .
\end{equation}
Chiral primaries in the NS sector have $m = l$ while antichirals have $m = -l$, Ramond sector ground states are obtained by spectral flow.

To each superconformal primary representation $\bH^{(\lambda)}_{l,m}$ we associate two characters
\begin{equation}\label{eq:char}
\begin{split}
    &\text{ch}^{(\lambda)}_{l,m}(\tau, z) = \text{Tr}_{\bH_{l,m}^{(\lambda)}}e^{2 i \pi \tau L_0}e^{2 i \pi z j_0}\\ & \widetilde{\text{ch}}^{(\lambda)}_{l,m}(\tau, z) = \text{Tr}_{\bH_{l,m}^{(\lambda)}}(-1)^Fe^{2 i \pi \tau L_0}e^{2 i \pi z j_0} = e^{-i \pi q^{(\lambda)}_{l,m}}\text{ch}^{(\lambda)}_{l,m}\left(\tau, z+\frac{1}{2}\right)
\end{split}
\end{equation}
where, since the fermionic modes have $U(1)_R$ charge $\pm 1$, we represented fermion parity as $(-1)^{j_0 -q^{(\lambda)}_{l,m}}$. To discuss $SL(2, \bZ)$ transformation is convenient to work with characters of the bosonic subalgebra, this is equivalent to realize the minimal model as the coset $\widehat{\mathfrak{su}}(2)_k \times \widehat{\mathfrak{u}}(1)_2 / \widehat{\mathfrak{u}}(1)_{k+2}$ which we will discuss momentarily. The bosonic sub-representations include only states with a fixed fermion number $\bmod \  2$, their characters are
\begin{equation}\label{eq:chichar1}
   \chi^{(\lambda)}_{l,m}(\tau, z) = \frac{1}{2}\left(\text{ch}_{l,m}^{(\lambda)}(\tau, z) + \widetilde{\text{ch}}^{(\lambda)}_{l,m}(\tau, z)\right)  
\end{equation}
which only contains states with even fermion numbers, and
\begin{equation}\label{eq:chichar2}
   \widetilde{\chi}^{(\lambda)}_{l,m}(\tau, z) = \frac{1}{2}\left(\text{ch}_{l,m}^{(\lambda)}(\tau, z) - \widetilde{\text{ch}}^{(\lambda)}_{l,m}(\tau, z)\right) \, .  
\end{equation}
Following \cite{Gray:2008je} we relabel the primaries using new variables $a = l, c = m-2\lambda$ and $b = [a+c] = -2 \lambda$, where here and in the following we define 
\begin{equation}
    [x] \equiv x \bmod \ 2.
\end{equation}
We have 
\begin{equation}
\begin{split}
    & h_{a,c} = \frac{a(a+2)-c^2}{4(k+2)} + \frac{[a+c]}{8} \\
    & q_{a,c} = \frac{c}{k+2}-\frac{[a+c]}{2}
\end{split}
\end{equation}
and the new variables take values in 
\begin{equation}
    (a,c) \in P'_k = \{(a,c)\, | \, a = 0,\ldots, k \, , \,  |c- [a+c]| \le a \}\, .
\end{equation}
Now we set\footnote{The notation
\begin{equation}
  \widetilde{\chi}^{(\lambda)}_{l,m}(\tau, z) = \chi_{k-a, c+k+2}(\tau, z) \, . 
\end{equation}
is justified noticing that the states of lowest weight surviving the projection are obtained acting with $G^{\pm}_{-1/2}$ in the NS sector and $G^{\pm}_{0}$ in the R sector. It is then easy to check that
\begin{equation}
\begin{split}
    & q_{a,c}\pm 1 =q_{k-a, c+k+2} \bmod 2 \\
    & h_{k-a, c+k+2}- h_{a,c} = \frac{a+c+1}{2} \bmod 1
\end{split}
\end{equation}
then $h_{k-a, c+k+2} \bmod 1$ is the eigenvalue of $T: \tau\mapsto \tau+1$ on $\chi_{k-a, c+k+2}(\tau, z)$.}
\begin{equation}
    \chi^{(\lambda)}_{l,m}(\tau, z) = \chi_{a,c}(\tau, z) \qquad \widetilde{\chi}^{(\lambda)}_{l,m}(\tau, z) = \chi_{k-a, c+k+2}(\tau, z) \, . 
\end{equation}
Notice that for $(a,c) \in P'_k$ the pair $(k-a, c+k+2)$ does not belong to $P'_k$, thus we have to enlarge the indexing set for characters to
\begin{equation}
    Q_k = P'_k \cup \{ (k-a, c+k+2), (a,c) \in P'_k\} = \{(a,c) \, |\, 0\le a \le k\, , \, 0\le c\le 2k+3\} . 
\end{equation}
It turns out that $\chi_{a,c}$ with $(a,c) \in Q_k$ do have nice modular properties and yield a unitary representation of $SL(2, \bZ)$ with $S$ and $T$ matrices
\begin{equation}\label{eq:ST}
\begin{split}
    & S_{ac; a'c'} = \frac{1}{k+2}\sin\left(\frac{\pi(a+1)(a'+1)}{k+2}\right)e^{i \pi \frac{cc'}{k+2}}e^{-i \pi \frac{[a+c][a'+c']}{2}}\\ 
    & T_{ac;a'c'} = e^{2 i \pi \left(h_{a,c}-\frac{c}{24}\right)}\delta_{a,a'}\delta_{c,c'}\, .
\end{split}
\end{equation}
As usual $S^{2} = C$, $C^{2} = \unit$ with 
\begin{equation}
\begin{split}
    &  C_{a'c'; ac} = \delta_{a', a^+}\delta_{c',c^+} \\ & 
    (a^{+},c^+) = \begin{cases}
        (a, -c \bmod 2(k+2)) \quad \text{if} \quad [a+c]=0 \\
        (k-a, k+2-c \bmod 2(k+2)) \quad \text{if} \quad [a+c]=1 \, .
    \end{cases}
\end{split}
\end{equation}
Notice that in this basis of half-characters  the $T$ matrix is diagonal, while this would not be the case if we were working with the full characters $\text{ch}^{(\lambda)}_{l,m}(\tau, z)$. 

By means of the Verlinde formula we can obtain the fusion coefficients for the bosonic subrepresentations
\begin{equation}
\begin{aligned}
& N_{a c ; a^{\prime} c^{\prime}}^{\alpha \gamma}=\sum_{(d, f) \in Q_k} \frac{S_{a c ; d f} S_{a^{\prime} c^{\prime} ; d f} S_{\alpha \gamma ; d f}^*}{S_{00 ; d f}} \\
& =\left\{\begin{array}{ll}
\left(N^{\widehat{\mathfrak{s u}(2)_k}}\right)_{a, a^{\prime}}^\alpha\left(N^{\widehat{\mathfrak{u}}(1)_{k+2}}\right)_{c, c^{\prime}}^\gamma, & \text { if }[a+c]\left[a^{\prime}+c^{\prime}\right]=0 \\
\left(N^{\widehat{\mathfrak{su}(2)_k}}\right)_{a, a^{\prime}}^{k-\alpha}\left(N^{\widehat{\mathfrak{u}}(1)_{k+2}}\right)_{c, c^{\prime}}^{\gamma+k+2}, & \text { if }[a+c]\left[a^{\prime}+c^{\prime}\right]=1
\end{array}\right\} \\
& \text { for }(a c),\left(a^{\prime} c^{\prime}\right),(\alpha \gamma) \in Q_k\, ,
&
\end{aligned}
\end{equation}
where
\begin{equation}
\begin{aligned}
& \left(N^{\widehat{\mathfrak{s u}}(2)_k}\right)_{a, a^{\prime}}^l=\delta\left(\left|a-a^{\prime}\right| \leq l \leq \min \left(a+a^{\prime}, 2 k-a-a^{\prime}\right)\right) \delta\left(a+a^{\prime} \equiv l \bmod 2\right) \\
& \left(N^{\widehat{\mathfrak{u}}(1)_{k+2}} \right)_{c, c^{\prime}}^n=\delta\left(c+c^{\prime} \equiv n \bmod 2 (k+2)\right)
&
\end{aligned}
\end{equation}
are the fusion coefficients for the $\widehat{\su}(2)_k$ and $\widehat{\mathfrak{u}}(1)_{k+2}$ Kac-Moody algebras. One can check that this result is consistent with the conservation of the $U(1)_R$ charge in the OPE. 

\subsection{Symmetries and Orbifolds}

In the rest of the paper we will only consider the minimal model with diagonal modular invariant, the torus partition function is
\begin{equation}
     Z(\tau, z)=\sum_{(a, c) \in Q_k} |\chi_{a,c}(\tau,z)|^2 = \sum_{\substack{(l,m)\in P_k \\ \lambda=0, -1/2}} \text{Tr}_{\bH_{l,m}^{(\lambda)}}\left((1+(-1)^{F_L+F_R})q^{L_0-\frac{c}{12}}\bar{q}^{\overline{L}_0-\frac{c}{12}}y^{j_0}\bar{y}^{\overline{j}_0}\right) \, \, ,
\end{equation}
with $q=e^{2i \pi \tau}, y = e^{2i \pi z}$. Requiring modular invariance automatically includes the GSO projection, and in the diagonal case the physical primaries have to be fermionic or bosonic on both sides. Notice in particular that among those the holomorphic and antiholomorphic supercurrents do not survive the projection, rather they can appear only when paired with another fermionic state on the other side.

We now construct the Verlinde lines of the theory, the simplest way to do so is bootstrap them from the partition function. In Appendix \ref{app:boundaries} we also give an alternative derivation using the folding trick. The idea is to make an ansatz for the action of the line on the physical primaries of the theory, and then constrain it imposing concistency of the twisted Hilbert spaces. This is a well known construction, we repeat it here as a warm up for the more involved case of Gepner models. In the diagonal theory the circle Hilbert space is
\begin{equation}
    \bH = \bigoplus_{(a,c)\in Q_k}\bH_{a,c}\otimes \overline{\bH}_{a,c}
\end{equation}
and a physical primary $\Phi_{ac}$ corresponds to the state $\ket{a,c}\otimes \overline{\ket{a,c}}$. A topological line $L_{r,s}$ commutes with the Virasoro generators, here we also require it to be supersymmetric, namely to commute with all the generators of the bosonic subalgebra. We parametrize the action on primaries as
\begin{equation}
    L_{r, s} \Phi_{a,c} = X^{a,c}_{r,s}\Phi_{a,c}\, 
\end{equation}
and constrain it imposing that the twisted partition function $Z_{r,s}(\tau, z)$ admits a decomposition in characters of the bosonic subalgebra with integer multiplicities. We have
\begin{equation}
    Z_{r,s}(\tau, z) = \sum_{\substack{(a,c), (a',c')\\ (a'',c'')}}X^{a,c}_{r,s}S_{ac;a'c'}S_{ac;a''c''}^*\chi_{a',c'}(q, y)\chi_{a'',c''}(\bar{q},\bar{y})\, 
\end{equation}
then we require
\begin{equation}
    \sum_{(a,c) \in Q_k}X^{a,c}_{r,s}S_{ac;a'c'}S_{ac;a''c''}^* \in \bN\, .
\end{equation}
A natural solution for the multiplicities is given by the fusion coefficients, namely
\begin{equation}
    X^{a,c}_{r,s} = \frac{S_{rs;ac}}{S_{00;ac}}\, ,  \qquad \sum_{(a,c) \in Q_k}X^{a,c}_{r,s}S_{ac;a'c'}S_{ac;a''c''}^* = N_{rs; a'c'}^{a''c''}
\end{equation}
corresponding to 
\begin{equation}
    Z_{r,s}(\tau, z) = \sum_{(a',c')(a'',c'')}N_{rs; a'c'}^{a''c''}\chi_{a',c'}(q, y)\chi_{a'',c''}(\bar{q},\bar{y})\, .
\end{equation}
The action by linking on physical primaries is the usual one
\begin{equation}
    L_{r, s} \Phi_{a,c} = \frac{S_{rs;ac}}{S_{00;ac}}\Phi_{a,c}\, 
\end{equation}
and fusion immediately follows from the Verlinde formula
\begin{equation}
    L_{rs}\times L_{r's'} = \sum_{(r'',s'')\in Q_k}N_{rs; r's'}^{r''s''}L_{r'' s''}\, , 
\end{equation}
we therefore have $|Q_k|= 2(k+1)(k+2)$ topological defect lines. 

The partition functions $Z_{r,s}(\tau, z)$ are traces over the twisted Hilbert spaces. The states in those spaces are mapped, by the state operator correspondence, to non-genuine local operators, namely local operators attached to the topological line $L_{r,s}$. When inserted in correlation functions these twist defects generically introduce branch cut singularities, corresponding to the action of the attached TDL. We also notice that, since $L_{r,s}$ acts non-trivially on the physical primaries of the theory the corresponding twisted Hilbert spaces cannot contain the identity operator, namely the ground state necessarily has positive Virasoro weights.

We first want to find the set of invertible lines. To this extent it is useful to recall that the fusion of a line and its orientation reversal always contains the identity. Reversing the orientation however is equivalent to act with charge conjugation on the labels of the line. 
From the explicit expression of the fusion coefficients it is simple to compute
\begin{equation}
    N^{rs}_{ac;a^+c^+} = \begin{cases}
        \delta(0\le r\le \text{min}(2a,2k-2a) )\,\delta(r=0 \bmod 2)\, \delta_{s,0} \quad \text{if} \quad [a+c]=0 \\
        \delta(0\le r\le k-|2a-k|)\,\delta(r=0 \bmod 2)\, \delta_{s,0} \quad \text{if} \quad [a+c]=1
    \end{cases}
\end{equation}
which consistently obey $N^{00}_{ac;a^+c^+}=1$. A line is invertible if \emph{only} the identity appears in the fusion channel with its charge conjugate, we then see that the invertible lines are $L_{0, s}$ and $L_{k,s};$
where $s= 0,\ldots, 2k+3$. Thus there are $4(k+2)$ invertible lines. The fusion among those is controlled by the coefficients
\begin{equation}
    N_{0s;0s'}^{ac} = \begin{cases}
        \delta_{a,0}\delta_{c,s+s'} \quad [s][s']=0 \\
        \delta_{a,k}\delta_{c,s+s'+k+2} \quad [s][s']=1
    \end{cases}
\end{equation}
\begin{equation}
    N_{ks;ks'}^{ac} = \begin{cases}
        \delta_{a,0}\delta_{c,s+s'} \quad [k+s][k+s']=0 \\
        \delta_{a,k}\delta_{c,s+s'+k+2} \quad [k+s][k+s']=1
    \end{cases}
\end{equation}
\begin{equation}
    N_{ks;0s'}^{ac} = \begin{cases}
        \delta_{a,k}\delta_{c,s+s'} \quad [k+s][s']=0 \\
        \delta_{a,0}\delta_{c,s+s'+k+2} \quad [k+s][s']=1
    \end{cases} = N_{0s';ks}^{ac} \, . 
\end{equation}
The group structure depends on $k$: 
\begin{itemize}
    \item $k$ even. It is easy to see that
    \begin{equation}
        L_{k,1}^{2n} = L_{0, 2n + n(k+2)} \qquad L_{k,1}^{2n+1} = L_{k, 2n+1 + n(k+2)}
    \end{equation}
    and in particular $L_{k,1}^{2(k+2)} = \unit$, therefore $L_{k,1}$ generates a $\bZ_{2(k+2)}$ group. We also have 
\begin{equation}
    L_{k,0}^{2} = \unit
\end{equation}
so $L_{k,0}$ generates a $\bZ_2$. By computing
\begin{equation}
    L_{k,0}\cdot L_{k,1} \cdot L_{k,0} = L_{k,1}
\end{equation}
we see that the symmetry structure is a direct product. One also checks that all invertible lines can be obtained fusing $L_{k,0}$ and $L_{k,1}$. Thus the invertible lines for $k$ even form a group $ \bZ_{2} \times \bZ_{2(k+2)}$ with the generators acting as
    \begin{equation}
        L_{k,0}\Phi_{ac} = (-1)^{a}\Phi_{ac}\, ,\qquad L_{k,1}\Phi_{ac} = (-1)^{a}e^{\frac{2i \pi}{2(k+2)}\left( c - \frac{k+2}{2}[a+c]\right)}\Phi_{ac}\, .
    \end{equation}

\item $k$ odd. In this case fusing $L_{0,1}$ with itself we can generate all lines, in particular
    \begin{equation}
        L_{0,1}^{n} =\begin{cases} L_{0,n} \qquad n= 0,1 \bmod 4\\
        L_{k,n+ (k+2)} \qquad n= 2,3 \bmod 4\\
        \end{cases}
    \end{equation}
    so $L_{0,1}^{4(k+2)}= \unit$. Thus the invertible lines form a $\bZ_{4(k+2)}$ group, with the generator acting as
    \begin{equation}
       L_{0,1}\Phi_{ac} = e^{\frac{2i \pi}{4(k+2)}\left( 2c -(k+2) [a+c]\right)}\Phi_{ac}\, . 
    \end{equation}
\end{itemize}
These match known symmetries of the minimal models see e.g.\cite{Greene:1996cy, Maldacena:2001ky}. 

For  any value of $k$ we have a $\bZ_{k+2}$ subgroup generated by $L_{0,2}$
\begin{equation}
    L_{0,2}\Phi_{ac} = e^{2i \pi \frac{c}{k+2}}\Phi_{ac}
\end{equation}
notice that under this subgroup a full superconformal family transforms with the same charge since
\begin{equation}
    L_{0,2}\Phi_{k-a;c+k+2} = e^{2i \pi \frac{c}{k+2}}\Phi_{k-a;c+k+2}\, .
\end{equation}
Another symmetry present for all values of $k$ is a $\bZ_2$ generated by $L_{k, k+2}$. This acts by
\begin{equation}
    L_{k,k+2}\Phi_{ac} = (-1)^{a+c}\Phi_{ac}
\end{equation}
i.e. it leaves invariant the NS sector primaries while giving a sign on the R ones, one way of interpreting this is as the symmetry dual to $(-1)^{F}$ which has been trivialized by the GSO projection. Both these $\bZ_{k+2}$ and $\bZ_{2}$ will play an important role in the construction of the Gepner models.

As a warm-up for the next section we compute the partition functions of the diagonal minimal model orbifolded by the $\bZ_{k+2}$ described above. To do so we need to twist and twine the partition function by the generator of $\bZ_{k+2}$ and then sum over the group elements. The twined partition function is simple to write down, acting with (a $s$-th power of) the symmetry operator on the Hilbert space we get
\begin{equation}
    Z(\tau,z,s) = \sum_{(a,c) \in Q_k}e^{2 i \pi \frac{s c}{k+2}}\chi_{ac}(q, y)\chi_{ac}(\bar{q},\bar{y}) \, . 
\end{equation}
The twisted partition function is obtained with a $S$ modular transformation
\begin{equation}
\begin{split}
    Z_{s}(\tau, z) &= Z\left(-\frac{1}{\tau},\frac{z}{\tau} \right) = \sum_{(a,c), (a',c'), (a'',c'') \in Q_k}e^{2 i \pi \frac{s c}{k+2}}S_{ac; a'c'}S_{ac;a''c''}^*\chi_{a'c'}(q, y)\chi_{a''c''}(\bar{q}, \bar{y})\\ & =   \sum_{(a',c'), (a'',c'') \in Q_k}N_{a'c';02s}^{a''c''}\chi_{a'c'}(q,y)\chi_{a''c''}(\bar{q}, \bar{y}) = \sum_{(a,c) \in Q_k}\chi_{ac}(q,y)\chi_{a,c+2s}(\bar{q}, \bar{y})
\end{split}
\end{equation}
Where we used the Verlinde formula and, in the last step, the explicit form of the fusion coefficients. To write down the orbifold partition function we need both twisting and twining, the simplest way to twine a twisted partition function is to use the $T$ transformation. We have 
\begin{equation}
   T\cdot Z_{s}(\tau,z)\sum_{(a,c) \in Q_k}e^{2 i \pi s\frac{2c+2s}{2(k+2)}}\chi_{ac}(q, y)\chi_{a,c+2s}(\bar{q}, \bar{y})\, . 
\end{equation}
therefore
\begin{equation}
    Z_{s}(\tau,z, r) = \sum_{(a,c) \in Q_k}e^{2 i \pi r\frac{2c+2s}{2(k+2)}}\chi_{ac}(q,y)\chi_{a,c+2s}(\bar{q}, \bar{y})\, . 
\end{equation}
The partition function of the gauged theory is then
\begin{equation}
    Z_{\bZ_{k+2}}(\tau, z) = \frac{1}{k+2}\sum_{s,r=0}^{k+1} Z_{s}(\tau, z,r) = \frac{1}{k+2}\sum_{s,r=0}^{k+1} \sum_{(a,c) \in Q_k}e^{2 i \pi r\frac{2c+2s}{2(k+2)}}\chi_{ac}(q,y)\chi_{a,c+2s}(\bar{q}, \bar{y})
\end{equation}
the sum over $r$ sets
\begin{equation}
    2c + 2s = 0 \bmod 2(k+2) \rightarrow c + 2s = - c \bmod 2(k+2) 
\end{equation}
thus
\begin{equation}
     Z_{\bZ_{k+2}}(\tau, z) = \sum_{(a,c) \in Q_k}\chi_{ac}(q, y) \chi_{a, -c}(\bar{q}, \bar{y})\, . 
\end{equation}
This partition function is modular invariant\footnote{One can quickly check this using that $S_{a-c;a'c'} =S_{ac;a'-c'}$} and defines a sensible SCFT. We now want to determine the symmetries of the orbifolded theory. The first thing we notice is that
\begin{equation}
    N_{0,2s;ac}^{a'c'} =\delta_{a,a'}\delta_{c',c+2s}= N_{ac;02s}^{a'c'}
\end{equation}
i.e. the fusion of the generic Verlinde line with the $\bZ_{k+2}$ symmetry is abelian
\begin{equation}
    L_{0,2s}L_{ac} L_{0,-2s} = L_{ac}
\end{equation}
therefore we may hope that $L_{ac}$ survives the gauging operation. Another independent way we have to study the symmetries of the orbifold is to use modular covariance. The Hilbert space of the gauged theory is
\begin{equation}
  \bH^{(\bZ_{k+2})} = \bigoplus_{(a,c) \in Q_k} \bH_{(a,c)}\otimes \overline{\bH}_{(a,-c)}
\end{equation}
then we make an ansatz for the action of some new TDL $\cL_{(a,c)}$ on it. Denoting again a physical primary as $\Phi_{ac}$ we set
\begin{equation}
    \cL_{(r,s)} \Phi_{ac} = X_{rs}^{ac}\Phi_{ac}\, ,
\end{equation}
clearly this action preserves the full chiral algebra. We then constraint this ansatz by requiring that the TDL $\cL_{(r,s)}$ gives a consistent twisted Hilbert space. Specifically, via a modular transformation of the twined partition function, we impose
\begin{equation}
    \sum_{(a,c) \in Q_k} X_{rs}^{ac}S_{ac;a'c'}S_{a,-c;a''c''}^* \in \bN\, .
\end{equation}
Using the symmetry $S_{a-c;a'c'} =S_{ac;a'-c'}$ we see that there is an obvious solution 
\begin{equation}
    X_{rs}^{ac} = \frac{S_{ac;rs}}{S_{00;ac}}
\end{equation}
corresponding to
\begin{equation}
    \sum_{(a,c)} X_{rs}^{ac}S_{ac;a'c'}S_{a,-c;a''c''}^* = N_{rs;a'c'}^{a'',-c''}\, . 
\end{equation}
This shows that the lines $\cL_{r,s}$ acting as
\begin{equation}
    \cL_{(r,s)} \Phi_{ac} = \frac{S_{ac;rs}}{S_{00;ac}}\Phi_{ac}\, ,
\end{equation}
yield a consistent twisted Hilbert space. Whenever we gauge a discrete symmetry we expect a dual one to show up in the gauged theory. Indeed one can easily show that this dual symmetry is generated by $\cL_{0,2}$, which acts by
\begin{equation}
    \cL_{0,2s}\Phi_{ac} = e^{2 i \pi s \frac{c}{k+2}}\Phi_{ac}.
\end{equation}
giving charge to the new physical primaries coming from the old twisted sectors. It is also easy to see that gauging this new $\bZ_{k+2}$ symmetry we get back to the original theory with the diagonal modular invariant. One can repeat a similar analysis for the $\bZ_{2}$ generated by $L_{k, k+2}$ and also for the product $\bZ_{2}\times \bZ_{k+2}$. In particular one can check that gauging the product we obtain the charge-conjugation invariant partition function
\begin{equation}
     Z_{\bZ_2\times \bZ_{k+2}}(\tau, z) = \sum_{(a,c) \in Q_k} \chi_{ac}(q, y)\chi_{a^+ c^+}(\bar{q}, \bar{y})^*\, .
\end{equation}
This property is well known in the literature \cite{Zamolodchikov:1986gh, Gepner:1986hr,Greene:1996cy} and is important in the construction of mirror manifolds.  

\subsection{The $3d$ TQFT}
A very useful realization of the $\cN=2$ minimal models, that we implicitly used throughout this section, is as the coset (see e.g. \cite{liu2019coset} for a complete list of references)
\begin{equation}\label{eq:coset}
    M_k = \frac{\widehat{\mathfrak{su}}(2)_k \times \widehat{\mathfrak{u}}(1)_2}{\widehat{\mathfrak{u}}(1)_{k+2}}\, .
\end{equation}
In particular when the minimal model is presented in this form it is immediate to write down the $3d$ TQFT corresponding to it. This is simply the Chern-Simons theory with gauge group
\begin{equation}\label{eq:gaugegroupCS}
   G_k= \frac{SU(2)_k \times U(1)_2 \times U(1)_{-(k+2)}}{\bZ_{2}^{(1)}}
\end{equation}
where $\bZ_{2}^{(1)}$ is the one-form symmetry deriving from common center of gauge group factors. The Wilson lines for $SU(2)_k \times U(1)_2 \times U(1)_{k+2}$ can be labelled by three integers $(a,s,c)$ with $a=0,\ldots,k$ for $SU(2)_k$, $s =0,1,2,3$ and $c=0,\ldots,2(k+2)-1$ for the two $U(1)$s. The fusion rules are
\begin{equation}
    \cL_{(a,s,c)}\times \cL_{(a',s',c')} = \sum_{\substack{a''=|a-a'|\\ a'' = a-a' \bmod 2}}^{\text{min}(a+a', 2k-a-a')}\cL_{(a'', s+s',c+c')}
\end{equation}
while the topological $S$ matrix and spins are
\begin{equation}
\begin{split}
    & S_{asc; a's'c'} = \frac{1}{k+2}\sin\left(\frac{\pi(a+1)(a'
    +1)}{k+2}\right)e^{2 i \pi \frac{s s'}{4}}e^{-2i \pi \frac{c c'}{2(k+2)}}\, . \\
    & \theta_{asc} = e^{i \pi \frac{a(a+2)}{2(k+2)}}e^{i \pi \frac{s^2}{2}}e^{i \pi \frac{m^2}{k+2}}
\end{split}
\end{equation}
Also the $F$-symbols are factorized, see \cite{Fuchs:2002cm} for explicit expressions. The $\bZ_{2}^{(1)}$ is generated by the line $(k,2,k+2)$, which has $\theta_{k,2,k+2} = 1$ for any $k$, hence can always be gauged. The eigenvalue of the action by linking on other lines is 
\begin{equation}
   \frac{S_{k,2,k+2;asc}}{S_{000;asc}} = (-1)^{a+c+s}
\end{equation}
while fusion is
\begin{equation}
    \cL_{(k,2,k+2)}\times \cL_{(a,s,c)} = \cL_{(k-a, s+2, c+k+2)} \, .
\end{equation}
Then in the theory with gauge group \eqref{eq:gaugegroupCS} we identify
\begin{equation}
     (a,s,c) \sim (k-a, s+2, c+k+2)
\end{equation}
and keep only gauge invariant lines with $a+c+s = 0 \,\bmod\, 2$. Notice that the action by fusion has no fixed points, therefore there is no doubled line. These of course match the field identifications and restrictions in the coset \eqref{eq:coset}. In our analysis above we have chosen, following\cite{Gray:2008je}, a particular gauge fixing in which $s=0,1 = [a+c]$, then anyons can be labelled by a pair of integers $(a,c) \in Q_k$. The $S$-matrix, fusion rules and $F$-symbols are well defined on the anyons of the gauged theory, one only needs to be careful with the gauge fixing chosen when writing down explicit expressions. 

When discussing the full CFT and not only a chiral half the coupled $3d$-$2d$ system consists of a $3d$ bulk given by a Chern-Simons theory with gauge group $G_k \times G_{-k}$ with two boundary conditions, a conformal one and a topological one. Anyons of the theory are labelled by four integers $(a_L,c_L,a_R,c_R) \in Q_{k}\times Q_k$ giving the labels of the Wilson lines for $G_k$ and $G_{-k}$ respectively. Lines can terminate on the conformal boundary giving rise to local, but not necessarily genuine, operators labeled by the same four integers $(a_L,c_L,a_R,c_R)$. As usual the physical spectrum is determined by the topological boundary condition \cite{Fuchs:2002cm}.

\section{Symmetries of Gepner Models}\label{sec:symgep}

By the Landau-Ginzburg/Calabi-Yau correspondence the Gepner model can be constructed as an orbifold of a tensor product of $\cN=2$ minimal models, in particular
\begin{equation}
    \left(\bigotimes_{i=1}^r M_{k_i}\right) \Big/\bZ_{H}
\end{equation}
where $M_{k_i}$ is a minimal model at level $k_i$ and $H= \text{lcm}\{k_i+2\}$, with the group $\bZ_{H}$ being generated by the line operator
\begin{equation}
 \bigotimes_{i=1}^{r}L_{0,2}\,\,\, .
\end{equation}
The levels are chosen to satisfy the Calabi-Yau condition \eqref{eq:CYcond} so that the total central charge is a multiple of $3$
\begin{equation}
    c = \sum_{i=1}^{r}\frac{3k_i}{k_i+2} = 3(r-2)\, .
\end{equation}
Before carrying out the $\bZ_{H}$ orbifold we have to perform the correct GSO projection. This has to be imposed simultaneously on all the minimal models, namely we are going to allow only states whose components along the single minimal models are all either in the NS or in R sector. We detail the construction of the Gepner model via subsequent gaugings starting from the product of GSO-projected minimal models in the first subsection. In the rest of this section we study the spectrum of the model as well as its symmetries. 

\subsection{Construction of the Model}

To construct the model we start from the product of GSO-projected minimal models, the partition function is simply 
\begin{equation}\label{eq:naivediag}
    Z(\tau, z) = \prod_{i=1}^{r}\sum_{(a_i,c_i)\in Q_{k_i}}\left|\chi_{a_i,c_i}(q, y)\right|^{2}\, , 
\end{equation}
and the physical primaries are
\begin{equation}
    \Phi_{\{a_i, c_i\}} = \bigotimes_{i=1}^{r}\Phi_{a_i,c_i}^{(i)}\, .
\end{equation}
Clearly the TDLs of the tensor product theory are just tensor products of the lines of each minimal model, therefore in total we have
\begin{equation}\label{eq:totlines}
     \prod_{i=1}^{r}2(k_i+1)(k_i+2)
\end{equation}
topological line defects. To achieve the correct GSO projection consider the lines
\begin{equation}
    L_{k_1,k_1+2}\otimes L_{k_i, k_i+2} \qquad i=2, \ldots r\, , 
\end{equation}
these generate a $\bZ_{2}^{r-1}$ group and act as
\begin{equation}
    L_{k_1,k_1+2}\otimes L_{k_i, k_i+2} \Phi_{\{a_i, c_i\}} = (-1)^{a_1+c_1+a_i+c_i}\Phi_{\{a_i, c_i\}}\, .
\end{equation}
Therefore gauging this $\bZ_{2}^{r-1}$ enforces the correct projection, and the addition of the corresponding twisted sectors ensures modular invariance. Instead of going through the gauging procedure via the insertion of defects a quicker way to obtain the correct expression is to consider the diagonal modular invariant partition function one would write using the full superconformal characters
\begin{equation}
    Z_{\text{GSO}}(\tau, z) =  \sum_{\substack{\{l_i, m_i\}\\ \lambda=0, -1/2}} \prod_{i=1}^{r}|\text{ch}^{(\lambda)}_{l_i, m_i}(q,y)|^2+\prod_{i=1}^{r}|\widetilde{\text{ch}}^{(\lambda)}_{l_i,m_i}(q,y)|^2\, , 
\end{equation}
in which the proper NS/R alignement is imposed by hand and is manifestly $S$ and $T$ invariant. Now, rewriting it in terms of the half-characters $\chi_{ac}$, we have
\begin{equation}\label{eq:ZGSO}
\begin{split}
   & Z_{\text{GSO}}(\tau, z) = \sum_{A \in \cS_r} Z_A(\tau,z) \\
   & Z_A(\tau,z)=\sum_{\{a_i, c_i\}} P^{\text{GSO}}_{\{a_i,c_i\}}\prod_{i\in A^{\perp}}\chi_{a_i,c_i}(q,y)\chi_{a_i,c_i}(\bar{q},\bar{y})\prod_{i\in A}\chi_{a_i,c_i}(q,y)\chi_{k_i-a_i,c_i+k_i+2}(\bar{q},\bar{y})
\end{split}    
\end{equation}
where $A$ is an ordered subset of $\{1,\ldots,r\}$ of even order, $A^{\perp}$ is its complement and
\begin{equation}
   P^{\text{GSO}}_{\{a_i,c_i\}}=\prod_{j>1}\frac{1+(-1)^{a_1+c_1+a_j+c_j}}{2}\, 
\end{equation}
enforces the proper projection. We have also denoted by $\cS_r$ the set of all ordered subsets of $\{1,\ldots,r\}$ of even order. The sum over $A$ is the sum over twisted sectors of the $\bZ^{r-1}_{2}$, indeed $|\cS_r| = 2^{r-1}$ and there are exactly $2^{r-1}-1$ non-empty ordered subsets of $\{1, \ldots,r\}$ of even order, one for each non-trivial element of $\bZ^{r-1}_{2}$\footnote{A generic element of $\bZ_{2}^{r-1}$ corresponds to the line 
\begin{equation}
    \bigotimes_{j\in A}L_{k_j, k_j+2} \qquad \text{with} \qquad A \in \cS_r\, . 
\end{equation}}.

We now repeat the bootstrapping analysis for the TDLs of this theory. The physical primaries are labelled as $\Phi_{\{a_i, c_i\}, A}$ and are subject to the NS/R alignement constraint $[a_1+c_1] = [a_i+c_i]$ $\forall i=2, \ldots,r$. A TDL $L_{\{r_i,s_i\}, B}$ acts by
\begin{equation}
    L_{\{r_i,s_i\}, B}\Phi_{\{a_i, c_i\}, A} = \left(\zeta_{AB}\prod_{i=1}^{r}X^{a_i,c_i}_{r_i,s_i}\right)\Phi_{\{a_i, c_i\}, A}\, ,
\end{equation}
here we have added an extra sign $\zeta_{AB}$ which parametrizes the quantum $\bZ_{2}^{r-1}$ symmetry acting on the twisted sectors, namely $\zeta_{AB}=(-1)^{\delta_{AB}}$ and $\zeta_{A \varnothing} = 1$. The constraint on multiplicities is
\begin{equation}\label{eq:GSOmult}
    \sum_{A \in \cS_r}\zeta_{AB} \sum_{\{a_i,c_i\}}P^{\text{GSO}}_{\{a_i,c_i\}} \prod_{i \in A^{\perp}}X^{r_i,s_i}_{a_i,c_i}S_{a_i,c_i; a_i', c_i'}S^*_{a_i,c_i; a_i'', c_i''}\prod_{i \in A}X^{r_i,s_i}_{a_i,c_i}S_{a_i,c_i; a_i', c_i'}S^*_{k_i-a_i,c_i+k_i+2; a_i'', c_i''}\in \bN\, .
\end{equation}
We first notice that
\begin{equation}\label{eq:PGSO}
\begin{split}
    & P^{\text{GSO}}_{\{a_i,c_i\}} = \frac{1}{2^{r-1}}\sum_{A' \in \cS_r}(-1)^{\sum_{i \in A'}a_i+c_i} =\frac{1}{2^{r-1}}\sum_{A'\in \cS_r} \prod_{i \in A'}\frac{S_{k_i, k_i+2; a_i, c_i}}{S_{0,0; a_i, c_i}}\, ,  \\ & S_{k_i-a_i,c_i+k_i+2; a'_i,c_i'} = (-1)^{a_i'+c_i'}S_{a_i, c_i; a'_i, c'_i}\, .
\end{split}
\end{equation}
Then 
\begin{equation}
    \begin{split}
        \eqref{eq:GSOmult}=\frac{1}{2^{r-1}}\sum_{A,A'\in \cS_r}(-1)^{\sum_{i\in A}a_i''+c_i''}\zeta_{AB}& \prod_{i \in A'}\left(\sum_{a_i,c_i}X^{r_i,s_i}_{a_i,c_i}\frac{S_{k_i, k_i+2; a_i, c_i}}{S_{0,0; a_i, c_i}} S_{a_i,c_i; a_i', c_i'}S^*_{a_i,c_i; a_i'', c_i''}\right) \times \\ & \prod_{i \in A'^{\perp}}\left(\sum_{a_i,c_i}X^{r_i,s_i}_{a_i,c_i} S_{a_i,c_i; a_i', c_i'}S^*_{a_i,c_i; a_i'', c_i''} \right)\, . 
    \end{split}
\end{equation}
Now, setting
\begin{equation}
    X^{r_i,s_i}_{a_i,c_i} = \frac{S_{r_i,s_i; a_i,c_i}}{S_{00;a_i,c_i}}
\end{equation}
and using that $X^{r_i,s_i}_{a_i,c_i}$ is a one-dimensional representation of the fusion ring\footnote{Concretely
\begin{equation}
    \frac{S_{r,s; a,c}}{S_{00; a,c}}\frac{S_{r',s'; a,c}}{S_{00; a,c}} = \sum_{(r'',s'') \in Q_k}N_{rs; r's'}^{r''s''}\frac{S_{r'',s''; a,c}}{S_{00; a,c}}\, .
\end{equation}
In the case at hand
\begin{equation}
    \frac{S_{r,s; a,c}}{S_{00; a,c}}\frac{S_{k,k+2; a,c}}{S_{00; a,c}} = \frac{S_{k-r,s+k+2; a,c}}{S_{00; a,c}}
\end{equation}
which is equivalent to the statement that the line $L_{k, k+2}$ is invertible in the fusion ring of a single minimal model.}
we find the multiplicities
\begin{equation}\label{eq:multiGSO}
    \begin{split}
        \frac{1}{2^{r-1}}\left(\sum_{A\in \cS_r}(-1)^{\sum_{i\in A}a_i''+c_i''}\zeta_{AB}\right)\sum_{A'\in \cS_r} \prod_{i \in A'^{\perp}}N_{r_i, s_i; a_i',c_i'}^{a_i'' c_i''}\prod_{i \in A'}N_{k_i-r_i, s_i+k_i+2; a_i',c_i'}^{a_i'' c_i''}\, .
    \end{split}
\end{equation}
The factor
\begin{equation}
    \frac{1}{2^{r-1}}\left(\sum_{A}(-1)^{\sum_{i\in A\in \cS_r}a_i''+c_i''}\zeta_{AB}\right)
\end{equation}
is again a projector, and the whole expression is always a positive integer. Clearly for $\zeta_{AB} = 1$ and $r_i=s_i=0$ we get back the original partition function. We then can conclude that the GSO-projected theory has TDLs $L_{\{r_i,s_i\}, B}$ acting as
\begin{equation}\label{eq:actionTDLGSOproj}
     L_{\{r_i,s_i\}, B}\Phi_{\{a_i, c_i\}, A} = \left(\zeta_{AB}\prod_{i=1}^{r}\frac{S_{r_i,s_i; a_i,c_i}}{S_{00;a_i,c_i}}\right)\Phi_{\{a_i, c_i\}, A}\, ,
\end{equation}
however because of the NS/R alignement condition on physical primaries the parametrization above is redundant. Namely a line with labels $\{r_i, s_i\}$ and one in which we replace $(r_j, s_j)\mapsto (k_j-r_j, s_j+k_j+2)$ for any even number of values of $j =1, \ldots, r$ act in the same way on physical primaries and can be identified. Nevertheless, because of the presence of the quantum symmetry, the total number of faithfully acting lines is preserved by the orbifold, and still equals \eqref{eq:totlines}. We shall comment further on these redundancies later.

Among the TDLs we found there is also the $\bZ_{H}$ group that we have to gauge to obtain the Gepner model. Indeed setting $(r_i, s_i) =(0,2)$ for every $i=1, \ldots r$ as well as $B= \varnothing$ we obtain a line $L_{\{0, 2\}, \varnothing}$ acting as
\begin{equation}
    L_{\{0, 2\}, \varnothing} \Phi_{\{a_i,c_i\}, A} = e^{2 i \pi \sum_{i=1}^{r}\frac{c_i}{k_i+2}}\Phi_{\{a_i,c_i\}, A}\, . 
\end{equation}
The final step is then to gauge this symmetry. We start by letting a line $ L_{\{0, 2s\}, \varnothing}$ act on the circle Hilbert space, namely we insert it along the space cycle of the torus. This gives us the twined partition function
\begin{equation}
\begin{split}
     Z_{\text{GSO}}(\tau, z, s) = & \sum_{A; \{a_i,c_i\}}P^{\text{GSO}}_{\{a_i,c_i\}}e^{2 i \pi s \sum_{i=1}^{r}\frac{c_i}{k_i+2}}\times \\ & \times \prod_{i\in A^{\perp}}\chi_{a_i,c_i}(q,y)\chi_{a_i,c_i}(\bar{q},\bar{y})\prod_{i\in A}\chi_{a_i,c_i}(q,y)\chi_{k_i-a_i,c_i+k_i+2}(\bar{q},\bar{y})
\end{split}
\end{equation}
where $s \in \bZ_{H}$ is the extra fugacity. By means of an $S$ transformation on the expression above, or directly employing the multiplicities \eqref{eq:multiGSO}, we obtain the twisted partition functions
\begin{equation}
\begin{split}
     Z_{\text{GSO}, x}(\tau, z) = & \sum_{A; \{a_i,c_i\}}P^{\text{GSO}}_{\{a_i,c_i\}}\times \\ & \times \prod_{i\in A^{\perp}}\chi_{a_i,c_i}(q,y)\chi_{a_i,c_i+2x}(\bar{q},\bar{y})\prod_{i\in A}\chi_{a_i,c_i}(q,y)\chi_{k_i-a_i,c_i+k_i+2+2x}(\bar{q},\bar{y})\, ,
\end{split}
\end{equation}
with $x \in \bZ_{H}$. To combine both the twisting and twining operations we need to know how the symmetry acts on twisted sectors. For invertible abelian symmetries however this action is already completely encoded in the modular transformations. Indeed the $T$ transformation mixes the two cycles of the torus, and acting with it on $Z_{\text{GSO}, x}$ twines the twisted partition function. Since $T$ is diagonal on the half-characters the result of its action is the phase
\begin{equation}
    \exp{\left[2 i \pi \left(\sum_{i\in A^{\perp}}h_{a_i, c_i}-h_{a_i,c_i+2x}+ \sum_{i\in A}h_{a_i, c_i}-h_{k_i-a_i,c_i+2x+k_i+2}\right)\right]}
\end{equation}
multiplying the characters. From the explicit expression of the weights we have
\begin{equation}
    \begin{split}
        &h_{a_i, c_i}-h_{a_i,c_i+2x} = x\frac{c_i+x}{k_i+2}\,\,  \bmod \, 1\, ,  \\ &h_{a_i, c_i}-h_{k_i-a_i,c_i+2x+k_i+2} = \frac{a_i+c_i+1}{2}+ x\frac{c_i+x}{k_i+2}\,\,  \bmod \, 1\, . 
    \end{split}
\end{equation}
Using the NS/R alignement constraint, the Calabi-Yau condition \eqref{eq:CYcond} and the fact that the order of $A$ is always even the phase simplifies to
\begin{equation}
    e^{2 i \pi x \sum_{i=1}^{r}\frac{c_i}{k_i+2}}\, .
\end{equation}
Acting multiple times with $T$ we obtain power of this phase, therefore the twisted and twined partition function is
\begin{equation}
\begin{split}
     Z_{\text{GSO}, x}(\tau, z,s) = & \sum_{A\in \cS_r; \{a_i,c_i\}}P^{\text{GSO}}_{\{a_i,c_i\}}e^{2 i \pi s \sum_{i=1}^{r}\frac{c_i}{ k_i+2}}\times \\ & \times \prod_{i\in A^{\perp}}\chi_{a_i,c_i}(q,y)\chi_{a_i,c_i+2x}(\bar{q},\bar{y})\prod_{i\in A}\chi_{a_i,c_i}(q,y)\chi_{k_i-a_i,c_i+k_i+2+2x}(\bar{q},\bar{y})\, .
\end{split}
\end{equation}
The Gepner model partition function is then obtained summing over $x, s \in \bZ_{H}$
\begin{equation}
    Z_{\text{Gep}}(\tau, z) = \frac{1}{H}\sum_{s,x \in \bZ_{H}} Z_{\text{GSO}, x}(\tau, z,s)
\end{equation}
The sum over $s$ produces a projector
\begin{equation}
   P^{\bZ_H}_{\{a_i,c_i\}} =  \frac{1}{H}\sum_{s\in \bZ_{H}}e^{2 i \pi s \sum_{i=1}^{r}\frac{c_i}{k_i+2}} = \delta\left(\sum_{i=1}^{r}\frac{c_i}{k_i+2} = 0 \bmod 1 \right)
\end{equation}
we might then express the total partition function as a sum over the twisted sector contributions
\begin{equation}\label{eq:ZGep}
        Z_{\text{Gep}}(\tau, z) = \sum_{x \in \bZ_{H}; A\in \cS_r}Z_{x, A}(\tau, z)
\end{equation}
with
\begin{equation}
   \begin{split}
        Z_{x,A}(\tau, z) = &\sum_{\{a_i, c_i\}} P^{\text{GSO}}_{\{a_i,c_i\}}  P^{\bZ_H}_{\{a_i,c_i\}} \times \\ & \times  \prod_{i\in A^{\perp}}\chi_{a_i,c_i}(q,y)\chi_{a_i,c_i+2x}(\bar{q},\bar{y})\prod_{i\in A}\chi_{a_i,c_i}(q,y)\chi_{k_i-a_i,c_i+k_i+2+2x}(\bar{q},\bar{y})\, .
   \end{split}
\end{equation}
This is the partition function of the Gepner model. 

Another method to construct Gepner models is via simple current extensions \cite{Fuchs:2000gv, Blumenhagen:2013fgp}. In particular starting from the tensor product of the GSO-projected minimal models \eqref{eq:naivediag} the simple current of interest are
\begin{equation}
\begin{split}
     &J_{i} = \Phi_{k_1, k_1+2} \otimes \unit \ldots \otimes \Phi_{k_i, k_i+2} \otimes \ldots \unit \qquad i=2,\ldots r \\
     &J_{\text{orb}} = \Phi_{k_1, k_1+4}\Phi_{k_1,k_1+2}^{3(r-2)} \bigotimes_{i=2}^{r}\Phi_{k_i, k_i+4}\, .
\end{split}
\end{equation}
Extending the diagonal theory by these currents means to gauge the corresponding Verlinde lines. The $J_{i}$ correspond to the $\bZ_{2}^{r-1}$ group imposing the GSO projection, while $J_{\text{orb}}$ corresponds to a line acting as
\begin{equation}
   \Phi_{\{a_i,c_i\}}\mapsto (-1)^{3r(a_1+c_1)}(-1)^{\sum_{i=1}^{r}(a_i+c_i)}e^{2i \pi \sum_{i}\frac{c_i}{k_i+2}}\Phi_{\{a_i,c_i\}}\, .
\end{equation}
In the theory before NS/R alignement this is not a $\bZ_{H}$ action, rather its order depends on the various levels $k_i$ \cite{Fuchs:2000gv}. However proceeding in steps as we did and performing first the $\bZ_{2}^{r-1}$ gauging it is immediate to check that, on physical primaries of the NS/R aligned theory, the action above turns exactly in the $\bZ_{H}$ we have considered. Another way to state this is that the identification of the $\bZ_{H}$ in the theory \eqref{eq:naivediag} is ambiguous because of the redundancies we discussed around \eqref{eq:actionTDLGSOproj}. The fact that the action of $J_{\text{orb}}$ and $\bZ_{H}$ coincide after GSO projection precisely means that the two are identified in the aligned theory.

\subsection{Spectrum and Exactly Marginal Deformations}

By inspection of the partition function we read off the circle Hilbert space
\begin{equation}
\begin{split}
    &\bH = \bigoplus_{x \in \bZ_H; A}\bH^{(x,A)}\\
    & \bH^{(x,A)} = \bigoplus_{\{a_i,c_i\}}\bigotimes_{i \in A^{\perp}}\bH_{a_i,c_i} \otimes \overline{\bH}_{a_i,c_i+2x} \bigotimes_{i \in A}\bH_{a_i,c_i} \otimes \overline{\bH}_{k_i-a_i,c_i+k_i+2+2x}
\end{split}
\end{equation} 
subject to the conditions
\begin{equation}
    \sum_{i}\frac{c_i}{k_i+2} = 0 \, \bmod \, 1\, ,  \qquad [a_1+c_1]=[a_2+c_2]=\ldots = [a_r+c_r]\, .
\end{equation}
The first condition ensures that, for all states at the Gepner point, both the left and right $U(1)_R$ charges are integer for states coming from the NS sector (and half-integer for R sector states). This integrality condition is crucial in string compactification as it ensures spacetime supersymmetry.  We start by noticing that this spectrum contains a single spectral flow operator. The vacuum state is unique and the spectral flow operator is obtained acting on it with $1/2$ units of spectral flow. Clearly doing this only on a subsets of the minimal models violates NS/R alignement, only acting on all minimal models simultaneously we get a state that's still in the spectrum. The spectral flow operator corresponds to the state
\begin{equation}
    \bigotimes_{i=1}^{r}\ket{0, 1}\otimes \overline{\ket{0, 1}} \in \bH^{0, \varnothing}
\end{equation}
which satisfies the constraints and has 
\begin{equation}
    h_L=h_R = \frac{c}{24}= \frac{r-2}{8} \qquad q_L=q_R = -\frac{c}{6}= -\frac{r-2}{2}\, . 
\end{equation}
We can also spectral flow by $-1/2$ unit obtaining a state with $h_L=h_R = (r-2)/8$ and $q_L=q_R=(r-2)/2$. Similarly with a $\pm 1$ unit of spectral flow we obtain states with $h_L=h_R=(r-2)/2$ and $q_L=q_R= \pm (r-2)$. The CY sigma model has an (extended) $\cN=(2,2)$ algebra, which at the Gepner point, is realized as the diagonal subalgebra of the tensor product of algebras of the single minimal models. Note that $h_L-h_R \in \bZ$ consistently with $T$ invariance of the partition function and that $q_L, q_R \in \frac{1}{2}\bZ$, with the half-integer charge state being those in the R sector. 

We are interested in the states corresponding to exactly marginal deformations, as those allow to probe the moduli space of the Calabi-Yau manifold. In order to preserve $\cN=2$ supersymmetry the deformation has to be BPS. The condition of marginality selects a subset of the BPS operators, the requirement being that the deformation preserves both the holomorphic and antiholomorphic $R$-symmetries. It turns out that the superconformal Ward identities imply that any $\cN=2$-supersymmetric marginal deformation is exactly marginal \cite{Yin:2018DH, Greene:1996cy}. Therefore, modulo complex conjugation, there are two classes of operators to consider:
\begin{itemize}
    \item Given a chiral-chiral primary $\phi(z,\bar{z})$ with $q_L = q_R = 1$ its descendant
    \begin{equation}\left(\overline{G}^{-}_{-1/2}G^{-}_{-1/2}\phi\right)(w,\bar{w})
    \end{equation}
   has $h_L= h_R=1$ and $q_L= q_R = 0$.
    \item Given an antichiral-chiral primary $\phi(z,\bar{z})$ with $q_L = -q_R = 1$ its descendant
    \begin{equation}
    \left(G^{+}_{-1/2}\overline{G}^{-}_{-1/2}\phi\right)(w,\bar{w})
    \end{equation}
    has $h_L= h_R=1$ and $q_L= q_R = 0$.
\end{itemize}
In both those cases, in order to have a real deformation one has to add the complex conjugate field given by (a descendant of) the antichiral-antichiral or chiral-antichiral primary. In terms of representations of the bosonic subalgebra notice that the exactly marginal deformation has the same total fermion parity of its primary state, thus to detect its presence in the spectrum it is enough to find the corresponding primary.

The BPS spectrum of the Gepner model, in the NS sector, consists of the four chiral rings. A primary of the Gepner model is chiral or antichiral if it is so under every $\cN=2$ subalgebra of to the single minimal models\footnote{To see this we consider the generator of the diagonal subalgebra
\begin{equation}
    G^{\pm}_{-s} = \sum_{i=1}^r\unit \otimes \ldots \otimes G^{(i); \pm}_{-s}\otimes \ldots \unit
\end{equation}
and let it act on a tensor product state
\begin{equation}
    G^{\pm}_{-s}\bigotimes_{i=1}^r \ket{\phi_i}
\end{equation}
assuming that the $\phi_i$ are orthonormal we see that the norm of such descendant state is
\begin{equation}
    \left|\left|  G^{\pm}_{-s}\bigotimes_{i=1}^r \ket{\phi_i}\right|\right| ^{2} = \sum_{i=1}^r  \left|\left|  G^{(i);\pm}_{-s}\ket{\phi_i}\right|\right| ^{2}\, . 
\end{equation}
Then the total state is annihilated if and only if all its components are.}. Let's then take the generic state in a sector with both $A,A^{\perp}\neq \varnothing$, requiring the holomorphic side to be chiral or antichiral we need to set $c_i= \pm a_i$ for all $i$. Requiring the antiholomorphic side to be of the same type as the holomorphic part we would need $c_i + 2x = \pm a_i$ for $i \in A^{\perp}$ and $c_i+k_i+2+2x=\pm (k_i-a_i)$ for $i \in A$. Clearly this sets $x=0$ and the remaining equation is
\begin{equation}
    \pm a_i +k_i+2 = \pm (k_i-a_i) \bmod 2(k_i+2)\quad \Rightarrow \quad 2 a_i + 2 = 0 \bmod 2(k_i+2)
\end{equation}
which has no solution in the range $a_i = 0,\ldots k_i$. The first extremal case to consider, that is present for any $r$, is $A = \varnothing$. Here it is easy to find the BPS states
\begin{equation}
    \bigotimes_{i=1}^r\ket{a_i,\pm a_i}\otimes \overline{\ket{a_i, \pm a_i}} \in \bH^{(0, \varnothing)}\, ,
\end{equation}
of course only those with integral $U(1)_R$ charges survive the projection. The other extremal case to consider is when $r$ is even and $A= \{1,\ldots r\}$. Here again $c_i = \pm a_i$ for all $i$, but we are no longer forced to set $x=0$, rather we need to solve the equations
\begin{equation}
    2(a_i+1 \pm x) = 0 \bmod 2(k_i+2)  \quad \forall i =1, \ldots r \,
\end{equation}
which fix 
\begin{equation}
\begin{split}
      & a_i = k_i +1 - x \, ,\\
      & a_i = x-1\, . 
\end{split}
\end{equation}
for $+$ and $-$ signs respectively. In the range $a_i= 0, \ldots, k_i$ we have $1\le s \le \text{min}(k_i)+1$. Therefore for $r$ even we find two extra conjugate BPS states for each non-zero $x \in \bZ_H$
\begin{equation}
\begin{split}
    &\bigotimes_{i=1}^{r} \ket{k_i+1-x, k_i+1-x} \otimes \overline{\ket{x-1,x-1}} \in \bH^{(s,\{1,\ldots r\})}\, \\
    &\bigotimes_{i=1}^{r} \ket{x-1,-x+1} \otimes \overline{\ket{k_i+1-x,-k_i-1+x}} \in \bH^{(x,\{1,\ldots r\})}\,
\end{split}
\end{equation}
which also satisfy the charge constraint. We conclude that the chiral-chiral or antichiral-antichiral states can only belong to the untwisted sector and, for $r$ even to the twisted sectors with $A= \{1,\ldots, r\}$. Marginal deformations correspond to those states with $|q_L|= |q_R|=1$, in the untwisted sector this constraint is
\begin{equation}
    \sum_{i=1}^{r}\frac{a_i}{k_i+2} = 1
\end{equation}
due to the CY condition the state with $a_i= 1$ for all $i$ are always a solution. The chiral-chiral and antichiral-antichiral states in $\bH^{(s, \{1,\ldots,r\})}$ have charges
\begin{equation}
\begin{split}
     & q_L = \sum_{i=1}^{r}\frac{k_i+1-x}{k_i+2} = \frac{c}{3} + 1 - x = r - 1 - x\, ;\,  \quad q_R = x-1 \\
     & q_L = 1 - x\, ;\,  \quad q_R =  - r + 1 + x\,  \\
\end{split}
\end{equation}
respectively. Those giving rise to a marginal deformation have
\begin{equation}
   r-1-x = x-1 = 1 \iff x=2\, \, , r = 4\, \, .
\end{equation}
We can repeat the analysis for the chiral-antichiral states and their conjugates. For $A,A^{\perp}\neq \varnothing$ the equations are
\begin{equation}
    \begin{split}
       & c_i = \pm a_i \, \, , \, \, \forall i \\
       & c_i + 2x = \mp a_i \, \, , \, \, i \in A^{\perp}\\ 
       & c_i + k_i + 2 + 2x = \mp (k_i-a_i)\,\, , \, \,  i \in A\, 
    \end{split}
\end{equation}
and one easily sees that there is no admissible solution. We again consider first the extremal case $A = \varnothing$, here we do not have to impose the third equation above, therefore we have
\begin{equation}
    c_i = \pm a_i \qquad  x= \mp a_i
\end{equation}
hence $a_i=a$ for all $i$. The corresponding states are
\begin{equation}
    \bigotimes_{i=1}^{r}\ket{a,\pm a} \otimes \overline{\ket{a, \mp a} } \in \bH^{(\mp a, \varnothing)}\, .
\end{equation}
The other extremal case is again $A = \{1,\ldots,r\}$ for $r$ even. Here we have $c_i = \pm a_i$ and $x = \pm 1$, the states are
\begin{equation}
    \bigotimes_{i=1}^{r}\ket{a_i, \pm a_i}\otimes\overline{\ket{k_i-a_i, \mp(k_i-a_i)}}\in \bH^{(\pm 1, \{1,\ldots, r\})}\, . 
\end{equation}
Thus we have an antichiral-chiral BPS state for every $\bH^{(x,\varnothing)}$ with appropriate $x$ and, if $r$ is even, we have more coming from the twisted sector $\bH^{(\pm 1, \{1,\ldots r\})}$. Among those in the $\bH^{(x,\varnothing)}$ sectors only for $x=\pm 1$ we have a marginal deformation. Instead the conditions on the $R$ charge of the states showing up for $r$ even are
\begin{equation}
   \sum_{i=1}^{r}\frac{a_i}{k_i+2} = 1\, ,  \qquad \sum_{i=1}^{r}\frac{k_i-a_i}{k_i+2}= 1 
\end{equation}
if the first condition is met the second requires
\begin{equation}
   1= \sum_{i=1}^{r}\frac{k_i-a_i}{k_i+2}= r-3
\end{equation}
i.e $r = 4$. Summarizing, the marginal deformations are:
\begin{itemize}
    \item chiral-chiral and antichiral-antichiral states. For any $r$ 
    \begin{equation}
        \bigotimes_{i=1}^r\ket{a_i,\pm a_i}\otimes \overline{\ket{a_i, \pm a_i}} \in \bH^{(0, \varnothing)} \qquad \sum_{i=1}^{r}\frac{a_i}{k_i+2} = 1\, . 
    \end{equation}
    For $r=4$ and $H >2$ we also have
    \begin{equation}
    \begin{split}
         & \bigotimes_{i=1}^4\ket{k_i-1,k_i-1}\otimes \overline{\ket{1,1}} \in \bH^{(2, \{1,\ldots,4\})} \\
         & \bigotimes_{i=1}^4\ket{1,-1}\otimes \overline{\ket{k_i-1,1-k_i}} \in \bH^{(2, \{1,\ldots,4\})} \, . 
    \end{split}
    \end{equation}
    \item antichiral-chiral and chiral-antichiral states. For any $r$ there's only one with the correct $R$-charges
    \begin{equation}
        \bigotimes_{i=1}^r\ket{1,\pm 1}\otimes \overline{\ket{1,\mp 1}} \in \bH^{(\mp 1, \varnothing)}. 
    \end{equation}
    For $r=4$ and $H >2$ we also have
    \begin{equation}
        \bigotimes_{i=1}^4\ket{a_i,\pm a_i}\otimes \overline{\ket{k_i-a_i,\mp (k_i-a_i)}} \in \bH^{(\pm 1, \{1,\ldots,4\})}\, ,  \qquad \sum_{i=1}^{r}\frac{a_i}{k_i+2}= 1\, . 
    \end{equation}
\end{itemize}
The case $r=4$ is evidently special as it admits particular marginal deformations. The theory has $c= 3(r-2) = 6$, corresponding to a sigma model on a $K3$ surface. We notice that for $r=4$ we have the same number of chiral-chiral and antichiral-chiral marginal deformations. It is also known that there is an enhancement of supersymmetry and the theory enjoys an $\cN=(4,4)$ superconformal algebra, therefore the marginal deformations moving along the moduli space should preserve the full supersymmetry. Indeed the degeneracy between chiral-chiral and antichiral-chiral states meets the expectation that $\cN=2$ BPS states pair up to form a BPS multiplet for the larger algebra. For $r=4$ we also have the states
\begin{equation}
    \bigotimes_{i=1}^{r}\ket{k_i, k_i}\otimes \overline{\ket{0,0}} \in \bH^{(1, \{1,\ldots 4\})} \qquad \bigotimes_{i=1}^{r}\ket{k_i, -k_i}\otimes \overline{\ket{0,0}} \in \bH^{(-1, \{1,\ldots 4\})}
\end{equation}
which correspond to holomorphic operators with $h_L=1$ and $q_L = \pm 2$. Similar states exist also for the antiholomorphic side. Together with the $R$-symmetry generator we have three currents that transform in the adjoint of $SU(2)_R$, the $R$-symmetry of the $\cN=4$ algebra. It is known, see e.g. \cite{Distler:1992gi}, that for any CY sigma model the chiral algebra is extended, for K3 the resulting symmetry is $\cN=4$ supersymmetry, while in general the algebra is the $\cN=2$ one extended by the square of the spectral flow operator \cite{Odake:1989dm}. One could then expect that also when $r$ is odd new holomorphic and antiholomorphic states corresponding to this operator show up in the spectrum, this however is not the case. Indeed if the complex dimension of the CY is odd this operator is a fermion with half-integer weight, therefore, due to the GSO projection, it cannot appear as a purely left-moving state, rather it can only appear tensored with a right-moving fermion. When $r$ is even instead these operators are bosonic, and appear tensored with the identity in the antiholomorphic sector precisely in the twisted sectors with $A = \{1,\ldots,r\}$.

\subsection{Topological Defect Lines and $3d$ TQFT}

We now investigate the symmetries of the model. Again our strategy is to bootstrap the action of line operators on physical primaries imposing consistency of the twisted Hilbert spaces. The physical primaries appearing in \eqref{eq:ZGep} can be labelled as $\Phi_{\{a_i,c_i\}, A, x}$ with the set of labels $\{a_i, c_i\}$ subject to 
\begin{equation}
    \sum_{i}\frac{c_i}{k_i+2} = 0 \, \bmod \, 1\, ,  \qquad [a_1+c_1]=[a_2+c_2]=\ldots = [a_r+c_r]\, .
\end{equation}
Since we obtained the Gepner model as a $\bZ_{H}$ orbifold of the GSO-projected theory we can factor out the dual $\bZ_{H}$ symmetry in our ansatz, we set
\begin{equation}
    \cL_{\{r_i, s_i\}, B, \eta}\Phi_{\{a_i,c_i\}, A, x} = \zeta_{AB} e^{2 i \pi \frac{\eta x}{H}}\left(\prod_{i=1}^{r}X^{a_i, c_i}_{r_i,s_i}\right)\Phi_{\{a_i,c_i\}, A, x}\, . 
\end{equation}
Inserting $ \cL_{\{r_i, s_i\}, B, \eta}$ as an operator acting on the Hilbert space and acting with an $S$ transformation we obtain the multiplicities in the twisted sector
\begin{equation}\label{eq:multGep}
    \begin{split}
       & \sum_{x\in \bZ_H; A\in \cS_r}\zeta_{AB}e^{2i \pi \frac{\eta x}{H}}\sum_{\{a_i, c_i\}}P^{\text{GSO}}_{\{a_i, c_i\}}P^{\bZ_{H}}_{\{a_i, c_i\}} \times \\ & \times \prod_{i\in A^{\perp}}X^{a_i, c_i}_{r_i,s_i}S_{a_ic_i; a_i'c_i'}S_{a_i,c_i+2x; a_i'',c_i''}^*\prod_{i\in A}X^{a_i, c_i}_{r_i,s_i}S_{a_ic_i; a_i'c_i'}S_{k_i-a_i, c_i+k_i+2+2x;a_i'',c_i''}^* \, .
    \end{split}
\end{equation}
To rewrite this in a more manageable form we use
\begin{equation}
    \begin{split}
        &S_{k-a, c+k+2+2x; a', c'} = (-1)^{a'+c'}e^{2i \pi \frac{x c'}{k+2}}S_{ac;a'c'} \\
        & P^{\bZ_{H}}_{\{a_i, c_i\}} = \frac{1}{H}\sum_{s\in \bZ_{H}}e^{2 i \pi s \sum_{i=1}^{r}\frac{c_i}{k_i+2}} = \frac{1}{H}\sum_{s\in \bZ_{H}}\prod_{i=1}^{r}\frac{S_{0,2s;a_ic_i}}{S_{00;a_ic_i}}
    \end{split}
\end{equation}
and also \eqref{eq:PGSO} for the GSO projector. We get
\begin{equation}
    \begin{split}
        & \eqref{eq:multGep} = \frac{1}{2^{r-1}H}\sum_{x\in \bZ_H; A}\zeta_{AB}(-1)^{\sum_{i \in A}a_i''+c_i''}e^{2i \pi x\left(\frac{\eta }{H}+ \sum_{i=1}^{r}\frac{c_i''}{k_i+2}\right)} \\ & \sum_{\substack{\{a_i, c_i\};\\ A', s\in \bZ_H}}\prod_{i\in A'^{\perp}}X^{a_i, c_i}_{r_i,s_i}\frac{S_{0,2s;a_ic_i}}{S_{00;a_ic_i}}S_{a_ic_i; a_i'c_i'}S_{a_i,c_i; a_i'',c_i''}^*\prod_{i\in A'}X^{a_i, c_i}_{r_i,s_i} \frac{S_{0,2s;a_ic_i}}{S_{00;a_ic_i}}\frac{S_{k_i, k_i+2; a_i, c_i}}{S_{0,0; a_i, c_i}} S_{a_ic_i; a_i'c_i'}S_{a_i, c_i;a_i'',c_i''}^* \, .
    \end{split}
\end{equation}
Again we see that there is a natural solution
\begin{equation}
    X^{a_i, c_i}_{r_i,s_i} = \frac{S_{r_i, s_i; a_i, c_i}}{S_{00; a_i, c_i}}
\end{equation}
that gives the integer multiplicities
\begin{equation}
 \begin{split}
        & N^{\text{Gep} \, \{a_i'', c_i''\}}_{\{r_i, s_i\}, \{a_i', c_i'\}} = \frac{1}{2^{r-1}H}\sum_{x\in \bZ_H; A\in \cS_r}\zeta_{AB}(-1)^{\sum_{i \in A}a_i''+c_i''}e^{2i \pi x\left(\frac{\eta }{H}+ \sum_{i=1}^{r}\frac{c_i''}{k_i+2}\right)} \\ & \sum_{ A'\in \cS_r, s\in \bZ_H}\prod_{i\in A'^{\perp}}N_{r_i, s_i+2s; a_i', c_i'}^{a_i'', c_i''}\prod_{i\in A'}N_{k_i-r_i, s_i+k_i+2+2s; a_i', c_i'}^{a_i'', c_i''} \, .
    \end{split}
\end{equation}
We conclude that the Gepner model enjoys line defects $\cL_{\{r_i, s_i\};B, \eta}$ acting as
\begin{equation}\label{eq:actionphysprim}
\begin{split}
    \cL_{\{r_i, s_i\};B, \eta}\Phi_{\{a_i, c_i\}; A, s}& =\zeta_{AB} e^{2 i \pi \frac{\eta x}{H}}\left(\prod_{i=1}^{r}\frac{S_{r_i, s_i; a_i, c_i}}{S_{00; a_i, c_i}}\right)\Phi_{\{a_i,c_i\}, A, x} \\ & = \zeta_{AB}e^{i \pi\left( \frac{2 \eta x}{H}+ \sum_{i} \frac{s_i c_i}{k_i+2} - \frac{[a_1+c_1][r_i+s_i]}{2}\right)}\prod_{i=1}^{r}\frac{\sin\left(\frac{\pi(r_i+1)(a_i+1)}{k_i+2}\right)}{\sin\left(\frac{\pi(a_i+1)}{k_i+2}\right)}\Phi_{\{a_i,c_i\}, A, x}\, . 
\end{split}    
\end{equation}
By our analysis of the lines of a single minimal model we see that any line for which at least one $r_i\neq 0, k_i$ is non-invertible. The fusion ring is simple to describe, we have
\begin{equation}
    \cL_{\{r_i, s_i\};B, \eta}\times \cL_{\{r_i', s_i'\};B', \eta'} = \sum_{\{r_i'', s_i''\}} \prod_{i=1}^{r}N_{r_i, s_i; r_i', s_i'}^{r_i''; s_i''}\cL_{\{r_i'', s_i''\};BB', \eta+ \eta'}
\end{equation}
with the dual symmetry labels following a group law. The remarks concerning the redundancy of our parametrization apply also in this case. Namely the set of labels of faithfully acting lines is the quotient of $\bigoplus_{i}Q_{k_i}$, where the $2r$-tuple $\{r_i,s_i\}$ takes values, with respect to the equivalence relations
\begin{equation}\label{eq:lineequiv}
\begin{split}
     &(r_j, s_j)\sim (k_j-r_j, s_j+k_j+2) \qquad \forall j \in A\, , \, A \in \cS_r \\
    & (r_i, s_i)\sim (r_i, s_i+2)\qquad \forall i=1,\ldots,r\, .
\end{split}
\end{equation}
Of course, accounting for the dual symmetry labels, the total number of lines still equals \eqref{eq:totlines}. If we identify these lines with those of the theory prior to GSO projection and orbifold the equivalences above derive by fusion with the gauged lines, which are invisible in the Gepner model. Both the multiplicities $N^{\text{Gep} \, \{a_i'', c_i''\}}_{\{r_i, s_i\}, \{a_i', c_i'\}}$ and the fusion coefficients are well defined on the quotient. In the multiplicities of the twisted sectors this is guaranteed by the sum over $s \in \bZ_{H}$ and $A' \in \cS_r$, as changing the representative of the line only reshuffles the terms in the sums. For the fusion coefficients, since the identifications follow from fusing with invertible lines, changing representatives of the lines on the left hand side does not affect the fusion coefficients or the equivalence class of the result of the fusion.

Also in this case we can give a $3$-dimensional description of this symmetry. Since we only performed orbifolds in $2d$ the $3d$ TQFT corresponding to the Gepner model is the same one corresponding to the product of GSO-projected minimal model. This is just the Chern-Simons theory with gauge group
\begin{equation}
    G_{\text{Gep}} = G_{k_1}\times \ldots \times G_{k_r} \times  G_{-k_1}\times \ldots \times G_{-k_r}
\end{equation}
with $G_{k_i}$ as in \eqref{eq:gaugegroupCS}. The MTC data of this TQFT can be computed from those of a single factor. In particular anyons are labelled, in our choice of gauge, by a $4r$-tuple $\{(r_i, s_i); (\bar{r_i}, \bar{s_i})\}$ with $(r_i, s_i), (\bar{r_i}, \bar{s_i}) \in Q_{k_i}$ and their fusion is 
\begin{equation}
    \cL_{\{(r_i, s_i); (\bar{r_i}, \bar{s_i})\} }\times \cL_{\{(r'_i, s'_i); (\bar{r_i}', \bar{s_i}')\}} = \sum_{\{(r_i'', s_i''); (\bar{r_i}'', \bar{s_i}'')\}}\prod_{i=1}^{r}N_{r_i, s_i; r_i', s_i'}^{r_i'', s_i''}N_{\bar{r_i}, \bar{s_i}; \bar{r_i}', \bar{s_i}'}^{\bar{r_i}'', \bar{s_i}''}\cL_{\{(r_i'', s_i''), (\bar{r_i}'', \bar{s_i}'')\}}\, . 
\end{equation}
From the TQFT perspective gauging a 0-form symmetry in the boundary amounts to change topological boundary condition.
The topological boundary condition for the tensor product of GSO-projected minimal models is the diagonal one, corresponding to the diagonal lagrangian algebra $\bigoplus_{\{(r_i, s_i)\}}\cL_{\{(r_i, s_i); (r_i, s_i)\}}$. In general we can read off the lagrangian algebra corresponding to a gapped boundary directly from the torus partition function. Consider the $3d$ TQFT on a solid torus with an inner and an outer torus boundaries, on the outer boundary we impose the conformal boundary condition, while on the inner one we set the topological boundary condition. Evaluating the path integral of the TQFT in this configuration produces the torus partition function of the CFT. Now, shrinking the inner torus leaves behind the lagrangian algebra corresponding to the chosen boundary condition, which we can write out as a sum of anyons, possibly with multiplicities. Since the path integral on a solid torus with the insertion of an anyon produces a character of the chiral algebra we can extract the anyons of the lagrangian algebra by comparing with the explicit expression of the partition function. For instance the lagrangian algebra imposing the GSO-projection is
\begin{equation}
   \mathfrak{L}_{\text{GSO}}= \bigoplus_{\{(r_i, s_i)\}, A \in \cS_r}P^{\text{GSO}}_{\{r_i, s_i\}}\cL_{\{(r_i, s_i); (r_i, s_i)\}} \times \cL_A
\end{equation}
where $\cL_A$ is the bulk line with $r_i=s_i=0$ $\forall i$ and $\bar{r}_j = k_j, \bar{s_j} = k_j+2$ $\forall j \in A$. And similarly one can write down an expression for the lagrangian algebra corresponding to the $\bZ_H$-orbifolded theory
\begin{equation}
   \mathfrak{L}_{\text{Gep}}= \bigoplus_{\{(r_i, s_i)\}, A \in \cS_r, x \in \bZ_{H}}P^{\text{GSO}}_{\{r_i, s_i\}}P^{\bZ_{H}}_{\{r_i, s_i\}}\cL_{\{(r_i, s_i); (r_i, s_i)\}; A, x} 
\end{equation}
with 
\begin{equation}
    \cL_{\{(r_i, s_i); (r_i, s_i)\}; A, x}=\cL_{\{(r_i, s_i); (r_i, s_i)\}} \times \cL_A \times \cL_{\{(0, 0); (0, 2x)\}}\, .
\end{equation}
Anyons participating in a lagrangian algebra can end on both the conformal and topological boundaries, therefore producing local operators in the CFT.
\begin{figure}[t]
\centering
\begin{tikzpicture}
\filldraw[color=white!90!red, opacity=0.75] (0, 0) -- (0.5, 1.5) -- (0.5, 4.5) -- (0,3) -- cycle;
    \filldraw[color=black, fill=black,] (0.25,2) circle (0.05);
    \draw[color=black, line width=1] (0.25, 2)--(5.25, 2);
    \node[above] at (1.75, 2) {$\cL_{\{(r_i, s_i)\},\{(\bar{r_i}, \bar{s_i})\}; A; x}$};
    \filldraw[color=white!90!green, opacity=0.75] (5, 0) -- (5.5, 1.5) -- (5.5, 4.5) -- (5,3) -- cycle;
    \filldraw[color=black, fill=black,] (5.25,2) circle (0.05);
    \draw[color= blue, line width = 1] (4,2.1) arc [x radius =0.2,y radius = 1,start angle=4,
        end angle=350];
    \node[above, color = blue] at (3.9, 3.1) {$\cL_{\{(\alpha_i,\beta_i); (\{\bar{\alpha_i}, \bar{\beta_i})\}}$};
    \node at (6.5, 2) {$= \gamma$};
    
    \filldraw[color=white!90!red, opacity=0.75] (7, 0)-- (7.5, 1.5) -- (7.5, 4.5) -- (7,3) -- cycle;
    \filldraw[color=black, fill=black,] (7.25,2) circle (0.05);
    \draw[color=black, line width=1] (7.25, 2)--(11.25, 2);
     \node[above] at (9.1, 2) {$\cL_{\{(r_i, s_i)\},\{(\bar{r_i}, \bar{s_i})\}; A; x}$};
    \filldraw[color=white!90!green, opacity=0.75] (11, 0)-- (11.5, 1.5) -- (11.5, 4.5) -- (11,3) -- cycle;
    \filldraw[color=black, fill=black,] (11.25,2) circle (0.05);
    
\end{tikzpicture}
\caption{An anyon of $\mathfrak{\cL}_{\text{Gep}}$ can end on both the (red) topological and the (green) conformal boundaries, upon shrinking of the bulk we obtain a local operator in the CFT. The symmetry of the boundary CFT is captured by linking in the bulk.}
\label{fig:symmact}
\end{figure}
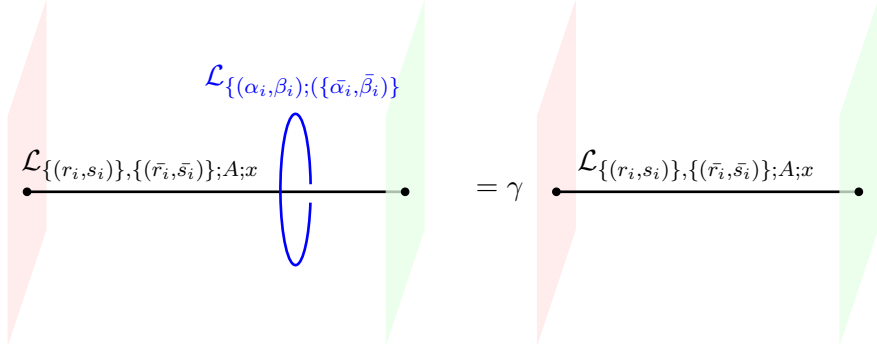

Bulk lines that are not condensed on the topological boundary generate the symmetry of the CFT. In the bulk the symmetry action is detected linking an anyon of the lagrangian algebra with a generic bulk line. In our case, for the gapped boundary corresponding to the Gepner model, we consider the configuration in Fig.\ref{fig:symmact}, then
\begin{equation}\label{eq:3dsymm}
\begin{split}
    &\gamma =\prod_{i=1}^{r}\frac{S_{\alpha_i, \beta_i; r_i, s_i}}{S_{00; r_i, s_i}}\prod_{i\in A^{\perp}}\frac{S_{\bar{\alpha_i}, \bar{\beta_i}; r_i, s_i+2x}^*}{S_{00; r_i, s_i+2x}}\prod_{i\in A}\frac{S_{\bar{\alpha_i}, \bar{\beta_i}; k_i-r_i, s_i+k_i+2+2x}^*}{S_{00;k_i- r_i, s_i+k_i+2+2x}}\\& = e^{-2i \pi x\left(\sum_{i=1}^{r}\frac{\bar{\beta_i}}{k_i+2}\right)}(-1)^{\sum_{i\in A}[\bar{\alpha_i}+\bar{\beta_i}]}\prod_{i=1}^{r}\frac{S_{\alpha_i, \beta_i; r_i,s_i}S_{\bar{\alpha_i},\bar{\beta_i}; r_i,s_i}^*}{S_{00; r_i,s_i}^{2}}\, .
\end{split}    
\end{equation}
In \eqref{eq:3dsymm} the two phases in front of the product give the action of the dual symmetries, with $\eta = H\sum_{i}\bar{\beta_i}/(k_i+2)$  and $\zeta_{AB} = (-1)^{\sum_{i\in A}[\bar{\alpha_i}+\bar{\beta_i}]}$. Due to the boundary condition many bulk lines will be mapped to the same boundary line, hence act on lines in $\mathfrak{L}_{\text{Gep}}$ with the same eigenvalue. As representatives of the faithfully acting lines we can take all those with $\alpha_i = \beta_i=0$ and impose the identifications implied by the projectors $P^{\text{GSO}}$ and $P^{\bZ_{H}}$, then \eqref{eq:3dsymm} matches what we found directly in the CFT.

\section{Symmetric Marginal Deformations and Selection Rules}\label{sec:symloci}

We found a wealth of non-invertible lines at the Gepner point of the Calabi-Yau sigma model. We now want to investigate the existence of continuous families of SCFTs which preserve some subcategory of lines. To this extent we look at marginal deformations invariant under the action of the lines described in the previous section. On general grounds, for an operator $\Phi$ of a $2d$ QFT to be invariant under the action of a TDL $\cL$, the two have to commute \cite{Chang:2018iay}. Shrinking $\cL$ produces a number, then the condition is
\begin{equation}
    \cL \Phi = \langle \cL \rangle \Phi
\end{equation}
where $\langle \cL \rangle$ is the quantum dimension of $\cL$, which can also be seen as the eigenvalue of $\cL$ on the identity operator (we are assuming a CFT with a unique vaccum). Therefore a physical primary $\Phi_{\{a_i, c_i\}, A, x}$ of the Gepner model is invariant under a line $\cL_{\{r_i,s_i\}, B, \eta}$ if
\begin{equation}
    \zeta_{AB}e^{2 i \pi \frac{s \eta}{H}}\prod_{i=1}^{r} \frac{S_{r_i,s_i; a_i,c_i}}{S_{00; a_i,c_i}}=\prod_{i=1}^{r} \frac{S_{r_i,s_i; 00}}{S_{00; 00}}\, . 
\end{equation}
Since we are going to consider only exactly marginal deformations we can take $\Phi_{\{a_i, c_i\}; A, x}$ to be in the NS sector, then the invariance condition is slightly simpler 
\begin{equation}\label{eq:commutationconstr}
    \zeta_{AB}e^{i \pi \left(\frac{2 s \eta}{H} + \sum_i \frac{s_i c_i}{k_i+2}\right)}\prod_{i=1}^{r}\frac{\sin\left(\frac{\pi(r_i+1)(a_i+1)}{k_i+2}\right)}{\sin\left(\frac{\pi(a_i+1)}{k_i+2}\right)} = \prod_{i=1}^{r}\frac{\sin\left(\frac{\pi(r_i+1)}{k_i+2}\right)}{\sin\left(\frac{\pi}{k_i+2}\right)}\, .
\end{equation} 
The operator we add to the action is a descendant of the BPS primaries we discussed in the previous section. To obtain it we act on those primaries with a product of the supercharges $G^{\pm}, \bar{G}^{\pm}$ of the diagonal $\cN=(2,2)$ algebra. In terms of the supercharges of each tensor factor the operators we are interested in are
\begin{equation}\label{eq:diagsuch}
    G^{x}\bar{G}^{y} = \sum_{i,j=1}^{r}\unit \otimes \ldots \otimes  G^{(i); x}  \otimes \ldots \otimes \bar{\unit} \otimes \ldots \otimes \bar{G}^{(j); y}\otimes \ldots \otimes \bar{\unit}\, ,  
\end{equation}
with $x,y = \pm$. As states in the Gepner model Hilbert space these operators correspond to 
\begin{equation}
    \ket{G^{x}\bar{G}^{y}} = \sum_{i,j=1}^{r} \ket{0,0}\otimes \ldots \otimes \ket{k_i, k_i+2} \otimes \ldots \overline{\ket{0,0}} \otimes \ldots \overline{\ket{k_j, k_j+2}}\otimes \ldots \otimes \overline{\ket{0,0}}\, .
\end{equation}
The summands belong to the twisted Hilbert spaces $\bH^{(0, A =\{i,j\})}$ if $i \neq j$, while are in the untwisted Hilbert space if $i=j$. A line $\cL_{\{r_i, s_i\}; B; \eta}$ acting on a component gives
\begin{equation}
\begin{cases}
 &(-1)^{r_i+s_i}\zeta_{\{ij\}, B} \qquad i \neq j\\
 & (-1)^{r_i+s_i} \qquad i = j\, . 
\end{cases}
\end{equation}
A single term in the sum \eqref{eq:diagsuch}, upon acting on the BPS primary, gives raise to an operator participating in the deformation. For each such component the commutation condition is 
\begin{equation}\label{eq:commutationconstrcompo}
    \zeta_{\{i,j\}, B}(-1)^{r_i+s_i}\zeta_{AB}e^{i \pi \left(\frac{2 s \eta}{H} + \sum_l \frac{s_l c_l}{k_l+2}\right)}\prod_{l=1}^{r}\frac{\sin\left(\frac{\pi(r_l+1)(a_l+1)}{k_l+2}\right)}{\sin\left(\frac{\pi(a_l+1)}{k_l+2}\right)} = \prod_{l=1}^{r}\frac{\sin\left(\frac{\pi(r_l+1)}{k_l+2}\right)}{\sin\left(\frac{\pi}{k_l+2}\right)}\, , \quad \forall i, j\, .
\end{equation} 
We can take the line $\cL_{\{r_i, s_i\}, ; B, \eta}$ to commute or anticommute with both the product of supercharges and the BPS primary. Requiring (anti-)commutativity with the supercharges we see that we have to set $(-1)^{r_i+s_i} = \pm 1$ for all $i=1, \ldots, r$ and $\zeta_{\{ij\}, B}= 1$.

Now, solving for the commutativity with the BPS primary in full generality is complicated. In order to continue our analysis we consider specific examples. In the following we shall only focus on chiral-chiral and antichiral-chiral deformations, as we can use charge conjugation to relate those to the antichiral-antichiral and chiral-antichiral ones. In the literature the usual way to denote Gepner models is as $\prod_{y}(k_y)^{m_y}$ where $m_y$ is the number of times $k_y$ appears in the list $(k_1,\ldots,k_r)$, we'll make use of this notation in the remainder of this section.

\subsection{Torus}

The simplest target space we can consider for the superconformal sigma model is the torus. The theory is free, and consists of a complex scalar $X$ parametrizing the torus, a complex left moving fermion $\psi$ and a complex right moving one $\lambda$. The conformal manifold is the Narain moduli space 
\begin{equation}
    \cM = O(2,2,\bZ)\backslash O(2,2, \bR)/O(2)\times O(2)\, . 
\end{equation}
A point in $\cM$ consist of a choice of metric $G_{ij}=G_{ji}$ and $B$-field $B_{ij}=-B_{ji}$ on the target torus, with $i,j=1,2$. The partition function of the theory on a torus worldsheet with modular parameter $\tau$ at a point $m \in \cM$ is the product of the bosonic one and the GSO-projected fermion partition function. 
The $R$-symmetry charges are
\begin{center}
\begin{tabular}{c c c }
 & $q$ & $\bar{q}$ \\ 
  $\psi$ & $1$ & $0$ \\
  $\lambda$ & $0$ & $1$ \\
  $X$ & $0$ & $0$
\end{tabular}
\end{center}
The tangent space of $\cM$ can be probed deforming the action by the operator 
\begin{equation}
        \cO=\left(\delta G_{ij}\delta^{\alpha \beta} + i \delta B_{ij}\epsilon^{\alpha \beta}\right)\partial_{\alpha} X^i\partial_{\beta}X^j
\end{equation}
where we wrote $X = X^1+ i X^2$ and introduced coordinates $\sigma^{\alpha}$ on a flat worldsheet. In terms of $X$ and $z = \sigma^1+ i \sigma^2$ we have
\begin{equation}
    \cO = g \partial X\bar{\partial}X + f \partial X\bar{\partial}\bar{X} + c.c.
\end{equation}
where we introduced the complex parameters
\begin{equation}
    g = \delta G_{11} - \delta G_{22} - 2 i \delta G_{12}\, ,  \qquad    f = \delta G_{11} + \delta G_{22} + 2 i \delta B_{12}\, .
\end{equation}
To compare $\cO$ with the exactly marginal deformations obtained from the $\cN=(2,2)$ algebra we write down the chiral rings using the sigma model fields. We have\footnote{Note that because of the GSO projection the fields $\psi, \lambda$ and their conjugates cannot appear in the partition function on their own, to survive the projection they need to be tensored with another fermionic state on the other holomorphic half.}
\begin{equation}
   \begin{split}
     &\text{(chiral-chiral)} = \{ \unit, \psi, \lambda, \psi \lambda\} \\
     & \text{(antichiral-chiral)}  = \{ \unit, \bar{\psi}, \lambda, \bar{\psi} \lambda\} \\
   \end{split}
\end{equation}
Using the explicit realization of the $\cN=(2,2)$ algebra generators in terms of the sigma model fields one can check that the two marginal operators $\partial X \bar{\partial}X$ and $\partial \bar{X} \bar{\partial}X$ are descendants of the bosonic BPS primaries $\psi \lambda$ and $ \bar{\psi} \lambda$ respectively. In other words $g$ couples to the chiral-chiral deformation while $f$ to the antichiral-chiral one.

On $\cM$ there are three Gepner points: $(1)^{3}$, $(2)^{2}$ or $(1)(4)$, each of which has total central charge $c=3$. The last two cases however, as presented, do not fulfill the CY condition. In particular, for both  $(2)^{2}$ and $(1)(4)$,
\begin{equation}
    \sum_{i=1}^{2}\frac{1}{k_i+2} = \frac{1}{2}\, . 
\end{equation}
We can solve this issue in both cases adding a minimal model with $k=0$. On its own this is the trivial (spin) theory, with a single ground state in both the NS and R sectors with vanishing charges and weights. However its presence is detected by the $\bZ_H$ orbifold: without it we would not project on NS sector states with integer $U(1)_R$ charges. The $(1)^3$ model is the least interesting as the Verlinde lines of the $k=1$ minimal model are all invertible. We shall then mainly focus on the $(2)^2(0)$ case. In this simple case one can check by hand that there is only one chiral-chiral marginal deformation corresponding to the state
\begin{equation}\label{eq:cctorus}
     (\ket{2,2})^{\otimes 2}\otimes\ket{0,0}  \otimes (\overline{\ket{2,2}})^{\otimes 2}\otimes\overline{\ket{0,0}} \in \bH^{(0, \varnothing)}\, . 
\end{equation}
Also there's a unique antichiral-chiral marginal operator corresponding to 
\begin{equation}
    (\ket{1,-1})^{\otimes 2}\otimes\ket{0,0}  \otimes (\overline{\ket{1,1}})^{\otimes 2}\otimes\overline{\ket{0,0}} \in \bH^{(3, \varnothing)}\, .
\end{equation}
Now let's look at what lines are preserved by these deformations.
\begin{itemize}
    \item A line $\cL_{\{r_i, s_i\}; B, \eta}$ (anti)commutes with the chiral-chiral operator if
        \begin{equation}
            e^{i \pi \frac{2s_1+2s_2}{4}}\frac{\sin\left(\frac{3\pi(r_1+1)}{4}\right)\sin\left(\frac{3\pi(r_2+1)}{4}\right)}{\sin\left(\frac{\pi(r_1+1)}{4}\right)\sin\left(\frac{\pi(r_2+1)}{4}\right)} = \pm 1 \, . 
        \end{equation}
        To eliminate the imaginary part we set $2s_1+2s_2 = 0 \bmod 4$, then 
        \begin{equation}
            \frac{\sin\left(\frac{3\pi(r_1+1)}{4}\right)\sin\left(\frac{3\pi(r_2+1)}{4}\right)}{\sin\left(\frac{\pi(r_1+1)}{4}\right)\sin\left(\frac{\pi(r_2+1)}{4}\right)} = \pm 1
        \end{equation}
        which are solved by
        \begin{equation}
            \begin{split}
                & (r_1, r_2)_{+}= (0,0),(2,0), (0, 2), (1,1), (2,2); \\
                 & (r_1, r_2)_{-}= (0,1),(1,0),(2,1), (1, 2)\, .
            \end{split}
        \end{equation}
       Any solution in which $r_1$ or $r_2$ is $1$ corresponds to a non-invertible line. It is easy to see that the antichiral-antichiral state conjugate to \eqref{eq:cctorus} is also invariant under those lines.  
 \item For $\cL_{\{r_i, s_i\}; B, \eta}$ to (anti-)commute with the antichiral-chiral operator we need instead
 \begin{equation}
      e^{i \pi \left(\frac{s_1+s_2}{4}+ \eta \right)}\prod_{i=1}^{2}\frac{\sin\left(\frac{2\pi(r_i+1)}{4}\right)}{\sin\left(\frac{\pi(r_i+1)}{4}\right)} = \pm 2\, .
 \end{equation}
 We can eliminate the imaginary part with $s_1+s_2= \eta \bmod 4$, then
 \begin{equation}
     \prod_{i=1}^{2}\frac{\sin\left(\frac{2\pi(r_i+1)}{4}\right)}{\sin\left(\frac{\pi(r_i+1)}{4}\right)} = \pm 2\, ,
 \end{equation}
 which are solved by
 \begin{equation}
     \begin{split}
         &(r_1,r_2)_+ = (0,0), (2,2); \\
         &(r_1,r_2)_- = (2,0), (0,2) \, .
     \end{split}
 \end{equation}
 Thus only invertible lines commute with this deformation.
\end{itemize}

By construction the symmetries of the chiral-chiral deformation form a fusion subcategory $\cS_{\cC \cC}$ of the full symmetry of the Gepner model. For instance, in $\cS_{\cC \cC}$ we have the line $\cD= \cL_{1,1,1,3; \varnothing, 0}$, this fuses as
\begin{equation}
    \cD \times  \cD^{\dagger} = \unit + \eta_1 + \eta_2 + \eta_1 \times \eta_2
\end{equation}
where $\eta_1= \cL_{0,2,0,0; \varnothing, 0}$ and $\eta_2 = \cL_{0,0,2,0; \varnothing, 0}$ are $\bZ_{2}$ generators. These are the fusion of a Tambara-Yamagami category over a group $\bZ_{2}\times \bZ_{2}$. The Gepner model $(2)^{2}(0)$ is known to sit at the point in $\cM$ corresponding to the square torus $G_{ij}=\delta_{ij}$, $B_{ij}=0$. Sitting at this point and deforming with a linear combination of the exactly marginal chiral-chiral and antichiral-antichiral operators, which we denote $\Phi_{\cC \cC}$ and $\Phi_{\cC \cC}^{\dagger}$ respectively,  we preserve $\cS_{\cC \cC}$. We first have to address the relation between $\Phi_{\cC \cC}$ and $\partial X \bar{\partial} X$. The latter is a superconformal descendant of the fermion product $\psi \lambda$, since the fermion action and the chiral rings are independent on the point in the conformal manifold it is natural to identify $\psi \lambda$ with the state \eqref{eq:cctorus}. The relation between $\Phi_{\cC \cC}$ and $\partial X \bar{\partial} X$ could still depend on the moduli through the supercharges. However the Gepner point corresponds to the square torus with no $B$-field, then the supercurrents at this point are simply those of the free field realization of the $\cN=(2,2)$ superconformal algebra, namely
\begin{equation}
    \begin{split}
      & G^{-}(z) = \frac{1}{2}\bar{\psi}\partial X\, \quad G^{+}(z) = \frac{1}{2}\psi\partial \bar{X}\\ &  \bar{G}^{-}(\bar{z}) = \frac{1}{2}\bar{\lambda}\bar{\partial} X\, \quad \bar{G}^{+}(\bar{z}) = \frac{1}{2}\lambda\bar{\partial} \bar{X}\, .
    \end{split}
\end{equation}
From these expressions and using the free field OPEs one easily gets that the marginal descendant of $\psi \lambda $ is $\partial X \bar{\partial} X$, thus leading us to the conclusion that $\Phi_{\cC \cC} = \partial X \bar{\partial} X$. Then consider the two real operators 
\begin{equation}
    \cO_1 = g_1\left(\Phi_{\cC \cC}+ \Phi^{\dagger}_{\cC \cC}\right)\, ,  \qquad  \cO_2 = -i g_2 \left(\Phi_{\cC \cC}- \Phi^{\dagger}_{\cC \cC}\right) 
\end{equation}
where $2g_1 = \delta G_{11}- \delta G_{22}$ and $g_2 = \delta G_{12}$. The values of the metric components as a function of $g_1$ and $g_2$ are
\begin{equation}
    G_{11} = 1+g_1\, ,  \qquad G_{22} = 1-g_1 \, , \qquad G_{12}= g_2\, .
\end{equation}
For $G_{ij}$ to be positive definite we have to restrict $g_{i}$ inside the disk $g_1^2+g_2^2 < 1$. As we change $g_1$ and $g_2$ we trace out a 2-dimensional submanifold of $\cM$ on which the symmetry $\cS_{\cC \cC}$ is preserved. A convenient presentation of $\cM$ is obtained introducing two complex parameters \cite{Polchinski:1998rq}
\begin{equation}
    \tau = \frac{1}{G_{11}}\left(G_{12}+ i \sqrt{\text{det}G}\right) \, , \qquad \rho = B_{12} + i \sqrt{\text{det}G}\, ; 
\end{equation}
corresponding respectively to the complex structure and complexified area of the target torus. Integer-valued change of coordinates on $G$, namely $G\mapsto M^{T}GM$ with $M \in SL(2, \bZ)$, induce $PSL(2, \bZ)$ transformations on $\tau$ leaving $\rho$ invariant. The $PSL(2, \bZ)$ acting on $\rho$ is generated by $T$-duality on both compact directions as well as shifts of the $B$-field by integers. $T$ duality on only one direction exchanges $\rho$ and $\tau$, parity in both the worldsheet and the target torus induce further $\bZ_{2}$ identifications. Then $\cM$ is a product of two copies of the fundamental domain of $SL(2, \bZ)$ in the upper-half plane with some extra discrete $\bZ_2$ identifications. In terms of $g_1$ and $g_2$ we have
\begin{equation}
    \tau = \frac{1}{1+g_1}\left(g_2 + i \sqrt{1-g_1^2-g_2^2}\right)\, ,  \qquad \rho = i\sqrt{1-g_1^2-g_2^2}\, . 
\end{equation}
or, using an $S$-transformation on both to have $\text{Im}\tau, \text{Im}\rho > 1$ 
\begin{equation}
    \tau = \frac{1}{1-g_1}\left(-g_2 + i \sqrt{1-g_1^2-g_2^2}\right)\, ,  \qquad \rho = \frac{i}{\sqrt{1-g_1^2-g_2^2}}\, . 
\end{equation}
Now for $g_i$ inside the disk we span the whole $\tau$ fundamental domain while moving along the complex direction in the $\rho$ plane from $i$ to $i \infty$.
\begin{figure}
    \centering
    \includegraphics[scale=0.8]{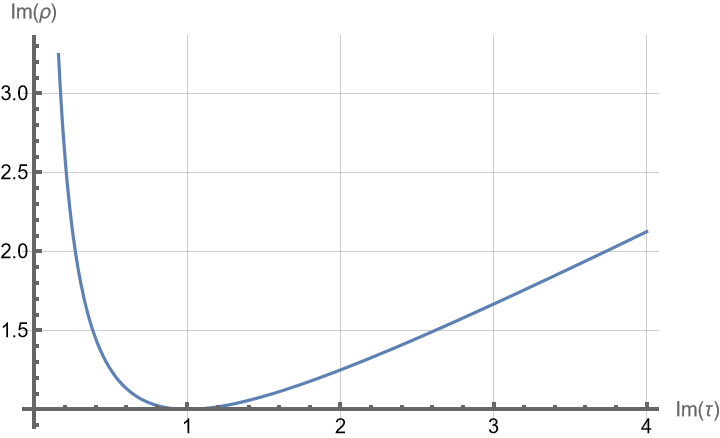}
    \caption{Setting $g_2=0$, so that $\text{Re}(\tau) = \text{Re}(\rho)=0$ we can consider $-1<g_1<1$. The submanifold on which $\cS_{\cC\cC}$ is preserved is given by the curve $
    \text{Im}(\rho) = \frac{1}{2}\left(\text{Im}(\tau) + \frac{1}{\text{Im}(\tau)}
\right)$ plotted here. }
    \label{fig:torus}
\end{figure}

\subsection{$K3$ surface}
The first interacting case we consider is a sigma model on a $K3$ surface with central charge $c=6$. The Gepner point can be realized as the product of four minimal models at level $k=2$, conventionally denoted $(2)^4$. We start looking in more detail to the BPS spectrum. The chiral-chiral states in the untwisted sector are
\begin{equation}
    \bigotimes_{i=1}^{4}\ket{a_i,a_i} \bigotimes_{i=1}^{4}\overline{\ket{a_i,a_i}}
\end{equation}
and $19$ of those satisfy the charge requirement 
\begin{equation}
    \sum_{i}\frac{a_i}{4} = 1\, .
\end{equation}
The values of $a_i$ for these BPS states are (non-trivial) permutations of the following
\begin{equation}\label{eq:chiralK3}
    (a_1, a_2, a_3, a_4) = (2, 2, 0,0); (2, 1, 1,0); (1,1,1,1)\, . 
\end{equation}
For $r=4$ we know that there is another chiral-chiral state with the appropriate $U(1)_R$ charges in the twisted component $\bH^{(2, \{1, \ldots, 4\})}$, this is given by
\begin{equation}\label{eq:ccK3}
    \left(\ket{1,1}\right)^{\otimes^4} \otimes \left(\overline{\ket{1,1}}\right)^{\otimes^4}\in \bH^{(2, \{1,\ldots, 4\})} \, .
\end{equation}
In total we have $20$ chiral-chiral states. Acting with charge conjugation on each of those states we obtain the $20$ antichiral-antichiral states corresponding to marginal deformations. Notice that the state \eqref{eq:ccK3} has the same weight and $R$-charges of the chiral-chiral state 
\begin{equation}\label{eq:ccK31}
    \left(\ket{1,1}\right)^{\otimes^4} \otimes \left(\overline{\ket{1,1}}\right)^{\otimes^4}\in \bH^{(0,\varnothing)} \, .
\end{equation}
appearing in the untwisted sector. These two however are not indistinguishable as the quantum numbers of the corresponding primaries under the dual symmetries are different.

The analysis of antichiral-chiral states corresponding to marginal deformations is essentially the same. In the twisted Hilbert spaces $\bH^{(x,\varnothing)}$, we have antichiral-chiral states of the form
\begin{equation}
    \left(\ket{a,-a}\right)^{\otimes^4}\otimes \left(\ket{a,a}\right)^{\otimes^4} \in \bH^{(a,\varnothing)}
\end{equation}
of those only the one with $a=1$ has left and right $R$-charges equal to $-1$ and $1$. The remaining antichiral-chiral states sit in $\bH^{3, \{1,\ldots 4\}}$ and are of the form
\begin{equation}\label{eq:acK3}
    \bigotimes_{i=1}^{4}\ket{a_i, -a_i}\otimes \overline{\ket{2-a_i, 2-a_i}} \in \bH^{3, \{1,\ldots 4\}}\, . 
\end{equation}
The charge constraint
\begin{equation}
    \sum_{i}\frac{a_i}{4} = 1
\end{equation}
selects 19 of them, with the values of $a_i$ again given by permutations of the 4-tuples in \eqref{eq:chiralK3}. We then have 20 chiral-chiral and 20 antichiral-chiral states (plus conjugates), which pair up into BPS multiplets for the $\cN=4$ algebra, consistently with $h^{1,1}= 20$ for $K3$. Now let's look at the symmetries. 
\begin{itemize}
    \item For chiral-chiral states we look at each vector \eqref{eq:chiralK3} separately.
    \begin{itemize}
        \item $(a_1,a_2,a_3,a_4)= (2,2,0,0)$. The invariance condition is
        \begin{equation}
            e^{i \pi \frac{2s_1+2s_2}{4}}\frac{\sin\left(\frac{3\pi(r_1+1)}{4}\right)\sin\left(\frac{3\pi(r_2+1)}{4}\right)}{\sin\left(\frac{\pi(r_1+1)}{4}\right)\sin\left(\frac{\pi(r_2+1)}{4}\right)} = \pm 1\, . 
        \end{equation}
        which is symmetric in $r_1$ and $r_2$. Note that this condition only constraints the action of the line on two minimal models, namely the parameters $r_3, s_3, r_4, s_4, \eta$ are not fixed by this condition. This already shows that the deformation preserves non-invertible lines. To eliminate the imaginary part we set $2s_1+2s_2 = 0 \bmod 4$, then 
        \begin{equation}
            \frac{\sin\left(\frac{3\pi(r_1+1)}{4}\right)\sin\left(\frac{3\pi(r_2+1)}{4}\right)}{\sin\left(\frac{\pi(r_1+1)}{4}\right)\sin\left(\frac{\pi(r_2+1)}{4}\right)} = \pm 1
        \end{equation}
        which are solved by
        \begin{equation}
            \begin{split}
                & (r_1, r_2)_{+}= (0,0), (0, 2), (1,1), (2,2)\\
                 & (r_1, r_2)_{-}= (0,1), (1, 2)\, .
            \end{split}
        \end{equation}
       Here we reported only one element of the orbit of the swap symmetry $r_1 \leftrightarrow r_2$ on the solutions. Any solution in which $r_1$ or $r_2$ is $1$ corresponds to a non-invertible line.  
        \item $(a_1,a_2,a_3,a_4)= (2,1,1,0)$. We impose
        \begin{equation}
            e^{i \pi \frac{2s_1+s_2+ s_3}{4}}\frac{\sin\left(\frac{3\pi(r_1+1)}{4}\right)\sin\left(\frac{2\pi(r_2+1)}{4}\right)\sin\left(\frac{2\pi(r_3+1)}{4}\right)}{\sin\left(\frac{\pi(r_1+1)}{4}\right)\sin\left(\frac{\pi(r_2+1)}{4}\right)\sin\left(\frac{\pi(r_3+1)}{4}\right)} = \pm 2
        \end{equation}
        hence for $2s_1+s_2+s_3 = 0 \bmod 4$ we have 
        \begin{equation}
            \frac{\sin\left(\frac{3\pi(r_1+1)}{4}\right)\sin\left(\frac{2\pi(r_2+1)}{4}\right)\sin\left(\frac{2\pi(r_3+1)}{4}\right)}{\sin\left(\frac{\pi(r_1+1)}{4}\right)\sin\left(\frac{\pi(r_2+1)}{4}\right)\sin\left(\frac{\pi(r_3+1)}{4}\right)}  = \pm 2
        \end{equation}
        with solutions
        \begin{equation}
            \begin{split}
                & (r_1, r_2, r_3)_{+}= (0,0,0); (0, 2,2); (1,0,2);(2,0,0); (2,2,2)\\
                 & (r_1, r_2, r_3)_{-}= (0,0,2), (1, 0,0); (1,2,2); (2,0,2)
            \end{split}
        \end{equation}
        where we again reported only one representative for the swap symmetry of $r_2$ and $r_3$. Again we see that there are solutions in which at least one among $r_1, r_2$ or $r_3$ is $1$, together with the freedom of choosing the parameter $r_4$, we see that there are preserved non-invertible TDLS along the deformations.
        \item $(a_1,a_2,a_3,a_4)= (1,1,1,1)$. This is the last chiral-chiral state we have to consider. Since it sits in a twisted component of the Hilbert space the invariance condition now also involve the dual symmetries and becomes
        \begin{equation}
            \zeta_{A; \{1,\ldots,4\}}e^{i \pi \left(\frac{s_1+s_2+ s_3+s_4}{4}+ \eta \right)}\prod_{i=1}^{4}\frac{\sin\left(\frac{2\pi(r_i+1)}{4}\right)}{\sin\left(\frac{\pi(r_i+1)}{4}\right)} = \pm 4\, .
        \end{equation}
        We can leave the dual symmetry labels $A, \eta$ free and only require $s_1+s_2+s_3 +s_4 = 0 \bmod 4$. Then
        \begin{equation}
            \prod_{i=1}^{4}\frac{\sin\left(\frac{2\pi(r_i+1)}{4}\right)}{\sin\left(\frac{\pi(r_i+1)}{4}\right)} = \pm 4
        \end{equation}
        with solutions
        \begin{equation}
            \begin{split}
                & (r_1, r_2, r_3, r_4)_{+}= (0,0,0, 0); (0,0, 2,2);(2,2,2,2)\\
                 & (r_1, r_2, r_3, r_4)_{-}= (0,0,0,2), (0,2,2,2);
            \end{split}
        \end{equation}
        where we again reported only one representative for the orbit of the permutation symmetry. This deformation is invariant only under invertible lines. 
    \end{itemize}
    Let's take stock. The majority of the chiral-chiral deformations preserve at least one non-invertible line. It is interesting to notice also that we can deform the Gepner model with multiple chiral-chiral operators and still preserve non-invertible lines. For instance we can turn on simultaneously  some of the deformations given by permutations of $(a_1, a_2, a_3, a_4) = (2,2, 0,0)$, still preserving some non-invertible lines.

    \item For antichiral-chiral states we again consider the vectors \eqref{eq:chiralK3} separately, each giving rise to a state of the form \eqref{eq:acK3}.
    \begin{itemize}
        \item $(a_1,a_2,a_3,a_4)= (2,2,0,0)$. The invariance condition is
        \begin{equation}
            \zeta_{B, \{1,\ldots,4\}}e^{i \pi \left(\frac{2s_1+2s_2}{4} +\frac{3 \eta}{2}\right)}\frac{\sin\left(\frac{3\pi(r_1+1)}{4}\right)\sin\left(\frac{3\pi(r_2+1)}{4}\right)}{\sin\left(\frac{\pi(r_1+1)}{4}\right)\sin\left(\frac{\pi(r_2+1)}{4}\right)} = \pm 1\, . 
        \end{equation}
        We can leave $B$ free and set $2 s_1+2s_2+ 2 \eta = 0 \bmod 4$. The remaining equation is
        \begin{equation}
            \frac{\sin\left(\frac{3\pi(r_1+1)}{4}\right)\sin\left(\frac{3\pi(r_2+1)}{4}\right)}{\sin\left(\frac{\pi(r_1+1)}{4}\right)\sin\left(\frac{\pi(r_2+1)}{4}\right)} = \pm 1
        \end{equation}
        Which is the same one we solved for the chiral-chiral deformation with $(a_1,a_2,a_3,a_4)= (2,2,0,0)$. Notice that the  lines preserving this deformation are not the same ones preserving the chiral-chiral one, but the two sets have a non-empty intersection. For instance lines with $\eta=0,4$ that preserve the chiral-chiral deformation will preserve also this one. 
         \item $(a_1,a_2,a_3,a_4)= (2,1,1,0)$. We impose
        \begin{equation}
            \zeta_{B, \{1,\ldots,4\}}e^{i \pi \left(\frac{2s_1+s_2+ s_3}{4}+ \frac{3\eta}{2}\right)}\frac{\sin\left(\frac{3\pi(r_1+1)}{4}\right)\sin\left(\frac{2\pi(r_2+1)}{4}\right)\sin\left(\frac{2\pi(r_3+1)}{4}\right)}{\sin\left(\frac{\pi(r_1+1)}{4}\right)\sin\left(\frac{\pi(r_2+1)}{4}\right)\sin\left(\frac{\pi(r_3+1)}{4}\right)} = \pm 2
        \end{equation}
        hence for $2s_1+s_2+s_3 + 2 \eta = 0 \bmod 4$ we have again an equation we solved in the chiral-chiral case. Also here to lines with $\eta=0,4$ will preserve both chiral-chiral and antichiral-chiral deformations.
       
        \item $(a_1,a_2,a_3,a_4)= (1,1,1,1)$. The invariance condition is now
        \begin{equation}
            e^{i \pi \left(\frac{s_1+s_2+ s_3+s_4}{4}+ \eta \right)}\prod_{i=1}^{4}\frac{\sin\left(\frac{2\pi(r_i+1)}{4}\right)}{\sin\left(\frac{\pi(r_i+1)}{4}\right)} = \pm 4\, .
        \end{equation}
        We can leave the dual symmetry label $\eta$ free and require $s_1+s_2+s_3 +s_4 = 0 \bmod 4$. Then
        \begin{equation}
            \prod_{i=1}^{4}\frac{\sin\left(\frac{2\pi(r_i+1)}{4}\right)}{\sin\left(\frac{\pi(r_i+1)}{4}\right)} = \pm 4
        \end{equation}
        which we have solved above. 
    \end{itemize}
    We see that the structure of the lines preserved by these antichiral-chiral deformations closely follows the one we found for chiral-chiral deformations. 
\end{itemize}

Note that lines that preserve both the chiral-chiral and antichiral-chiral deformations given by the same vector in \eqref{eq:chiralK3}, will also preserve the exactly marginal $\cN=4$ deformation we obtain taking their sum. Using what we found above we can also enlarge the submanifold of the moduli space that enjoys non-invertible symmetries. Consider both the chiral-chiral and antichiral-chiral deformations given by $(a_1,a_2,a_3,a_4)= (2,2,0,0), (0,0,2,2)$. These preserve the line with
\begin{equation}
\begin{split}
     &r_{i}=1\, ,  \qquad i=1,\ldots, 4\\
     & s_1=s_3 = 1\, , s_2=s_4=3
\end{split}
\end{equation}
as well as $\eta = 0,4$. Therefore we have a $4$-dimensional subspace of the moduli space with a non-invertible symmetry. More precisely denoting $\cD = \cL_{1,1,1,3,1,1,1,3; \varnothing,0}$ we have
\begin{equation}    \cD\times \cD^{\dagger} =  \unit+\sum_{i=1}^{4}\eta_{i}+\sum_{i<j}\eta_{i}\times \eta_{j}\sum_{i<j<k}\eta_{i}\times \eta_{j}\times \eta_k+\eta_{1}\times \eta_{2}\times_{3}\times \eta_4
\end{equation}
where
\begin{equation}
    \begin{split}
        &\eta_{1} = \cL_{2,0,0,0,0,0,0,0; \varnothing,0} \, , \qquad \eta_{2} = \cL_{0,0,2,0,0,0,0,0; \varnothing,0}\\ &  \eta_{3} = \cL_{0,0,0,0,2,0,0,0; \varnothing,0} \, , \qquad \eta_{4} = \cL_{0,0,0,0,0,0,2,0; \varnothing,0}
    \end{split}
\end{equation}
generate a $\bZ_{2}^{4}$ group. Including $\cD$ we have a $\bZ_{2}^{4}$ Tambara-Yamagami symmetry.

\subsection{Quintic Threefold}

For the Quintic threefold the Gepner point is $(3)^5$, namely we take five minimal models with $k=3$. The central charge is $c=9$. Again we start with chiral-chiral primaries, the states of the form
\begin{equation}
     \bigotimes_{i=1}^5\ket{a_i,a_i} \otimes \bigotimes_{i=1}^5\overline{\ket{a_i,a_i}}
\end{equation}
for $a_i=0,1,2,3$ are $4^5=1024$, of those the charge constraint
\begin{equation}
    \sum_i \frac{a_i}{5} = 1
\end{equation}
selects $101$. This matches $h^{2,1}=101$ counting $(2,1)$ forms on the quintic. These correspond to complex structure deformations of the underlying Calabi-Yau. For each of those states the vector $(a_1,a_2,a_3,a_4,a_5)$ is a  permutation of one of the following five 
\begin{equation}\label{eq:ccquitnic}
\begin{split}
   (a_1,a_2,a_3,a_4,a_5)= (3,2,0,0,0), (3,1,1,0,0), (2,2,1,0,0),(2,1,1,1,0), (1,1,1,1,1)\, .
\end{split}
\end{equation}
The antichiral-chiral states are of the form 
\begin{equation}
    \left(\ket{a,-a}\right)^{\otimes^5}\otimes  \left(\overline{\ket{a,a}}\right)^{\otimes^5} \in \bH^{(a,\varnothing)}
\end{equation}
but of those only the one with $a=1$ has left $R$ charge $-1$. Thus there is only one complex Khaler modulus, which agrees with $h^{1,1}=1$ for the quintic. Now we look at the symmetries.
\begin{itemize}
    \item For chiral-chiral states it is enough to find the solution of the invariance constraint (which is invariant under permutations) for each of the five vectors \eqref{eq:ccquitnic}. We have
    \begin{itemize}
        \item $(a_1,a_2,a_3,a_4,a_5)=(3,2,0,0,0)$. A line $\cL_{\{r_i,s_i\},B, \eta}$ leaving invariant the deformation has to satisfy
        \begin{equation}
            e^{i \pi \frac{3s_1+2s_2}{5}}\frac{\sin\left(\frac{4 \pi (r_1+1)}{5}\right)\sin\left(\frac{3\pi(r_2+1)}{5}\right)}{\sin\left(\frac{\pi(r_1+1)}{5}\right)\sin\left(\frac{\pi(r_2+1)}{5}\right)} = \pm \frac{1}{2}\left(1+ \sqrt{5}\right)
        \end{equation}
        with all labels other than $r_1, s_1$ and $r_2, s_2$ free.
        By choosing $3s_1+2s_2 = 0 \bmod 5$ we can look for solutions of
        \begin{equation}
           \frac{\sin\left(\frac{4 \pi (r_1+1)}{5}\right)\sin\left(\frac{3\pi(r_2+1)}{5}\right)}{\sin\left(\frac{\pi(r_1+1)}{5}\right)\sin\left(\frac{\pi(r_2+1)}{5}\right)} = \pm \frac{1}{2}\left(1+ \sqrt{5}\right)\, . 
        \end{equation}
        We find 
        \begin{equation}
        \begin{split}
           & (r_1,r_2)_+ = (0,0); (0,3); (2,0);(2,3) \, \qquad 3s_1+2s_2 = 0 \bmod 10 \\ & (r_1,r_2)_- = (1,0); (1,3); (3,0);(3,3)  \, \qquad 3s_1+2s_2 = 5 \bmod 10\,
        \end{split}
        \end{equation}
        for the two signs. Therefore for each pair of $(r_1,r_2)_{\pm}$ we choose $(s_1,s_2)$ accordingly while all other labels are free. Here, as long as $r_i$ is either $1$ or $2$ the corresponding TDL is non-invertible.
         \item $(a_1,a_2,a_3,a_4,a_5)=(3,1,1,0,0)$. Invariance requires 
         \begin{equation}
          e^{i \frac{\pi}{5}(3s_1+s_2+s_3)}\frac{\sin\left(\frac{4 \pi (r_1+1)}{5}\right)\sin\left(\frac{2\pi(r_2+1)}{5}\right)\sin\left(\frac{2\pi(r_3+1)}{5}\right)}{\sin\left(\frac{\pi(r_1+1)}{5}\right)\sin\left(\frac{\pi(r_2+1)}{5}\right)\sin\left(\frac{\pi(r_3+1)}{5}\right)} =\pm  \frac{1}{2}\left(3+ \sqrt{5}\right)
         \end{equation}
         thus for $3s_1+s_2+s_3= 0 \bmod 5$ we need to solve
         \begin{equation}
             \frac{\sin\left(\frac{4 \pi (r_1+1)}{5}\right)\sin\left(\frac{2\pi(r_2+1)}{5}\right)\sin\left(\frac{2\pi(r_3+1)}{5}\right)}{\sin\left(\frac{\pi(r_1+1)}{5}\right)\sin\left(\frac{\pi(r_2+1)}{5}\right)\sin\left(\frac{\pi(r_3+1)}{5}\right)} = \pm \frac{1}{2}\left(3+ \sqrt{5}\right)\, .
         \end{equation}
         Note that this is symmetric in $r_2$ and $r_3$, in the solutions below we write only one solution per orbit of this swap symmetry. We find
        \begin{equation}
        \begin{split}
           & (r_1,r_2, r_3)_+ = (0,0,0); (0,3,3); (1,0,3);(2,0,0); (2,3,3); (3,0,3); (3,3,0)\\ & 
           (r_1,r_2, r_3)_- = (0,0,3); (1,0,0); (1,3,3);(2,0,3); (3,0,0); (3,3,3) \, .
        \end{split}
        \end{equation}
        \item $(a_1,a_2,a_3,a_4,a_5)=(2,2,1,0,0)$. Invariance requires 
         \begin{equation}
          e^{i \frac{\pi}{5}(2s_1+2s_2+s_3)}\frac{\sin\left(\frac{3 \pi (r_1+1)}{5}\right)\sin\left(\frac{3\pi(r_2+1)}{5}\right)\sin\left(\frac{2\pi(r_3+1)}{5}\right)}{\sin\left(\frac{\pi(r_1+1)}{5}\right)\sin\left(\frac{\pi(r_2+1)}{5}\right)\sin\left(\frac{\pi(r_3+1)}{5}\right)} =\pm \left( 2+ \sqrt{5}\right)
         \end{equation}
         thus for $2s_1+2s_2+s_3= 0 \bmod 5$ we need to solve
         \begin{equation}
             \frac{\sin\left(\frac{3 \pi (r_1+1)}{5}\right)\sin\left(\frac{3\pi(r_2+1)}{5}\right)\sin\left(\frac{2\pi(r_3+1)}{5}\right)}{\sin\left(\frac{\pi(r_1+1)}{5}\right)\sin\left(\frac{\pi(r_2+1)}{5}\right)\sin\left(\frac{\pi(r_3+1)}{5}\right)} = \pm \left(2+ \sqrt{5}\right)\, .
         \end{equation}
         Note that this is symmetric in $r_1$ and $r_2$, in the solutions below we write only one solution per orbit of this swap symmetry. We find
        \begin{equation}
        \begin{split}
           & (r_1,r_2, r_3)_+ = (0,0,0); (0,3,0); (3,3,0)\\ & 
           (r_1,r_2, r_3)_- = (0,0,3); (0,3,3); (3,3,3) \, .
        \end{split}
        \end{equation}
        We see that a line preserving this deformation necessarily acts invertibly on it, altough it may act non-invertibly on other operators of the theory.
        \item $(a_1,a_2,a_3,a_4,a_5)=(2,1,1,1,0)$. Invariance requires 
         \begin{equation}
         \begin{split}
          &e^{i \frac{\pi}{5}(2s_1+s_2+s_3+s_4)}\frac{\sin\left(\frac{3 \pi (r_1+1)}{5}\right)\sin\left(\frac{2\pi(r_2+1)}{5}\right)\sin\left(\frac{2\pi(r_3+1)}{5}\right)\sin\left(\frac{2\pi(r_4+1)}{5}\right)}{\sin\left(\frac{\pi(r_1+1)}{5}\right)\sin\left(\frac{\pi(r_2+1)}{5}\right)\sin\left(\frac{\pi(r_3+1)}{5}\right)\sin\left(\frac{\pi(r_4+1)}{5}\right)} \\  & \qquad \qquad \qquad \qquad = \pm \frac{1}{2}\left(7+3 \sqrt{5}\right)
         \end{split} 
         \end{equation}
         thus for $2s_1+s_2+s_3+s_4= 0 \bmod 5$ we need to solve
         \begin{equation}
             \frac{\sin\left(\frac{3 \pi (r_1+1)}{5}\right)\sin\left(\frac{2\pi(r_2+1)}{5}\right)\sin\left(\frac{2\pi(r_3+1)}{5}\right)\sin\left(\frac{2\pi(r_4+1)}{5}\right)}{\sin\left(\frac{\pi(r_1+1)}{5}\right)\sin\left(\frac{\pi(r_2+1)}{5}\right)\sin\left(\frac{\pi(r_3+1)}{5}\right)\sin\left(\frac{\pi(r_4+1)}{5}\right)} = \pm \frac{1}{2}\left(7+3 \sqrt{5}\right)\, .
         \end{equation}
         Note that this is symmetric in $r_2$, $r_3$ and $r_4$, in the solutions below we write only one solution per orbit of this permutation symmetry. We find
        \begin{equation}
        \begin{split}
           & (r_1,r_2, r_3, r_4)_+ = (0,0,0,0); (0,0,3,3); (3,0,0,0);(3, 3, 3, 0)\\ & 
           (r_1,r_2, r_3, r_4)_- = (0,0,0,3); (0,3,3,3); (3,3,0,0); (3,3,3,3) \, .
        \end{split}
        \end{equation}
        Again a line preserving this deformation necessarily acts invertibly on it.
         \item $(a_1,a_2,a_3,a_4,a_5)=(1,1,1,1,1)$. Invariance requires 
         \begin{equation}
          e^{i \frac{\pi}{5}(s_1+s_2+s_3+s_4+s_5)}\prod_{i=1}^{5}\frac{\sin\left(\frac{2 \pi (r_i+1)}{5}\right)}{\sin\left(\frac{\pi(r_i+1)}{5}\right)} = \pm \frac{1}{2}(11+ 5\sqrt{5})
         \end{equation}
         thus for $s_1+s_2+s_3+s_4+s_5= 0 \bmod 5$ we need to solve
         \begin{equation}
            \prod_{i=1}^{5}\frac{\sin\left(\frac{2 \pi (r_i+1)}{5}\right)}{\sin\left(\frac{\pi(r_i+1)}{5}\right)} = \pm \frac{1}{2}(11+ 5\sqrt{5})
         \end{equation}
         Note that this is symmetric in all $r_i$, in the solutions below we write only one solution per orbit of this permutation symmetry. We find
        \begin{equation}
        \begin{split}
           & (r_1,r_2, r_3, r_4, r_5)_+ = (0,0,0,0,0); (0,0,0,3,3); (0,3,3,3,3);\\ & 
           (r_1,r_2, r_3, r_4, r_5)_- = (0,0,0,0,3); (0,0,3,3,3); (3,3,3,3,3) \, .
        \end{split}
        \end{equation}
        We see that a line preserving this deformation is necessarily invertible.
        
    \end{itemize}

    \item The symmetries preserved by the Khaler structure deformation obey 
    \begin{equation}
        \quad e^{i \frac{\pi}{5}\left(2\eta -\sum_{i}s_i\right)}\prod_{i=1}^5 \frac{\sin\left(\frac{2 \pi (r_i+1)}{5}\right)\sin\left(\frac{\pi}{5}\right)}{\sin\left(\frac{ \pi (r_i+1)}{5}\right)\sin\left(\frac{2 \pi}{5}\right)} = \pm 1
    \end{equation}
    which, after having picked $2\eta = \sum_i s_i \bmod 5$, is the same equation as the last case among the chiral-chiral deformations. Again only invertible symmetries preserve this operator.
\end{itemize}
Also in this example we can look for higher dimensional submanifolds preserving a non-invertible symmetry. As a simple illustration consider the chiral-chiral deformations with 
\begin{equation}
    (a_1, a_2, a_3, a_4, a_5) = (3,2,0,0,0);(3,0,2,0,0);(3,0,0,2,0);(3,0,0,0,2) \, ,
\end{equation}
from our analysis we see that each one of those commutes with the line $W = \cL_{2,0} \otimes (\unit)^{\otimes 4}$, whose fusion rule is
\begin{equation}
    W \times W = \unit + W\, .
\end{equation}
Thus on this 4-dimensional submanifold we have at least a Fibonacci category symmetry. We can also turn on the deformations  
\begin{equation}
\begin{split}
    (a_1, a_2, a_3, a_4, a_5)  = & (3,1,1,0,0);(3,1,0,1,0);(3,1,0,0,1); \\ & (3,0,1,1,0);(3,0,1,0,1);(3,0,0,1,1) \, , 
\end{split}
\end{equation}
and $W$ is still preserved, enlarging the Fibonacci-symmetric submanifold of the moduli space to 10 dimensions. We can also consider the submanifolds obtained turning on the deformations above with $a_1 \leftrightarrow a_i$ for $i =2,3,4,5$. On each of those 10-dimensional subspaces we have a different Fibonacci category symmetry. 

\subsection{Selection Rules}\label{sec:select}

The presence of these topological defects at the Gepner point and along certain submanifolds of the moduli space imposes constraints on the dynamics of the theory. The presence of a fusion category symmetry in a $2d$ QFT implies degeneracies between twisted and untwisted sectors, as non-invertible lines transform local operators in twist defects. This is properly addressed using the tube algebra built out of the fusion category, see e.g. \cite{Lin:2022dhv, Bartsch:2023wvv}. The elements of such algebra correspond to the lasso actions \cite{Chang:2018iay}
\bea    
\begin{tikzpicture}     \draw[fill=black] (0,0) circle (0.05);
     \draw[color=red, line width=1, decoration={markings, mark=at position 0.625 with {\arrow{<}}},
        postaction={decorate}] (0,0) circle (1);
     \draw[color=blue, line width=1, ->] (0,1) -- (0,1.5);
     \draw[color=blue, line width=1] (0,1.49) -- (0,2);
     \draw[fill=black] (0,1) circle (0.05);
     \node[below] at (0,0) {$\Phi$};
     \node[above] at (0,2) {$\cL'$};
     \node[left] at (-1,0) {$\cL$};
     \node[right] at (1.2,0) {=};
    \end{tikzpicture} \begin{tikzpicture}     
     \draw[color=blue, line width=1, ->] (0,0) -- (0,1);
     \draw[color=blue, line width=1] (0,0.99) -- (0,2);
     \draw[fill=black] (0,0) circle (0.05);
     \node[below] at (0,0) {$\hat{\cL}_{\cL'}\cdot\Phi$};
     \node[above] at (0,2) {$\cL'$};
     \node[right] at (1.2,0) {.};
    \end{tikzpicture}\, 
\eea
In general one also needs to specify a junction vector in $\text{Hom}(\cL \times \bar{\cL}, \cL')$, however in our case all junction spaces are at most one-dimensional and we can omit this extra label. More general lasso actions can be obtained acting on operators in twisted sectors. The total Hilbert space, which includes the twisted Hilbert spaces for all the topological defect lines of the theory, splits in representations of the Tube algebra. Therefore operators, both local and twisted, will also be organized in such representations. Representations of the tube algebra are in canonical one-to-one correspondence with anyons of the Drinfeld center of the fusion category, with the fusion rules of the $3d$ TQFT anyons coinciding with tensor products of Tube algebra representations. Moreover the representation in which a local operator of the CFT transforms is determined by the $3d$ bulk anyon ending on it (see Fig.\ref{fig:symmact}). The most immediate consequence of the symmetry are selection rules. In case of a fusion category symmetry these state that a correlation function can be non-zero only if the tensor product of the Tube algebra representations of all operators contains the identity\cite{Lin:2022dhv}.

A subsets of interesting selection rules however can be accessed without employing the full power of the Tube algebra. Two important observables on the conformal manifold of a CFT are the two and three point functions of the exactly marginal deformations. The former gives the Zamolodchikov metric of the conformal manifold, while the latter encode information about the curvature \cite{Distler:1992gi}. We shall consider the CFT on a genus zero surface, this allows us to nucleate a non-invertible line defect linking with all operators in the correlator at the price of dividing by its quantum dimension. Let $\Phi_1$ and $\Phi_2$ be exactly marginal operators,
and consider its their two point function. Opening an $\cL$ loop and dividing by $\langle \cL \rangle$ the correlator is unchanged, namely
\begin{equation}
     \langle \Phi_1 \Phi_2\rangle  = \frac{1}{\langle \cL \rangle}\langle \,\, \begin{tikzpicture} [baseline={([yshift=-2ex]current bounding box.center)},vertex/.style={anchor=base,
    circle,fill=black!25,minimum size=12pt,inner sep=2pt}]    \draw[fill=black] (-0.5,0) circle (0.05);
     \draw[fill=black] (0.5,0) circle (0.05);
     \draw[color=red, line width=1, decoration={markings, mark=at position 0.625 with {\arrow{<}}},
        postaction={decorate}] (0,0) circle (1);
     \node[below] at (0.5,0) {$\Phi_2$};
     \node[below] at (-0.5,0) {$\Phi_1$};
     \node[above] at (0,1) {$\cL$};
    \end{tikzpicture} \,\, \rangle\, . 
\end{equation}
Now, pinching the line in between the locations of the two local operators and fusing we get\footnote{The coefficient $\frac{\sqrt{\langle \cL' \rangle}}{\langle \cL \rangle}$ ensures the proper normalization of the completeness relation, see e.g. \cite{Barkeshli:2014cna}}
\begin{equation}
\begin{split}
     \langle \Phi_1 \Phi_2\rangle  & = \frac{1}{\langle \cL \rangle}\sum_{\cL'}\frac{\sqrt{\langle \cL' \rangle}}{\langle \cL \rangle}\langle \,\, \begin{tikzpicture} [baseline={([yshift=-2ex]current bounding box.center)},vertex/.style={anchor=base,
    circle,fill=black!25,minimum size=12pt,inner sep=2pt}]    \draw[fill=black] (-1.5,0) circle (0.05);
     \draw[fill=black] (1.5,0) circle (0.05);
     \draw[color=red, line width=1, decoration={markings, mark=at position 0.625 with {\arrow{<}}},
        postaction={decorate}] (-1.5,0) circle (0.7);
        \draw[color=red, line width=1, decoration={markings, mark=at position 0.625 with {\arrow{<}}},
        postaction={decorate}] (1.5,0) circle (0.7);
     \draw[color=blue, line width=1,->] (-0.8,0)-- (0,0);
      \draw[color=blue, line width=1] (-0.01,0)-- (0.8,0);
        \draw[color=red, line width=1, decoration={markings, mark=at position 0.625 with {\arrow{<}}},
        postaction={decorate}] (1.5,0) circle (0.7);
         \draw[fill=black] (-0.8,0) circle (0.05);
          \draw[fill=black] (0.8,0) circle (0.05);
     \node[below] at (1.5,0) {$\Phi_2$};
     \node[below] at (-1.5,0) {$\Phi_1$};
     \node[above] at (-1.5,0.7) {$\cL$};
     \node[above] at (1.5,0.7) {$\cL$};
     \node[above] at (0,0.1) {$\cL'$};
    \end{tikzpicture} \,\, \rangle \\ & =   \frac{1}{\langle \cL \rangle}\sum_{\cL'}\frac{\sqrt{\langle \cL' \rangle}}{\langle \cL \rangle}\langle \,\, \begin{tikzpicture} [baseline={([yshift=-2ex]current bounding box.center)},vertex/.style={anchor=base,
    circle,fill=black!25,minimum size=12pt,inner sep=2pt}]    \draw[fill=black] (-1.5,0) circle (0.05);
     \draw[fill=black] (1.5,0) circle (0.05);
     \draw[color=blue, line width=1,->] (-1.5,0)-- (0,0);
      \draw[color=blue, line width=1] (-0.01,0)-- (1.5,0);
     \node[below] at (1.5,0) {$\hat{\cL}_{\bar{\cL'}}\cdot\Phi_2$};
     \node[below] at (-1.5,0) {$\hat{\cL}_{\cL'}\cdot\Phi_1$};
     \node[above] at (0,0.1) {$\cL'$};
    \end{tikzpicture} \,\, \rangle 
\end{split}    
\end{equation}
where the sum over $\cL'$ runs over the lines appearing in the fusion channel $\cL \times \bar{\cL}$. Recall that this channel always contains the identity, so that
\begin{equation}\label{eq:selrule2pt}
    \langle \Phi_1 \Phi_2\rangle = \frac{1}{\langle \cL \rangle^{2}}\langle \hat{\cL}\cdot \Phi_1 \hat{\cL}\cdot\Phi_2\rangle  + \frac{1}{\langle \cL \rangle}\sum_{\cL' \neq \unit}\frac{\sqrt{\langle \cL' \rangle}}{\langle \cL \rangle}\langle \,\, \begin{tikzpicture} [baseline={([yshift=-2ex]current bounding box.center)},vertex/.style={anchor=base,
    circle,fill=black!25,minimum size=12pt,inner sep=2pt}]    \draw[fill=black] (-1.5,0) circle (0.05);
     \draw[fill=black] (1.5,0) circle (0.05);
     \draw[color=blue, line width=1,->] (-1.5,0)-- (0,0);
      \draw[color=blue, line width=1] (-0.01,0)-- (1.5,0);
     \node[below] at (1.5,0) {$\hat{\cL}_{\bar{\cL'}}\cdot\Phi_2$};
     \node[below] at (-1.5,0) {$\hat{\cL}_{\cL'}\cdot\Phi_1$};
     \node[above] at (0,0.1) {$\cL'$};
    \end{tikzpicture} \,\, \rangle \, .
\end{equation}
This is the general selection rule implied by the presence of a non-invertible line on two point functions, we see that it relates correlators of local operators to those of the twisted sectors. From this expression one can show that if, say, $\Phi_1$ is invariant under $\cL$, i.e. $\hat{\cL}\cdot \Phi_1 = \langle \cL \rangle \Phi_1$, then all non-trivial lasso actions annihilate $\Phi_1$ \cite{Yin:2018DH}. The proof is simple enough, take $\Phi_2 = \Phi_1$, then the first term on the right hand side of \eqref{eq:selrule2pt} already saturates the left hand side. The remaining terms then have to give zero, but since each can be interpreted as the norm squared of a vector in a twisted Hilbert space (we are assuming a unitary theory), they each vanish separately, implying that the image vector of $\Phi_1$ under the lasso $\hat{\cL}_{\cL'}$ is null. In other words 
\begin{equation}
    \begin{tikzpicture} [baseline={([yshift=-2ex]current bounding box.center)},vertex/.style={anchor=base,
    circle,fill=black!25,minimum size=12pt,inner sep=2pt}]    \draw[fill=black] (0,0) circle (0.05);
     \draw[color=red, line width=1, decoration={markings, mark=at position 0.625 with {\arrow{<}}},
        postaction={decorate}] (0,0) circle (1);
     \node[below] at (0.0,0) {$\Phi$};
     \node[above] at (0,1) {$\cL$};
    \end{tikzpicture} = \langle \cL \rangle \Phi \, \, \,  \Rightarrow \begin{tikzpicture}[baseline={([yshift=-1ex]current bounding box.center)},vertex/.style={anchor=base,
    circle,fill=black!25,minimum size=12pt,inner sep=2pt}]      \draw[fill=black] (0,0) circle (0.05);
     \draw[color=red, line width=1, decoration={markings, mark=at position 0.625 with {\arrow{<}}},
        postaction={decorate}] (0,0) circle (1);
     \draw[color=blue, line width=1, ->] (0,1) -- (0,1.5);
     \draw[color=blue, line width=1] (0,1.49) -- (0,2);
     \draw[fill=black] (0,1) circle (0.05);
     \node[below] at (0,0) {$\Phi$};
     \node[above] at (0,2) {$\cL'$};
     \node[left] at (-1,0) {$\cL$};
    \end{tikzpicture}= 0\, .
\end{equation}
Now take the selection rule \eqref{eq:selrule2pt} with $\Phi_1$ and $\Phi_2$ different and suppose $\Phi_1$ is invariant. By the argument above all contributions from twisted sectors vanish and we have 
\begin{equation}
    \langle \Phi_1 \Phi_2\rangle = \frac{1}{\langle \cL \rangle^{2}}\langle \hat{\cL}\cdot \Phi_1 \hat{\cL}\cdot\Phi_2\rangle  = \frac{1}{\langle \cL \rangle}\langle \Phi_1 \hat{\cL}\cdot\Phi_2\rangle \, , 
\end{equation}
thus for the correlator to be non-zero also $\Phi_2$ has to be invariant. This implies that, as we move away from the Gepner point preserving some non-invertible line, the mixed components of the Zamolodchikov metric involving the perturbation and any other marginal operator not invariant under the preserved lines vanish. A similar selection rule can be derived for three-point function on the sphere
\begin{equation}
\begin{split}
    \label{eq:selrule3pt}
    \langle \Phi_1 \Phi_2 \Phi_3\rangle  & = \frac{1}{\langle \cL \rangle^{2}}\langle \hat{\cL}\cdot \Phi_1 \hat{\cL}\cdot\Phi_2\hat{\cL}\cdot\Phi_3\rangle  +\\ & + \frac{1}{\langle \cL \rangle^{2}}\sum_{\substack{\cL' \neq \unit \\ \cL'' \neq \unit}}\sqrt{\langle \cL' \rangle\langle \cL'' \rangle}\langle \,\, \begin{tikzpicture} [baseline={([yshift=-0.5ex]current bounding box.center)},vertex/.style={anchor=base,
    circle,fill=black!25,minimum size=12pt,inner sep=2pt}]    \draw[fill=black] (-1,0) circle (0.05);
     \draw[fill=black] (1,0) circle (0.05);
     \draw[fill=black] (3,0) circle (0.05);
     \draw[color=blue, line width=1,->] (-1,0)-- (0,0);
     \draw[color=blue, line width=1] (-0.01,0)-- (1,0);
      \draw[color=green, line width=1, ->] (0.99,0)-- (2,0);
      \draw[color=green, line width=1] (1.99,0)-- (3,0);
     \node[below] at (1,0) {$\hat{\cL}_{\bar{\cL'}}\cdot\Phi_2$};
     \node[below] at (-1,0) {$\hat{\cL}_{\cL'}\cdot\Phi_1$};
     \node[below] at (3,0) {$\hat{\cL}_{\cL'}\cdot\Phi_3$};
     \node[above] at (0,0.1) {$\cL'$};
     \node[above] at (2,0.1) {$\cL''$};
    \end{tikzpicture} \,\, \rangle \, .
\end{split}    
\end{equation}
Now, if two out of three operators are invariant the correlator is non vanishing only if the also the third operator commutes with the line $\cL$:
\begin{equation}
    \langle \Phi_1 \Phi_2 \Phi_3\rangle   = \frac{1}{\langle \cL \rangle}\langle  \Phi_1 \Phi_2\hat{\cL}\cdot\Phi_3\rangle \neq 0 \rightarrow \hat{\cL}\cdot\Phi_3 = \langle \cL \rangle \Phi_3\, .
\end{equation}
When the $\Phi_i$ are BPS operators this selection rule can be translated as a constraint on the moduli dependence of the chiral ring coefficients. It implies that, as we move along submanifolds of the moduli space, certain chiral ring coefficients are forced to vanish.

\subsection{Constraints on Conformal Perturbation Theory}\label{sec:perturb}

Besides selection rules in the deformed theory we can use the full symmetry at the Gepner point to simplify the use of conformal perturbation theory to compute corrections to conformal weights and 3-point function coefficients. For concreteness let us focus on 2-point correlators, but the method applies to higher point functions as well. Suppose we turn on the marginal deformation $\cO$, in our case this will be expressed as $\cO = G \bar{G} \Phi$ where $\Phi$ is a BPS primary and $G, \bar{G}$ are the appropriate supercharges. Two-point functions of the deformed theory can be written
\begin{equation}
     \langle \Phi_1 \Phi_2\rangle_{\lambda}=\langle \Phi_1 \Phi_2 e^{\lambda \int d^{2}w \cO(w)}\rangle
\end{equation}
and the corrections to the  weights of $\Phi_1$ and $\Phi_2$ form a power series in $\lambda$
\begin{equation}
    h(\lambda) = \sum_{n=0}^{\infty} h^{(n)}\lambda^{n}\, , 
\end{equation}
with the $n$-th term determined by the integrated correlation function
\begin{equation}
    \int d^{2}w_1 \ldots d^{2}w_n \langle \Phi_1 \Phi_2 \cO(w_1) \ldots \cO(w_n) \rangle
\end{equation}
computed at the Gepner point. By using the selection rules implied by the Tube algebra we can find patterns of zeros in the series of correction. Consider for example the case of the $K3$ sigma model and take the deformation to be the one deriving from the chiral-chiral state
\begin{equation}
    (\ket{2,2})^{\otimes 2}\otimes(\ket{0,0})^{\otimes 2} \otimes  (\overline{\ket{2,2}})^{\otimes 2}\otimes(\overline{\ket{0,0}})^{\otimes 2}\, .
\end{equation}
To use the selection rules implied by the Tube algebra we have to compute the tensor products of all the representations of the Tube algebra associated to the operator insertions. In the following we will always indicate a representation $\Gamma$ of the Tube algebra by the corresponding $3d$ anyon, in particular we will write
\begin{equation}
    \Gamma = \left(\vec{\mu}, \vec{\bar{\mu}}\right)
\end{equation}
where $\vec{\mu}, \vec{\bar{\mu}}$ are $2r$-components vectors containing the Wilson lines labels
\begin{equation}
    \vec{\mu} = ((a_1, c_1), \ldots , (a_r, c_r)); \qquad (a_i, c_i) \in Q_{k_i}
\end{equation}
Let's start from discussing the representation of $\cO$. The primary $\Phi$ is in a representation
\begin{equation}
    \Gamma_0 = \left(\vec{\mu}_0, \vec{\mu}_0\right)\qquad \vec{\mu}_0 = ((2,2), (2,2), (0,0), (0,0)) 
\end{equation}
which corresponds to an invertible line in $3d$. The supercharges of the diagonal superalgebra instead are in a reducible representation of the Tube algebra. More precisely, for a general Gepner model,  
\begin{equation}
    \Gamma_G = \bigoplus_{i=1}^{r}\left(\vec{\mu}_i, \vec{0}\right) \qquad \Gamma_{\bar{G}} = \bigoplus_{i=1}^{r}\left(\vec{0},\vec{\mu}_i\right)
\end{equation}
where
\begin{equation}
    \vec{\mu}_i = \left((0,0), \ldots (k_i, k_i+2), \ldots, (0,0)\right)\, .
\end{equation}
Now, in a correlator with $n$ insertion of $\cO$ we have to compute the $n$-th tensor powers of the three representations $\Gamma_0,\Gamma_G$ and $\Gamma_{\bar{G}}$. This is greatly simplified by the fact that the irreducible factors in $\Gamma_G$ and $\Gamma_{\bar{G}}$ corresponds to lines that are not only invertible but also of order $2$. We find
\begin{equation}
\begin{split}
    & \Gamma_G^{\otimes n} = \bigoplus_{k_1+\ldots +k_4 = n}\binom{n}{k_1 \dots k_4}\left(\vec{\mu}_{k_1, \ldots, k_4},\vec{0}\right)\, ,  \\ & \vec{\mu}_{k_1,\ldots, k_4}= \left((2[k_1], 4[k_1]), (2[k_2], 4[k_2]), (2[k_3], 4[k_3]), (2[k_4], 4[k_4]),\vec{0}\right)
\end{split}
\end{equation}
and similarly for $\Gamma_{\bar{G}}$. It is also easy to compute tensor powers of $\Gamma_0$, we have
\begin{equation}
    \Gamma_0^{\otimes n} = \left(\vec{\mu}_0^{\otimes n}, \vec{\mu}_0^{\otimes n}\right) \qquad \vec{\mu}_0^{\otimes n} = ((2[n], 2n), (2[n], 2n), (0,0), (0,0))\, , 
\end{equation}
notice that only $n \bmod 4$ matters as $\Gamma_0^{\otimes 4} = \unit$. Then the representations entering in the correlators are
\begin{equation}
    \Gamma_{\cO}^{\otimes n} = \Gamma_G ^{\otimes n} \otimes \Gamma_{\bar{G}}^{\otimes n}\otimes \Gamma_{0}^{\otimes n} = \bigoplus_{k_i, \bar{k_i}}\binom{n}{k_1 \ldots k_4}\binom{n}{\bar{k_1} \ldots \bar{k_4}}\left( \vec{\mu}_{\cO^n}, \vec{\bar{\mu}}_{\cO^n}\right)
\end{equation}
with 
\begin{equation}
    \vec{\mu}_{\cO^n}= ((2[k_1 +n], 4[k_1]+ 2n), (2[k_2 +n], 4[k_2]+ 2n), (2[k_3], 4[k_3]), (2[k_4], 4[k_4]))
\end{equation}
and similarly for $\vec{\bar{\mu}}_{\cO^n}$. To give a concrete example we consider the $12$ lightest non-BPS primaries corresponding to the states $\ket{\phi_{ij}}$, these are all of the form
\begin{equation}
    \ket{\phi_{12}} = \ket{1,1}\otimes \ket{1,-1} \otimes (\ket{0,0})^{\otimes 2} \otimes \overline{\ket{1,1}}\otimes \overline{\ket{1,-1}} \otimes (\overline{\ket{0,0}})^{\otimes 2}
\end{equation}
with $i$ and $j$ denoting the tensor factor with $\ket{1,1}$ and $\ket{1, -1}$ respectively. Note that $\phi_{ij}^{\dagger} = \phi_{ji}$. These are all degenerate operators with $h=\bar{h}=1/4$ and vanishing $R$-charges. We are interested in the two point functions $\langle \phi_{ij}^{\dagger} \phi_{lk}\rangle_{\lambda}$. The associated tube algebra representations $\Gamma_{\phi_{ij}}$ are of the form
\begin{equation}
    \Gamma_{\phi_{12}} = \left(\vec{\mu}_{\phi_{12}}, \vec{\mu}_{\phi_{12}}\right) \qquad \vec{\mu}_{\phi_{12}} = ((1,1), (1, -1), (0,0),(0,0))\, ,  
\end{equation}
and correspond to non-invertible lines. The tensor products $\vec{\mu}_{\phi_{ji}}\otimes \vec{\mu}_{\phi_{kl}}$ contain the identity if and only if $k=i, l=j$, thus the only non-zero correlators at the Gepner point are $\langle \phi_{ji}\phi_{ij}\rangle$. Turning on $\lambda$ we can have mixing among the operators, which is constrained by the selection rules. Since  $\Gamma_{\cO^n}$ only contains invertible anyons we see that a necessary condition for the identity to appear in $\vec{\mu}_{\phi_{ji}}\otimes \vec{\mu}_{\phi_{kl}}\otimes \vec{\mu}_{\cO^{n}}$ is that $\vec{\mu}_{\phi_{ji}}\otimes \vec{\mu}_{\phi_{kl}}$ contains at least one invertible line. This immediately shows that the only non-vanishing 2-point functions are those of the form $\langle \phi_{ji}\phi_{ij}\rangle_{\lambda}$ or $\langle \phi_{ij}\phi_{ij}\rangle_{\lambda}$. For those correlators the relevant representations are of the form
\begin{equation}
\begin{split}
    & \Gamma_{12}\otimes \Gamma_{12} = \left(\vec{\mu}_{\phi_{12}}^{\otimes 2},\vec{\mu}_{\phi_{12}}^{\otimes 2}\right)\, , \qquad\vec{\mu}_{\phi_{12}}^{\otimes 2} = ((0,2) \oplus (2,2), (0, -2) \oplus (2, -2), (0,0), (0,0))\, \\ & \Gamma_{12}\otimes \Gamma_{21} = \left(\vec{\mu}_{\phi_{12}}\otimes\vec{\mu}_{\phi_{21}},\vec{\mu}_{\phi_{12}}\otimes\vec{\mu}_{\phi_{21}}\right)\, , \\ &  \qquad\qquad\qquad\qquad \qquad\qquad\vec{\mu}_{\phi_{12}}\otimes\vec{\mu}_{\phi_{21}} = ((0,0) \oplus (2,0), (0, 0) \oplus (2, 0), (0,0), (0,0))\, . 
\end{split}
\end{equation}
This conclusion holds for any deformation such that $\vec{\mu}_{\cO^{n}}$ contains only invertible lines, in our specific example however we can do better. Indeed also all correlators of the form $\langle \phi_{ij}\phi_{ij}\rangle_{\lambda}$ vanish whenever $i$ or $j$ is different than $1$ or $2$. This is because when tensoring, say $\vec{\mu}_{\phi_{13}}^{\otimes 2}$ with $\vec{\mu}_{\cO^{n}}$ we get terms that in the third tensor factor have the pair $(2[k_3], 4[k_3]+2)$, which never trivializes allowing
the singlet representation. It follows that the correlators to consider are $\langle \phi_{12}\phi_{12}\rangle_{\lambda}$ and $\langle \phi_{ji}\phi_{ij}\rangle_{\lambda}$, namely the only mixing allowed by the deformation is between $\phi_{12}$ and itself. We can also study the power series in $\lambda$ in more detail, starting from $\langle \phi_{12}\phi_{12}\rangle_{\lambda}$.
The selection rule requires the product $\Gamma_{\phi_{12}}^{\otimes 2} \otimes \Gamma_{\cO}^{\otimes n}$ to contain the identity, this forces us to choose $k_3, \bar{k_3}$ and $k_4, \bar{k_4}$ even, then any representation appearing in the decomposition in the tensor product is of the form $(\vec{\mu}_{\phi}^{\otimes 2} \otimes \vec{\mu}_{\cO^{n}},\vec{\mu}_{\phi}^{\otimes 2} \otimes \vec{\bar{\mu}}_{\cO^{n}})$ with
\begin{equation}
\begin{split}
\vec{\mu}_{\phi}^{\otimes 2} \otimes \vec{\mu}_{\cO^{n}} = & \Big((2[k_1+n], 4[k_1]+2n+2)\oplus (2[k_1+n+1], 4[k_1]+2n+2),\\ & (2[k_1+n], 4[k_2]+2n-2)\oplus (2[k_2+n+1], 4[k_2]+2n-2), (0,0), (0,0)\Big) \, .
\end{split}
\end{equation}
Now we notice that when $n=0,2 \bmod 4$ there is no value of $k_1$ or $k_2$ such that
\begin{equation}
    4[k_1]+2n + 2 = 4[k_1]\pm 2 = 0 \bmod 8\, 
\end{equation}
and the singlet representation appears only when $n=1,3 \bmod 4$. Therefore, the power series in only contains the odd powers $\lambda^{2m+1}$. Another two-point function we can consider is $\langle \phi_{21}\phi_{12}\rangle_{\lambda}$. In this case, for the $n$-th order correction we find 
\begin{equation}
\begin{split}
     \vec{\mu}_{\phi}\otimes \vec{\mu}_{\phi^{\dagger}} \otimes \vec{\mu}_{\cO^{n}} = & \Big((2[k_1+n], 4[k_1]+2n)\oplus (2[k_1+n+1], 4[k_1]+2n),\\ & (2[k_1+n], 4[k_2]+2n)\oplus (2[k_2+n+1], 4[k_2]+2n), (0,0), (0,0)\Big) \, 
\end{split}
\end{equation}
which shows that for $n=1, 3 \bmod 4$ the identity does not appear in the tensor product, and the series is in even powers $\lambda^{2m}$. These result are compatible with those of \cite{Keller:2023ssv}.
Similar computations can be repeated for other operators, more complicated correlation functions or more general Gepner models.


\section*{Acknowledgements}

We thank Diego Garc\'{i}a-Sep\'{u}lveda for helpful conversations. CC acknowledges support from the Simons Collaboration on Global Categorical Symmetries, the US Department of Energy Grant 5-29073, and the Sloan Foundation. GR is supported by the ERC-COG grant NP-QFT No.~864583 ``Non-perturbative dynamics of quantum fields: from new deconfined phases of
matter to quantum black holes”,
by the MIUR-FARE grant EmGrav No.~R20E8NR3HX ``The Emergence of Quantum Gravity from Strong Coupling Dynamics'',
and by the INFN ``Iniziativa Specifica ST\&FI”. GR thanks the Enrico Fermi Institute and Kadanoff Center for Theoretical Physics at the University of Chicago for hospitality during the development of this work. 

\section*{Conflict of Interest}
There are no potential sources of conflict of interest.

\appendix

\section{Superconformal Symmetry}\label{app:conv}
In this Appendix we collect several facts on the $\cN=2$ superconformal algebra and summarize our conventions. The $\cN=2$ multiplet containing the stress energy tensor $T_B(z)$ also includes two fermionic supercurrents $T_F^{\pm}(z)$ as well as a $U(1)_R$ current $J(z)$. The non zero OPEs are (equality below is up to regular terms)
\begin{equation}
    \begin{split}
        & T_B(z)T_B(0) = \frac{c}{2 z^4} + \frac{2}{z^2}T_B(0) + \frac{1}{z}\partial T_B(0)\\
        & T_B(z)T_F^{\pm}(0) = \frac{3}{2z^2}T_F^{\pm}(0) + \frac{1}{z}\partial T_F^{\pm}(0)\\
        & T_B(z)J(0) = \frac{1}{z^2}J(0) + \frac{1}{z}\partial J(0)\\
        & T_F^{+}(z)T_F^{-}(0) = \frac{2c}{3 z^2} + \frac{2}{z^2}J(0) + \frac{2}{z}\partial T_B(0)+ \frac{1}{z}\partial J(0)\\
        & J(z) T_F^{\pm}(0) = \pm \frac{1}{z}T_F^{\pm}(0)\\
        & J(z)J(0) = \frac{c}{3 z^2}
    \end{split}
\end{equation}
Besides the regular $T_B$ OPE these tell us that $T_F^{\pm}$ are (Virasoro) primary fields with weight $3/2$ and $U(1)_R$ charge $\pm 1$. As usual the conserved $R$-current $J(z)$ is a primary of weight $1$. On the cylinder we can decompose these fields in Fourier modes, then mapping back to the punctured plane we have
\begin{equation}
   \begin{split}
        & T_B(z) = \sum_{n \in \bZ} \frac{L_n}{z^{n+2}}\, , \qquad J(z) = \sum_{n \in \bZ} \frac{j_n}{z^{n+1}}\, , \qquad T_F^{\pm}(z) = \sum_{r \in \bZ \pm \nu } \frac{G_r^{\pm}} {z^{r+ 3/2}}\, .
   \end{split}
\end{equation}
where $\nu$ depends on the spin structure chosen: $\nu = 0$ corresponds to antiperiodic boundary conditions for the fermions (Ramond sector) while $\nu = 1/2$ gives periodic fermions (NS sector). The algebra of modes is 
\begin{equation}
    \begin{split}
        & \left[L_m, G_r^{\pm}\right] = \left(\frac{m}{2} - r\right)G_{m + r}^{\pm}\, , \qquad \left[L_m, L_n\right]=(m-n)L_{m+n} + \frac{c}{12}(m^3-m)\delta_{m+n,0}\\
        &\left[L_m, j_n\right]= -n j_{m+n}\, , \qquad \left[j_m, j_n\right] = \frac{c}{3}m\delta_{m+n,0}\, , \qquad \left[j_m, G^{\pm}_r\right] = \pm G_{m+r}^{\pm}\, , \\ &
        \left\{G_{r}^{+}, G_s^{-}\right\} = 2 L_{r+s} + (r-s)j_{r+s} + \frac{c}{3}\left(r^2 - \frac{1}{4}\right)\delta_{r+s,0}\, .
    \end{split}
\end{equation}
A convenient choice for the Cartan subalgebra is the pair $L_0, j_0$ so that states in representation spaces are labelled by both their conformal weight $h$ and the $U(1)_R$ charge $q$
\begin{equation}
    L_0\ket{h,q} = h \ket{h,q} \qquad  j_0\ket{h,q} = q \ket{h,q}\, . 
\end{equation}
Irreducible representations of this algebra are lowest weight representations (LWR) built on top of a superconformal primary state $\ket{h,q}$ such that
\begin{equation}
    L_n\ket{h,q} = j_n\ket{h,q} = G_{r}^{\pm}\ket{h,q}= 0 \qquad \forall \, n,r > 0\, .
\end{equation}
A unitary representation is one in which we have an hermitian conjugation operation with respect to which $(L_m)^{\dagger} = L_{-m}$, $(G^{+}_{r})^{\dagger} = G^{-}_{-r}$ and $(j_n)^{\dagger} = j_{-n}$. Unitarity puts strong constraints on the spectrum of allowed weights and $U(1)_R$ charges for a given central charge. A simple unitarity bound in the NS sector is obtained imposing 
\begin{equation}\label{eq:unitarity}
    0 \leq \bra{h,q}\left\{G^{+}_{\mp 1/2}, G^{-}_{\pm 1/2}\right\}\ket{h,q} = \bra{h,q}\left(2L_0 \mp j_0\right)\ket{h,q} = 2 h \mp q 
\end{equation}
that is states in a unitary representation in the NS sector obey $h \ge |q|/2$. For more details see e.g. \cite{Eguchi:1988af,Boucher:1986bh,Lin:2016gcl}. Since the superconformal algebra includes Virasoro as a subaglebra we can split its representations in Virasoro irreps. This basically amounts to find states annihilated only by the positive Virasoro modes. Let's consider a superconformal primary $\ket{h,q}$ and its fermionic descendants $G_{-r}^{\pm}\ket{h,q}$ with $r>0$. We have
\begin{equation}
    L_m G_{-r}^{\pm}\ket{h,q} = [L_m, G_{-r}^{\pm}]\ket{h,q} + G_{-r}^{\pm}L_m\ket{h,q} = \left(\frac{m}{2}+ r\right)G_{m-r}^{\pm}\ket{h,q}
\end{equation}
which vanishes only for $m>r$. Therefore these states are not Virasoro primaries. We can obtain further Virasoro primaries considering states obtained acting on $\ket{h,q}$ with products of fermionic generators $G_{-r}^{\pm}$ with different values of $r$. For instance in the NS sector one easily sees that $G_{-1/2}^{\pm}\ket{h,q}$ are Virasoro primaries while $G_{-3/2}^{\pm}\ket{h,q}$ are not. The next lowest weight Virasoro primary are instead $G_{-1/2}^{\pm}G_{-3/2}^{\pm}\ket{h,q}$, indeed
\begin{equation}
\begin{split}
    L_mG_{-1/2}^{\pm}G_{-3/2}^{\pm}\ket{h,q} &= [L_m,G_{-1/2}^{\pm}]G_{-3/2}^{\pm}\ket{h,q}+ G_{-1/2}^{\pm}[L_m,G_{-3/2}^{\pm}]\ket{h,q}  \\ & = \frac{m+1}{2}G_{m-1/2}^+G_{-3/2}^+\ket{h,q} +  \frac{m+3}{2}G_{-1/2}^+G_{m-3/2}^+\ket{h,q}
\end{split}
\end{equation}
which vanishes for all $m > 1$ due to $\ket{h,q}$ being a primary while for $m= 1$ because the state $G^{+}_{-1/2}G^{+}_{-1/2}\ket{h,q}$ is actually null (as one would expect). Thus in general a superconformal family includes an infinite number of conformal ones, with all possible values of the $U(1)_R$ charge.

\subsection{Chiral Ring and Spectral Flow}

There are two useful features of the $\cN =2$ superconformal symmetry, the first is the existence of shortened representations whose lowest weight state is called \textit{chiral primary}, the second is the presence of an external automorphism of the algebra, the \textit{spectral flow}. Chiral primaries and their ring are associated to the NS sector, here one defines chiral states as those such that
\begin{equation}
    G^{+}_{-1/2}\ket{h,q} = 0\, . 
\end{equation}
In a $\cN= (2,2)$ SCFT we have left and right chirals. If $\ket{h,q}$ is also a superconformal primary \eqref{eq:unitarity} shows that chiral primaries saturate the unitarity bound and have $h=q/2$. Interstingly one can also show the converse, thus for a state $\ket{h,q}$ being a chiral primary is equivalent to having $h = q/2$. Now consider the OPE of two chiral primaries $\phi_a$ and $\phi_b$, this has the general form
\begin{equation}
    \phi_a(z)\phi_b(w) = \sum_{c}\sum_{n\in \bN_0}\frac{\partial^n\phi_c}{(z-w)^{h_a+h_b- h_c-n}}\, .
\end{equation}
Since the $R$ charge has to be conserved any operator appearing in the OPE must have $q_c = q_a + q_b$ and thus the unitarity bound implies
\begin{equation}
    h_a + h_b - h_c = \frac{q_a+ q_b}{2} -h_c = \frac{q_c}{2}- h_c \le 0 
\end{equation}
hence the OPE of two chiral primaries is free of singular terms. We can then define a product as the limit of coincident points of the OPE
\begin{equation}
    \left(\phi_a \cdot \phi_b\right)(z) = \lim_{w\rightarrow z}\phi_a(z)\phi_b(w) = \sum_{c}C_{ab}^c \phi_c(z)\, . 
\end{equation}
The rhs of the product cannot contain terms with derivatives, indeed it can only involve operators with $q_c/2 = h_c$, i.e. other chiral primaries. This product the closes on chiral primaries and endows them with a ring structure. In an $\cN=(2,2)$ theory we have four of these rings depending on wheter we take a chiral or antichiral state on the left or on the right. 

The other interesting feature of the $\cN=2$ algebra is the spectral flow. This is the following one parameter deformation of the generators
\begin{equation}
\begin{split}
    & L_n' = L_n + \eta j_n + \frac{\eta^{2}}{6 }c \delta_{n,0}\\
    & j_n' = j_n + \frac{c}{3}\eta \delta_{n,0}\\
    & G_{r}^{\pm \, '} = G_{r\pm \eta}^{\pm} \, , 
\end{split}
\end{equation}
one easily checks that the primed generators satisfy the same algebra of the unprimed ones. Notice also that the flow changes the moding of the fermionic generators, so, for $\eta \in \bZ/2$ it interpolates between NS and R sectors. Since this is an automorphism of the algebra it maps representations one into the other. Introducing a unitary operator $U_{\eta}$ that implements the flow as
\begin{equation}
\begin{split}
    & L_n' = U_\eta L_n U_\eta^\dagger \\
    & j_n' = U_\eta j_n U_\eta^\dagger\\
    & G_{r}^{\pm \, '} = U_\eta G_r^{\pm} U_\eta^\dagger \, , 
\end{split}
\end{equation}
we can spectrally flow a representation acting with $U_\eta$ on the various states. In particular a state $\ket{h,q}$ is mapped to $U_{\eta}\ket{h,q}$, combining the relations above it is easy to show that 
\begin{equation}\label{eq:spflo}
\begin{split}
    &  L_0 U_\eta \ket{h,q} = \left(h - \eta q + \frac{\eta^2 c}{6}\right)U_\eta \ket{h,q}\\
    & j_0  U_\eta \ket{h,q} = \left( q- \frac{\eta c }{3}\right)U_\eta \ket{h,q}
\end{split}
\end{equation}
thus states in the spectrally flowed representations are still eigenstates of $L_0$ and $j_0$. Given a superconformal primary $\ket{h,q}$ we see that
\begin{equation}
    L_m U_\eta \ket{h,q} = j_m U_{\eta}\ket{h,q} = 0 \qquad \forall \, m > 0
\end{equation}
while 
\begin{equation}
    G_{r}^{\pm}U_\eta\ket{h,q} = U_\eta G^{\pm}_{r \mp \eta}\ket{h,q}
\end{equation}
which vanishes only for $r \mp \eta >0$. Thus a LWR representation will be mapped to another LWR as long as we allows various moding of the fermionic generators. As an example let's consider the spectral flow with $\eta = 1/2$ of a chiral primary representation. The chiral primary $\ket{q/2,q}$ flows to a state with weight $c/24$ and charge $q- c/6$ annihilated by all positive modes of $T_B$ and $J$ as well as the $G_r^{\pm}$ with $r \in \bN$, i.e. a superconformal primary in the Ramond sector. As we choose different chiral primaries to flow we obtain degenerate states that differ for their $R$-charge. It is also easy to show that these are ground states in the R sector, we compute 
\begin{equation}\label{eq:unitarityR}
    0 \le |G^{+}_0\ket{h,q}|^2 + |G^{-}_0\ket{h,q}|^2 = \bra{h,q}\{G_0^+, G_0^-\}\ket{h,q} = 2 \left( h-\frac{c}{24}\right)|\ket{h,q}|^2
\end{equation}
so unitarity implies $h \ge c/24$ and the ground states above saturate the bound. We can also define a spectral flow operator looking at the image under $U_\eta$ of the NS vacuum $\ket{0,0}$, this then has weight $\eta^2 c/6$ and charge $-\eta c/3$. 

\section{ Verlinde Formulas, and Modularity for Superconformal Primaries}\label{app:sVerl}

In this Appendix we write down the modular transformations of the characters of the full superconformal representations and derive Verlinde like formulas for their fusion. From the modular $S$-matrix of the half-character is easy to derive the modular transformations of the full characters
\begin{equation}
\begin{split}
   & S\cdot \text{ch}^{(\text{NS})}_{l,m} = \sum_{l',m'}S^{\text{NSNS}}_{lm; l'm'}\text{ch}^{(\text{NS})}_{l',m'} \qquad \qquad\qquad
    S\cdot \text{ch}^{(\text{R})}_{l,m} = \sum_{l',m'}S^{\text{R}\widetilde{\text{NS}}}_{lm; l'm'}\widetilde{\text{ch}}^{(\text{NS})}_{l',m'} \\
   & S\cdot \widetilde{\text{ch}}^{(\text{NS})}_{l,m} = \sum_{l',m'}S^{\widetilde{\text{NS}}\text{R}}_{lm; l'm'}\text{ch}^{(\text{R})}_{l',m'}  \qquad \qquad\qquad S\cdot \widetilde{\text{ch}}^{(\text{R})}_{l,m} = \sum_{l',m'}S^{\widetilde{\text{R}}\widetilde{\text{R}}}_{lm; l'm'}\widetilde{\text{ch}}^{(\text{R})}_{l',m'} \\
\end{split}
\end{equation}
where
\begin{equation}
    \begin{split}
        & S^{\text{NSNS}}_{lm; l'm'} = \frac{2}{k+2}\sin\left(\frac{\pi(l+1)(l'+1)}{k+2}\right)e^{i \pi \frac{m m'}{k+2}} \\
        & S^{\text{R}\widetilde{\text{NS}}}_{lm; l'm'} = \frac{2}{k+2}\sin\left(\frac{\pi(l+1)(l'+1)}{k+2}\right)e^{i \pi \frac{(m+1) m'}{k+2}} \\
        & S^{\widetilde{\text{NS}}\text{R}}_{lm; l'm'} = \frac{2}{k+2}\sin\left(\frac{\pi(l+1)(l'+1)}{k+2}\right)e^{i \pi \frac{m (m'+1)}{k+2}} \\
        & S^{\widetilde{\text{R}}\widetilde{\text{R}}}_{lm; l'm'} = -\frac{2 i}{k+2}\sin\left(\frac{\pi(l+1)(l'+1)}{k+2}\right)e^{i \pi \frac{(m+1) (m'+1)}{k+2}} \, .
    \end{split}
\end{equation}
which correctly mimic the action of $SL(2, \bZ)$ on spin structures. One can verify that those $S$ matrices are unitary and furnish a representation of $SL(2, \bZ)$, i.e. $S^4 = \unit$. In verifying this last property one should be careful in taking into account the action on spin structures. In particular it makes no sense to square $S^{\text{R}\widetilde{\text{NS}}}$ or $S^{\widetilde{\text{NS}}\text{R}}$, rather the charge conjugation matrices in the R and $\widetilde{\text{NS}}$ sectors are, respectively, $C^{\text{R}}=S^{\text{R}\widetilde{\text{NS}}}S^{\widetilde{\text{NS}}\text{R}}$ and $C^{\widetilde{\text{NS}}}=S^{\widetilde{\text{NS}}\text{R}}S^{\text{R}\widetilde{\text{NS}}}$, while $C^{\text{NS}} = S^{\text{NSNS}}S^{\text{NSNS}}$ and $C^{\widetilde{\text{R}}}=S^{\widetilde{\text{R}}\widetilde{\text{R}}}S^{\widetilde{\text{R}}\widetilde{\text{R}}}$. Neither of these matrices is the identity, but they all square to it, thus at least one representations in each sector is not self-conjugate. 

Knowing the fusion coefficients $N_{ac; a'c'}^{a''c''}$ of the half-families we can extract the fusion coefficients for the full superconformal families as simply
\begin{equation}
    \widehat{N}_{ac;a'c'}^{\alpha, \gamma}=N_{a c ; a^{\prime} c^{\prime}}^{\alpha \gamma} + N_{a c ; a^{\prime} c^{\prime}}^{k-\alpha \gamma+k+2} 
\end{equation}
where now $(a,c), (a',c'), (\alpha, \gamma) \in P'_k$ label a superconformal primary rather than an half-family. We now want to separate out the NS and R sectors explicitly and write down Verlinde formulas in the various sectors. We first notice that
\begin{equation}
     N_{a c ; a^{\prime} c^{\prime}}^{\alpha \gamma}=\sum_{(d, f) \in P'_k} \frac{S_{a c ; d f} S_{a^{\prime} c^{\prime} ; d f} S_{\alpha \gamma ; d f}^*}{S_{00 ; d f}}(1 + (-1)^{a+c+a'+c'+\alpha + \gamma})
\end{equation}
and
\begin{equation}
    N^{k-\alpha \, \gamma + k+2}_{ac;a'c'} = \sum_{(d, f) \in P'_k} \frac{S_{a c ; d f} S_{a^{\prime} c^{\prime} ; d f} S_{\alpha \gamma ; d f}^*}{S_{00 ; d f}}(1 + (-1)^{a+c+a'+c'+\alpha + \gamma})(-1)^{d+f}
\end{equation}
so the fusion coefficients are non-zero only when $a+c+a'+c'+\alpha + \gamma = 0 \bmod 2$. Now switching to the $(l,m, \lambda)$ parametrization we find
\begin{equation}
\begin{split}
     \widehat{N}^{l'',m'',\lambda''}_{lm\lambda; l'm'\lambda'} & = \left(1+(-1)^{2(\lambda+\lambda'+\lambda'')}\right) \sum_{(r,s) \in P_k; x= 0,-1/2}\frac{S_{lm\lambda; rsx}S_{l'm'\lambda'; rsx}S^*_{l''m''\lambda'';rsx}}{S_{000;rsx}}(1+(-1)^{2x})\\
     & = \left(1+(-1)^{2(\lambda+\lambda'+\lambda'')}\right) \sum_{(r,s) \in P_k}\frac{S_{lm\lambda; rs0}S_{l'm'\lambda'; rs0}S^*_{l''m''\lambda'';rs0}}{S_{000;rs0}} \, . 
\end{split}
\end{equation}
Notice that the modular matrix $S^{\widetilde{\text{R}}\widetilde{\text{R}}}$ can never appear in these expressions. Recalling that $\lambda = 0$ is NS and $\lambda = -1/2$ is R we see that there are four fusion channels 
\begin{equation}
    \begin{split}
        & \text{NS}\times \text{NS} = \text{NS} \qquad \text{R}\times \text{R} = \text{NS}\\
         & \text{R}\times \text{NS} = \text{R} \qquad \text{NS}\times \text{R} = \text{R} \, ,
    \end{split}
\end{equation}
for each of those we have a Verlinde formula
\begin{itemize}
    \item $\text{NS}\times \text{NS} = \text{NS}$
    \begin{equation}\label{eq:superver1}
        \widehat{N}^{l''m'', \text{ NSNS}}_{lm; l'm'} = \sum_{(r,s) \in P_k} \frac{S^{\text{NSNS}}_{lm; rs}S^{\text{NSNS}}_{l'm'; rs}\left(S^{\text{NSNS}}_{l''m''; rs}\right)^*}{S^{\text{NSNS}}_{00; rs}}
    \end{equation}
    \item $\text{R}\times \text{NS} = \text{R}$
    \begin{equation}\label{eq:superver2}
        \widehat{N}^{l''m'', \text{ RNS}}_{lm; l'm'} = \sum_{(r,s) \in P_k} \frac{S^{\text{R}\widetilde{\text{NS}}}_{lm; rs}S^{\text{NSNS}}_{l'm'; rs}\left(S^{\text{R}\widetilde{\text{NS}}}_{l''m''; rs}\right)^*}{S^{\text{NSNS}}_{00; rs}}
    \end{equation}
    \item $\text{NS}\times \text{R} = \text{R}$ 
    \begin{equation}\label{eq:superver3}
        \widehat{N}^{l''m'', \text{ NSR}}_{lm; l'm'} = \sum_{(r,s) \in P_k} \frac{S^{\text{NSNS}}_{lm; rs}S^{\text{R}\widetilde{\text{NS}}}_{l'm'; rs}\left(S^{\text{R}\widetilde{\text{NS}}}_{l''m''; rs}\right)^*}{S^{\text{NSNS}}_{00; rs}}
    \end{equation}
    \item $\text{R}\times \text{R} = \text{NS}$ 
    \begin{equation}\label{eq:superver4}
        \widehat{N}^{l''m'', \text{ RR}}_{lm; l'm'} = \sum_{(r,s) \in P_k} \frac{S^{\text{R}\widetilde{\text{NS}}}_{lm; rs}S^{\text{R}\widetilde{\text{NS}}}_{l'm'; rs}\left(S^{\text{NSNS}}_{l''m''; rs}\right)^*}{S^{\text{NSNS}}_{00; rs}}\, .
    \end{equation}
\end{itemize}
from the explicit formulas of the $S$-matrices we see that $S^{\text{R}\widetilde{\text{NS}}}_{lm; rs} = e^{i \pi \frac{s}{k+2}}S^{\text{NSNS}}_{lm,rs}$ then
\begin{equation}
    \widehat{N}^{l''m'', \text{ NSNS}}_{lm; l'm'} =  \widehat{N}^{l''m'', \text{ RNS}}_{lm; l'm'} =\widehat{N}^{l''m'', \text{ NSR}}_{lm; l'm'} \, .
\end{equation}
The positive integers $\widehat{N}^{l''m'', \text{ RR}}_{lm; l'm'}$ can also be related to $ \widehat{N}^{l''m'', \text{ NSNS}}_{lm; l'm'}$ albeit in a less trivial way. In examples we have checked that there exist a permutation of the labels of primaries $\sigma : P_k \rightarrow P_k$ such that
\begin{equation}
    \widehat{N}^{l''m'', \text{ RR}}_{lm; l'm'} =  \widehat{N}^{\sigma(l''m''), \text{ NSNS}}_{\sigma(lm); \sigma(l'm')}\, .
\end{equation}
These integers have the interpretation of fusion coefficients for superconformal primaries. There are however other integers we can construct out of the $S$ matrices by considering
\begin{equation}
    \widehat{M}^{\alpha \gamma}_{ac;a'c'} = N^{\alpha \gamma}_{ac; a'c'} - N^{k-\alpha \gamma +k+2}_{ac; a'c'}
\end{equation}
those are manifestly integers although not necessarily positive. However they obey Verlinde-like formulas:
\begin{itemize}
    \item $\text{NS}\times \text{NS} = \text{NS}$
    \begin{equation}\label{eq:superver1a}
        \widehat{M}^{l''m'',\text{NSNS}}_{lm; l'm'} = \sum_{(r,s) \in P_k} \frac{S^{\widetilde{\text{ NS}}\text{R}}_{lm; rs}S^{\widetilde{\text{ NS}}\text{R}}_{l'm'; rs}\left(S^{\widetilde{\text{ NS}}\text{R}}_{l''m''; rs}\right)^*}{S^{\widetilde{\text{NS}}\text{R}}_{00; rs}}
    \end{equation}
    \item $\text{R}\times \text{NS} = \text{R}$
    \begin{equation}\label{eq:superver2a}
        \widehat{M}^{l''m'', \text{ RNS}}_{lm; l'm'} = \sum_{(r,s) \in P_k} \frac{S^{\widetilde{\text{R}}\widetilde{\text{R}}}_{lm; rs}S^{\widetilde{\text{NS}}\text{R}}_{l'm'; rs}\left(S^{\widetilde{\text{R}}\widetilde{\text{R}}}_{l''m''; rs}\right)^*}{S^{\widetilde{\text{NS}}\text{R}}_{00; rs}}
    \end{equation}
    \item $\text{NS}\times \text{R} = \text{R}$ 
    \begin{equation}\label{eq:superver3a}
        \widehat{M}^{l''m'', \text{ NSR}}_{lm; l'm'} = \sum_{(r,s) \in P_k} \frac{S^{\widetilde{\text{NS}}\text{R}}_{lm; rs}S^{\widetilde{\text{R}}\widetilde{\text{R}}}_{l'm'; rs}\left(S^{\widetilde{\text{R}}\widetilde{\text{R}}}_{l''m''; rs}\right)^*}{S^{\widetilde{\text{NS}}\text{R}}_{00; rs}}
    \end{equation}
    \item $\text{R}\times \text{R} = \text{NS}$ 
    \begin{equation}\label{eq:superver4a}
        \widehat{M}^{l''m'', \text{ RR}}_{lm; l'm'} = \sum_{(r,s) \in P_k} \frac{S^{\widetilde{\text{R}}\widetilde{\text{R}}}_{lm; rs}S^{\widetilde{\text{R}}\widetilde{\text{R}}}_{l'm'; rs}\left(S^{\widetilde{\text{NS}}\text{R}}_{l''m''; rs}\right)^*}{S^{\widetilde{\text{NS}}\text{R}}_{00; rs}}
    \end{equation}\, .
\end{itemize}
Again, noticing that $S^{\widetilde{\text{R}}\widetilde{\text{R}}}_{lm;rs} = e^{i \pi \frac{s+1}{k+2}}S^{\widetilde{\text{NS}}\text{R}}_{lm; rs}$ one checks that
\begin{equation}
     \widehat{M}^{l''m'', \text{ NSNS}}_{lm; l'm'} =  \widehat{M}^{l''m'', \text{ RNS}}_{lm; l'm'} =\widehat{M}^{l''m'', \text{ NSR}}_{lm; l'm'} \, .
\end{equation}
Also in this case there exist a permutation relating $\widehat{M}^{l''m'', \text{ RR}}_{lm; l'm'}$ to $\widehat{M}^{l''m'', \text{ NSNS}}_{lm; l'm'} $.

\section{Supersymmetric Boundaries in Minimal Models and Folding Trick}\label{app:boundaries}
In this Appendix we derive the supersymmetric Verlinde lines of a single minimal model using the folding trick \cite{Drukker:2010jp}. The first step is to determine the supersymmetric boundary condition, see \cite{Cardy:2004hm,Ishibashi:1988kg,Recknagel:1997sb,Blumenhagen:2009zz,Recknagel:2013uja} for more details on boundary conditions in CFT.
Supersymmetric boundaries in $\cN=1$ minimal models have been worked out in \cite{Nepomechie:2001bu}. The $\cN=2$ superconformal algebra has an outer automorphism called mirror map
\begin{equation}
    \Omega_M :\begin{cases}
  j_n \rightarrow -j_n   \\
 G_r^{\pm}\rightarrow G^{\mp}_r
  \end{cases}
\end{equation}
thus there are two types of boundary states, the untwisted ones, or B-type
\begin{equation}
    \begin{aligned}
& (L_n-\bar{L}_{-n})\ket{\cB_i}\rangle_B=(j_n+\bar{j}_{-n})\ket{\cB_i}\rangle_B=0 \\
&(G_r^{+}+i \eta \bar{G}_{-r}^{+}\ket{\cB_i}\rangle_B=(G_r^{-}+i \eta \bar{G}_{-r}^{-})\ket{\cB_i}\rangle_B=0
\end{aligned}
\end{equation}
and the twisted ones, or A-type
\begin{equation}
    \begin{aligned}
    & (L_n-\bar{L}_{-n})\ket{\cB_i}\rangle_A=(j_n-\bar{j}_{-n})\ket{\cB_i}\rangle_A=0 \\
&(G_r^{+}+i \eta \bar{G}_{-r}^{-}\ket{\cB_i}\rangle_A=(G_r^{-}+i \eta \bar{G}_{-r}^{+})\ket{\cB_i}\rangle_A=0\, .
\end{aligned}
\end{equation}
In the S-dual channel the boundary conditions are 
\begin{equation}\label{eq:ABbc}
    \begin{split}
        & \text{A-type:} \qquad J(z) = - \bar{J}(\bar{z}) \qquad  G^{\pm}(z) = \eta \bar{G}^{\mp}(\bar{z}) \\
        & \text{B-type:} \qquad J(z) = + \bar{J}(\bar{z}) \qquad  G^{\pm}(z) = \eta \bar{G}^{\pm}(\bar{z}) \qquad    \, .   
    \end{split}
\end{equation}
Here $\eta$ can be any phase in general, choosing $\eta = \pm 1$ one can see that both types of boundary conditions preserve an $\cN=1$ subalgebra. As usual these boundary conditions preserve only one copy of the $\cN=2$ algebra. For the B-type boundary conditions the preserved copy is the diagonal of the holomorphic and antiholomorphic algebras. The parameter $\eta$ labels a continuous family of boundary conditions, let's be more precise about it. Consider an $\cN=2$ SCFT on the upper half-plane and impose the boundary conditions
\begin{equation}\label{eq:bcfermion}
    \begin{split}
        &G^{\pm}(z) = \bar{G}^{\pm}(\bar{z}) \qquad z= \bar{z} >0 \\
        & G^{\pm}(z) = \eta\bar{G}^{\pm}(\bar{z}) \qquad z= \bar{z} <0\, .
    \end{split}
\end{equation}
As in the doubling trick we can construct an holomorphic field on the whole complex plane by 
\begin{equation}
    \mathfrak{G}^{\pm}(z) = \begin{cases}
        G^{\pm}(z) \qquad \text{Im}(z) >0 \\
        \bar{G}^{\pm}(\bar{z}) \qquad \text{Im}(z)<0
    \end{cases}
\end{equation}
which is not single valued in the complex plane as it obeys
\begin{equation}
    \mathfrak{G}^{\pm}(e^{2 i \pi }z) = \eta \mathfrak{G}^{\pm}(z)\, . 
\end{equation}
This can be interpreted as the insertion at the origin of a twist defect for the $U(1)_R$ symmetry, which comes with an attached topological defect line $L_\eta$ implementing $\eta \in U(1)$. Since having different boundary conditions on the positive and negative real axes is interpreted as the insertion of a boundary changing operator at the origin, we see that, for the boundary conditions above, the boundary changing operator corresponds to a twist defect for the $U(1)_R$ symmetry. With more general boundary conditions
\begin{equation}\label{eq:generalfermionbc}
    \begin{split}
        &G^{\pm}(z) = \eta'\bar{G}^{\pm}(\bar{z}) \qquad z= \bar{z} >0 \\
        & G(z)^{\pm} = \eta\bar{G}^{\pm}(\bar{z}) \qquad z= \bar{z} <0\, .
    \end{split}
\end{equation}
the extended field obeys
\begin{equation}
    \widetilde{\mathfrak{G}}(e^{2 i \pi }z) =\frac{\eta}{\eta'}\widetilde{\mathfrak{G}}(z)\, . 
\end{equation}
Thus the boundary changing operator is a twist defect attached to the line $L_{\eta/\eta'}$. Since the boundary conditions preserve the $U(1)_R$ symmetry, there exist well defined topological junctions between the $U(1)_R$ lines and the boundary. Therefore we can have, in the upper-half plane, TDLs homotopic to semi-circles stretching across the positive and negative real axes (eventually with trivalent junctions involving the boundary changing operator twist line). On the strip this configuration corresponds to a network of $U(1)_R$ lines, with a $U(1)_R$ line connecting the two boundaries and one running along the non-compact direction. 

By the equations above we see that the boundary parameter $\eta$ determines the mode expansion of the extended fermionic fields. Since those modes are used to construct the Hilbert space of the theory we see that having different values of $\eta$ on the positive and negative axis leads to twisted interval Hilbert spaces. In particular when $\eta = -1$ the associated topological defect line implements $(-1)^F$, which is a $\bZ_2$ subgroup of $U(1)_R$, and hence the theory with boundary conditions $\eta=1$ and $\eta'=-1$ has a Ramond sector Hilbert space on the interval. The line $L_\gamma$ stretching between the two boundaries instead acts on this Hilbert space.

The general case in the upper-half plane is to consider two boundary conditions $B_\eta$ and $B_{\eta'}$, related by the twist defect of $L_{\eta/\eta'}$, as well as another TDL $L_{\gamma}$ stretching between the boundaries. We map this configuration on the strip and compactify the extended direction, resulting in a finite cylinder. If we interpretet the compact direction as time (open sector) we have a trace over the Hilbert space with boundary conditions  $B_{\eta}$ and $B_{\eta'}$, i.e. a twisted Hilbert space, with an insertion of $L_\gamma$. In the S-dual channel  (closed sector), with a periodic space direction, $L_{\eta/\eta'}$ acts on the boundary states while $L_{\gamma}$ twists the Hilbert space, meaning that the boundary states will have components not in the vanilla circle Hilbert space $\bH$ but in the twisted one $\bH^{(\gamma)}$. In formulas
\begin{equation}
    \bra{B_\eta^{(\gamma)}}\widetilde{q}^{L_0-\frac{c}{24}}L_{\eta/\eta'}\ket{B_{\eta'}^{(\gamma)}} = \text{Tr}_{\bH^{(\eta, \eta')}}L_{\gamma}q^{L_0 - \frac{c}{24}}\, .
\end{equation}
Therefore choosing boundary conditions with $\eta \neq \eta'$ inevitably lead to a closed sector overlap involving a $L_{\eta/\eta'}$ insertion, or, equivalently, to an open sector tracing over a twisted Hilbert space. Similarly enriching the trace in the open sector with a fugacity for $U(1)_R$ can only correspond to an overlap of boundary states with components in a twisted Hilbert space. Another important fact is that boundary conditions preserving the superconformal algebra are invariant under $U(1)_R$, i.e. $L_{\eta'}\ket{B_\eta^{(\gamma)}} = \ket{B_\eta^{(\gamma)}}$, therefore we can forget about the insertion of $L_{\eta/\eta'}$ in the closed sector and simply write
\begin{equation}
    \bra{B_\eta^{(\gamma)}}\widetilde{q}^{L_0-\frac{c}{24}}\ket{B_{\eta'}^{(\gamma)}} = \text{Tr}_{\bH^{(\eta, \eta')}}L_{\gamma}q^{L_0 - \frac{c}{24}}
\end{equation}
Now recall that $(-1)^F$ is actually a subgroup of $U(1)_R$, so that the associated twist defect is the boundary changing operator between the boundary conditions with $\eta$ and $\eta' = -\eta$. For simplicity let's stick to $\eta ,\eta', \gamma = \pm 1$, then invariance under the $S$ transformation requires
\begin{equation}\label{eq:susyC}
\begin{split}
    & \bra{B_\pm^{(\text{NS})}}\widetilde{q}^{L_0-\frac{c}{24}}\ket{B_{\mp}^{(\text{NS})}} = \text{Tr}_{\bH^{\text{(R}, \pm)}}q^{L_0 - \frac{c}{24}} \\
    &\bra{B_\pm^{(\text{R})}}\widetilde{q}^{L_0-\frac{c}{24}}\ket{B_{\pm}^{(\text{R})}} = \text{Tr}_{\bH^{\text{(NS}, \pm)}}(-1)^Fq^{L_0 - \frac{c}{24}}\\
    & \bra{B_\pm^{(\text{NS})}}\widetilde{q}^{L_0-\frac{c}{24}}\ket{B_{\pm}^{(\text{NS})}} = \text{Tr}_{\bH^{\text{(NS}, \pm)}}q^{L_0 - \frac{c}{24}}\\
    &  \bra{B_\pm^{(\text{R})}}\widetilde{q}^{L_0-\frac{c}{24}}\ket{B_{\mp}^{(\text{R})}} = \text{Tr}_{\bH^{\text{(R}, \pm)}}(-1)^Fq^{L_0 - \frac{c}{24}}
\end{split}
\end{equation}
which generalise Cardy's equations in an $\cN=2$ supersymmetric setting. To solve this conditions we need to introduce Ishibashi states. We can construct a unique Ishibashi state $\ket{\cB_{i, \pm}^{(X)}}\rangle$ solving a given boundary constraint with $\eta= \pm 1$ for any irrep $\bH_i^{(X)}$. The components of the states are elements of $\bH_i^{(X)} \otimes \bH_{\omega(i^+)}^{(X)}$. Here we are using a single label $i$ to denote representations of the susy algebra and $\omega$ is the action of the automorphism defining the boundary conditions on the representations of the chiral algebra. Thus $\omega = \unit$ for the B-type and $\omega=C$ for A-type. As usual we use Ishibashi states to construct physical boundary states that solve Cardy's condition. We set
\begin{equation}\label{eq:physbry}
    \ket{B_{a, \pm}^{(X)}} = \sum_{i \in I^{(X)}_{\Omega}} B^{i, (X)}_{a, \pm} \ket{\cB_{i, \pm}^{(X)}}\rangle
\end{equation}
where $X=$NS, R and $I_{\Omega}^{(X)}$ labels the representations in the $X$ sector that can be used to construct the Ishibashi states, namely it contains only those $i$ for which $\bH_{i}\otimes \bH_{\omega(i^+)}$ appears in the circle Hilbert space. The overlaps of Ishibashi states are 
\begin{equation}
\begin{split}
    & \langle\bra{\cB_{i,\pm}^{(X)}}\widetilde{q}^{L_0 - \frac{c}{24}}\ket{\cB_{j,\pm}^{(X)}}\rangle = \delta_{ij}\text{ch}_i^{(X)}(\widetilde{q})\\ 
    & \langle\bra{\cB_{i,\pm}^{(X)}}\widetilde{q}^{L_0 - \frac{c}{24}}\ket{\cB_{j,\mp}^{(X)}}\rangle = \delta_{ij}\widetilde{\text{ch}}_i^{(X)}(\widetilde{q})\\ 
\end{split}
\end{equation}
as one can check from their explicit definition. The interval Hilbert spaces with supersymmetric boundary conditions labeled by $a,b$ are representations of the superconformal algebra, thus
\begin{equation}
\begin{split}
    & \text{Tr}_{\bH^{\text{(NS}, \pm)}_{ab}}q^{L_0 - \frac{c}{24}}= \sum_{i \in I^{(\text{NS})}} n_{ab; \pm}^{i}\text{ch}_i^{\text{(NS)}}(q)\\
    & \text{Tr}_{\bH^{\text{(R}, \pm)}_{ab}}q^{L_0 - \frac{c}{24}}= \sum_{i \in I^{(\text{R})}} m_{ab; \pm}^{i}\text{ch}_i^{\text{(R)}}(q) \\
    &\text{Tr}_{\bH^{\text{(NS}, \pm)}_{ab}}(-1)^Fq^{L_0 - \frac{c}{24}}= \sum_{i \in I^{(\text{NS})}} \widetilde{n}_{ab;\pm}^{i}\widetilde{\text{ch}}_i^{\text{(NS)}}(q)\\
    &\text{Tr}_{\bH^{\text{(R}, \pm)}_{ab}}(-1)^Fq^{L_0 - \frac{c}{24}}= \sum_{i \in I^{(\text{R})}} \widetilde{m}_{ab; \pm}^{i}\widetilde{\text{ch}}_i^{\text{(R)}}(q)\, . 
\end{split}
\end{equation}
Imposing \eqref{eq:susyC} on the physical boundary states \eqref{eq:physbry} we then obtain 
\begin{equation}
\begin{split}
     \sum_{i \in I^{(\text{NS})}_{\Omega}} B^{i, \text{ (NS)}}_{a, \pm}B^{i, \text{ (NS)}}_{b, \pm} S_{ij}^{\text{NSNS}} &= n^{j}_{ab;\pm} \\
     \sum_{i \in I^{(\text{R})}_{\Omega}} B^{i, \text{ (R)}}_{a, \pm}B^{i, \text{ (R)}}_{b, \pm} S_{ij}^{\text{R}\widetilde{\text{NS}}} &= \widetilde{n}^{j}_{ab;\pm} \\
     \sum_{i \in I^{(\text{NS})}_{\Omega}} B^{i, \text{ (NS)}}_{a, \pm}B^{i, \text{ (NS)}}_{b, \mp} S_{ij}^{\widetilde{\text{NS}}\text{R}} &= m^{j}_{ab;\pm} \\   
     \sum_{i \in I^{(\text{R})}_{\Omega}} B^{i, \text{ (R)}}_{a, \pm}B^{i, \text{ (R)}}_{b, \mp} S_{ij}^{\widetilde{\text{R}}\widetilde{\text{R}}} &= \widetilde{m}^{j}_{ab;\pm} \, . 
\end{split}
\end{equation}
We now want to find solutions to these constraints for some special modular invariant, in particular one that guarantees that we have as many Ishibashi states as there are primaries in the theory, so that the sums over $i$ in the above equations run over all primaries. Notice also that the numbers $\widetilde{n}^{j}_{ab},\widetilde{m}^{j}_{ab}$ are not multiplicities of some primary representation and therefore are not restricted to be positive, this allows us to use the Verlinde formulas for the $\widehat{M}$ coefficients derived in Appendix \ref{app:sVerl}. Using the properties\footnote{These derive from $S^{\widetilde{\text{NS}}\text{R}} S^{\text{R}\widetilde{\text{NS}}} = S^{\text{NSNS}} S^{\text{NSNS}}=C^{\text{NSNS}}$ and $\left(S^{\widetilde{\text{NS}}\text{R}}\right)^T =S^{\text{R}\widetilde{\text{NS}}}$. }
\begin{equation}
\begin{split}
   & S^{\text{NSNS}}_{lm;rs} = \left(S^{\text{NSNS}}_{r^+s^+;lm}\right)^*\\
   & S^{\widetilde{\text{NS}}\text{R}}_{lm;rs}=\left(S^{\widetilde{\text{NS}}\text{R}}_{lm;r^+s^+}\right)^*= \left(S^{\text{R}\widetilde{\text{NS}}}_{r^+s^+; lm}\right)^*\\
   &S^{\text{R}\widetilde{\text{NS}}}_{lm;rs}=\left(S^{\text{R}\widetilde{\text{NS}}}_{lm;r^+s^+}\right)^*= \left(S^{\widetilde{\text{NS}}\text{R}}_{r^+s^+; lm}\right)^*
\end{split}
\end{equation}
we can write down the solutions
\begin{equation}
\begin{split}
    & B^{lm, \text{ (NS)}}_{a_1 a_2, +} = \frac{S^{\text{NSNS}}_{a_1 a_2; lm}}{\sqrt{S^{\text{NSNS}}_{00;lm}}} \qquad B^{lm, \text{ (NS)}}_{a_1 a_2, -} = \frac{S^{\text{R}\widetilde{\text{NS}}}_{a_1 a_2; lm}}{\sqrt{S^{\text{NSNS}}_{00;lm}}} \\
    & B^{lm, \text{ (R)}}_{a_1 a_2, +} = \frac{S^{\widetilde{\text{NS}}\text{R}}_{a_1 a_2; lm}}{\sqrt{S^{\widetilde{\text{NS}}\text{R}}_{00;lm}}} \qquad B^{lm, \text{ (R)}}_{a_1 a_2, -} = \frac{S^{\widetilde{\text{R}}\widetilde{\text{R}}}_{a_1 a_2; lm}}{\sqrt{S^{\widetilde{\text{NS}}\text{R}}_{00;lm}}}\, . 
\end{split}
\end{equation}
The corresponding multiplicities are
\begin{equation}
    \begin{split}
        &n^{r s}_{a_1 a_2 ; b_1 b_2; +} = \widehat{N}^{r^+ s^+, \text{ NSNS}}_{a_1 a_2 ; b_1 b_2}  \qquad n^{r s}_{a_1 a_2 ; b_1 b_2; -} = \widehat{N}^{r^+ s^+, \text{ RR}}_{a_1 a_2 ; b_1 b_2}\\ & \widetilde{n}^{r s}_{a_1 a_2 ; b_1 b_2;+} = \widehat{M}^{r^+ s^+, \text{ NSNS}}_{a_1 a_2 ; b_1 b_2} \qquad \widetilde{n}^{r s}_{a_1 a_2 ; b_1 b_2; -} = \widehat{M}^{r^+ s^+, \text{ RR}}_{a_1 a_2 ; b_1 b_2} \\
        &m^{r s}_{a_1 a_2 ; b_1 b_2; +} = \widehat{N}^{r^+ s^+, \text{ NSR}}_{a_1 a_2 ; b_1 b_2}  \qquad
        m^{r s}_{a_1 a_2 ; b_1 b_2; -} = \widehat{N}^{r^+ s^+, \text{ RNS}}_{a_1 a_2 ; b_1 b_2} \\ & \widetilde{m}^{r s}_{a_1 a_2 ; b_1 b_2;+} = \widehat{M}^{r^+ s^+, \text{ NSR}}_{a_1 a_2 ; b_1 b_2;} \qquad \widetilde{m}^{r s}_{a_1 a_2 ; b_1 b_2;-} = \widehat{M}^{r^+ s^+, \text{ RNS}}_{a_1 a_2 ; b_1 b_2;}  \, . 
    \end{split}
\end{equation}
There is also another family of solutions that we can obtain reversing the signs of the R sector coefficients, since those affect only $\widetilde{n}^{j}_{ab},\widetilde{m}^{j}_{ab}$ it is still a consistent family of solutions both within itself and with the family of solutions described above. All in all we found the physical boundary states
\begin{equation}\label{eq:phybs1}
\begin{split}
     &\ket{B_{a_1,a_2; f, +}} =\sum_{(l,m) \in P_k}\left(\frac{S^{\text{NSNS}}_{a_1 a_2; lm}}{\sqrt{S^{\text{NSNS}}_{00;lm}}}\ket{\cB_{lm,+}^{\text{(NS)}}}\rangle + f\frac{S^{\widetilde{\text{NS}}\text{R}}_{a_1 a_2; lm}}{\sqrt{S^{\widetilde{\text{NS}}\text{R}}_{00;lm}}}\ket{\cB_{lm,+}^{\text{(R)}}}\rangle \right)\\
     &\ket{B_{a_1,a_2; f, -}} =\sum_{(l,m) \in P_k}\left(\frac{S^{\text{R}\widetilde{\text{NS}}}_{a_1 a_2; lm}}{\sqrt{S^{\text{NSNS}}_{00;lm}}}\ket{\cB_{lm,-}^{\text{(NS)}}}\rangle + f\frac{S^{\widetilde{\text{R}}\widetilde{\text{R}}}_{a_1 a_2; lm}}{\sqrt{S^{\widetilde{\text{NS}}\text{R}}_{00;lm}}}\ket{\cB_{lm,-}^{\text{(R)}}}\rangle \right)
\end{split}
\end{equation}
with $f = \pm 1$. Thus for pair $(a_1,a_2) \in P_k$ we can construct four boundary states, this means that we have two for each superconformal primary. For the untwisted B-type boundary conditions these solutions exist upon choosing the charge conjugation invariant partition function, while for the A-type boundary conditions, since the mirror map maps a representation in its charge conjugate, we need to pick the diagonal modular invariant.

\subsection{Minimal Model Boundary States, Another Perspective}

Perhaps a simpler way to study superconformal boundary states directly in the minimal model case is to employ the separation in half-families that proved useful for modular invariance. These half-families are labelled by $(a,c) \in Q_k$ and can be thought of as representations of the bosonic subalgebra. This subalgebra is really a subalgebra of the universal enveloping algebra and is obtained keeping all generators with even fermion number. This also includes products of an even number of fermion generators. In this set-up the study of conformal boundaries goes as in the bosonic case. To each of the subrepresentations $\bH_{ac}$ we associate an Ishibashi state $\ket{\cB_{(a,c)}}\rangle$ with components in $\bH_{(a,c)}\oplus\bH_{\omega((a^+,c^+))}$ such that
\begin{equation}
    \langle \bra{\cB_{(a',c')}}q^{L_0 - \frac{c}{24}}\ket{\cB_{(a,c)}}\rangle = \delta_{a,a'}\delta_{c,c'}\chi_{ac}(q)\, .
\end{equation}
Then we expand the annulus partition function as
\begin{equation}
    Z_{\alpha \beta}(q) = \sum_{(a,c) \in Q_{k}}n^{ac}_{\alpha \beta} \chi_{ac}(q)
\end{equation}
while the boundary states as
\begin{equation}
    \ket{B_{\alpha}} = \sum_{(a,c) \in Q_{k}^{\Omega}}B_{\alpha}^{ac}\ket{\cB_{(a,c)}}\rangle\, .
\end{equation}
The Cardy condition now simply reads
\begin{equation}
    n^{a'c'}_{\alpha \beta} = \sum_{(a,c) \in Q_{k}^{\Omega}} B_{\alpha}^{ac}B_{\beta}^{ac}S_{ac,a'c'}
\end{equation}
and, if $Q_k^{\Omega} = Q_k$, is solved by the Cardy states
\begin{equation}\label{eq:phybs2}
   \ket{B_{\alpha_1 \alpha_2}}= \sum_{(a,c) \in Q_k} \frac{S_{\alpha_1 \alpha_2; ac}}{\sqrt{S_{00; ac}}}\ket{\cB_{ac}}\rangle
\end{equation}
with multiplicities
\begin{equation}
    n^{ac}_{\alpha_1 \alpha_2 ; \beta_1 \beta_2} = N^{a^+ c^+}_{\alpha_1 \alpha_2; \beta_1 \beta_2}\, .
\end{equation}
We can check that those boundary states are the same ones we found in the previous subsection, just written in another basis. First notice that, consistently, there is a boundary state for each half-family, since there are two of these for each superconformal primary the total number of boundary states agrees with what we found above. To match with the previous notation we have to divide $Q_k$ in $P'_k$ and the set of images under $P'_k \ni (a,c) \mapsto (k-a, c+k+2)$, we then define two families of boundary states
\begin{equation}
\begin{split}
    & \ket{X_{\alpha_1 \alpha_2}} = \sum_{(a,c) \in P'_k} B^{ac}_{\alpha_1 \alpha_2}\left(\ket{\cB_{ac}}\rangle + (-1)^{a+c}\ket{\cB_{k-a,c+k+2}}\rangle\right)\\
    & \ket{Y_{\alpha_1 \alpha_2}} = \sum_{(a,c) \in P'_k}(-1)^{a+c} B^{ac}_{\alpha_1 \alpha_2}\left(\ket{\cB_{ac}}\rangle + (-1)^{\alpha_1+\alpha_2}\ket{\cB_{k-a,c+k+2}}\rangle\right)
\end{split}
\end{equation}
where now $(\alpha_1, \alpha_2) \in P'_k$. Passing to the $(l,m,\lambda)$ labeling it is now straightforward to verify that
\begin{equation}
    \begin{split}
        &\ket{X_{lm0}} = \ket{B_{lm;1,+}} \qquad \ket{X_{lm-1/2}} = \ket{B_{lm;1,-}} \\
        &\ket{Y_{lm0}} = \ket{B_{lm;-1,+}} \qquad \ket{Y_{lm-1/2}} = \ket{B_{lm;-1,-}}\, . 
    \end{split}
\end{equation}

\subsection{Folding Trick and Topological Lines}

From a conformal boundary condition we can construct a topological defect line via the folding trick. A TDL is a topological interface between the CFT and itself, the topologicity condition being encoded in the vanishing commutator with both the holomorphic and anti-holomorphic energy momentum tensors
\begin{equation}
    [T(z), L]= [\bar{T}(\bar{z}),L]=0
\end{equation}
the operator $L$ is supported on a closed curve and depends only on its homotopy class (which accounts also for other operator insertions). Locally a topological defect can always be interpreted as a topological interface separating two copies of the same CFT. Of course among the possible interfaces there's also the trivial one, for which the theories on the two sides are completely decoupled. The folding trick consists in folding the theory along the interface determined by the TDL, so that the defect is mapped to a boundary condition for the doubled theory $\text{CFT} \times \overline{\text{CFT}}$. The bar in $\overline{\text{CFT}}$ represents the fact that folding acts by parity on the CFT and exchanges the holomorphic and anti-holomorphic sectors. Among the boundaries of the doubled CFT there are also those that are a direct product of a boundary for CFT and one for $\overline{\text{CFT}}$, these do not glue the stress energy tensor of CFT with that of $\overline{\text{CFT}}$ and hence, upon unfolding, the theories on the two sides of the interface are decoupled. Therefore non-trivial topological defect lines correspond to boundary conditions for the doubled theory that mix the two stress energy tensors, these are called permutation boundaries. In a generic CFT classifying TDL's is then as hard as classifying boundary conditions, the problem becomes tractable only in RCFTs requiring the boundary conditions to preserve the full chiral algebra. In this part of the appendix we shall consider a generic bosonic RCFT, and only later specify to the bosonic subalgebra of the $\cN=2$ superconformal symmetry.

\subsubsection{TDLs from Untwisted Boundary Conditions}

The TDLs we study will commute not only with Virasoro but also with all the other generators of the chiral algebra $W^{i}$
\begin{equation}
    [W^i(z), L] = [\overline{W}^i(\bar{z}), L]=0 \, . 
\end{equation}
In the doubled theory these commutation relations become equations defining the permutation boundary states, indeed 
\begin{equation}\label{eq:permbs}
\left(W_n^{i\,1}-(-1)^{h_i} \overline{W}_{-n}^{i\,2}\right)\ket{B}=\left(W_n^{i\, 2}-(-1)^{h_i} \overline{W}_{-n}^{i\,1}\right)\ket{B}=0
\end{equation}
where $1,2$ refer to the two copies of the CFT and the overline for the anti-holomorphic sector. For instance the modes of the two stress energy tensors satisfy
\begin{equation}
\left(L_n^{1}-\overline{L}_{-n}^{2}\right)\ket{B}=\left(L_n^{ 2}-\overline{L}_{-n}^{1}\right)\ket{B}=0
\end{equation}
or, in the open channel,
\begin{equation}
    T^{1}(z) = \overline{T}^2(\bar{z})\qquad  T^{2}(z) = \overline{T}^1(\bar{z}) \quad \text{at} \quad z= \bar{z}\, .
\end{equation}
For concreteness we work on the upper half plane, then unfolding the theory we identify $T^1$ and $ \overline{T}^1$ with the holomorphic and anti-holomorphic tensors in the upper half plane while $T^2$ and $ \overline{T}^2$ with the anti-holomorphic and holomorphic components of the energy momentum tensor on the lower half plane. In formulas
\begin{equation}
\begin{split}
    & T^{1}(z) = T(z) \qquad \overline{T}^{1}(\bar{z}) = \overline{T}(\bar{z}) \quad \text{Im}(z)>0 \\
    & T^{2}(z) = \overline{T}(\bar{z}) \qquad \overline{T}^{2}(\bar{z}) = T(z) \quad \text{Im}(z)<0 \\
\end{split}
\end{equation}
then the gluing conditions ensure that both $T(z)$ and $\overline{T}(\bar{z})$ are continuous across the defect. Now, suppose we have chosen a certain modular invariant Hilbert space for the CFT 
\begin{equation}
    \bH = \bigoplus_{i,j} M_{ij}\bH_i \otimes \overline{\bH}_j
\end{equation}
in the doubled theory the total Hilbert space is then
\begin{equation}
    \bH^{(tot)} = \bigoplus_{i,j,k,l} M_{ij}M_{k,l}\bH_i \otimes \overline{\bH}_j\otimes \bH_l \otimes \overline{\bH}_k\, .
\end{equation}
Again we use Ishibashi states $\ket{\cB_{ij}}\rangle$ to express physical boundary states, these are now labelled by pairs of representations of the Chiral algebra $\cA$ and have components in $\bH_{i}\otimes \overline{\bH}_{i^+} \otimes \bH_{j}\otimes \overline{\bH}_{j^+} $. Notice that the equations defining permutation boundary states require the representation content of the states to be equal for CFT$_1$ and CFT$_2$. We can see this from the fact that a generic physical boundary state
\begin{equation}
    \ket{B_{\alpha \beta}} = \sum_{ij}B^{ij}_{\alpha \beta}\ket{\cB_{ij}}\rangle
\end{equation}
can satisfy both \eqref{eq:permbs} only if
\begin{equation}
    B_{\alpha \beta}^{ij} = \delta^{ij} B_{\alpha \beta}^{i}\, . 
\end{equation}
That is permutation boundary states are expanded only using the (available) diagonal Ishibashi states
\begin{equation}
     \ket{B_{\alpha}} = \sum_{i}B^{i}_{\alpha}\ket{\cB_{ii}}\rangle\, .
\end{equation}
The overlap in the doubled theory is not difficult to compute. On the cylinder we have, for any boundary state satisfying \eqref{eq:permbs} 
\begin{equation}
    \bra{B}e^{-\frac{2 \pi }{t}\left(L_0^1 + L_0^2 +\overline{L}_0^1 + \overline{L}_0^2 -\frac{c}{6}\right)}\ket{B} = \bra{B}e^{-\frac{4 \pi }{t}\left(L_0 +\overline{L}_0-\frac{c}{12}\right)}\ket{B}=  \bra{B}q^{L_0 - \frac{c}{24}} \bar{q}^{\overline{L_0}- \frac{c}{24}}\ket{B}
\end{equation}
where we noticed that the central charge doubles in the folded theory and we used the boundary conditions. Then
\begin{equation}
    \langle \bra{\cB_{ii}} q^{L_0 - \frac{c}{24}} \bar{q}^{\overline{L_0}- \frac{c}{24}} \ket{\cB_{jj}}\rangle = \delta_{ij}\chi_{i}(q) \chi_{i^+}(\bar{q})\, .
\end{equation}
In the open channel, since the boundary conditions preserve the $\cA \times \cA$ symmetry, we expand the partition function as
\begin{equation}
    \text{Tr}_{\bH_{\alpha \beta}}\widetilde{q}^{L_0 - \frac{c}{24}} \bar{\widetilde{q}}^{\overline{L_0}- \frac{c}{24}} = \sum_{i,j}n^{ij}_{\alpha \beta} \chi_{i}(\widetilde{q})\chi_j(\bar{\widetilde{q}})\, . 
\end{equation}
Then Cardy's condition reads
\begin{equation}
\sum_{i}B^{i}_{\alpha}B^{i}_{\beta}S_{ij}S_{jk} = n^{jk}_{\alpha \beta}\, .
\end{equation}
Since the sum over $i$ runs over the available diagonal Ishibashi states it is difficult to provide a general solution. However when all Ishibashi states can be used we can set
\begin{equation}\label{eq:opaction}
    B^{i}_{\alpha} = \frac{S_{i \alpha}}{S_{0i}}
\end{equation}
using that the ratios $S_{ai}/S_{0i}$ furnish a 1-dimensional representation of the fusion algebra togheter with the Verlinde formula it is not difficult to show that the resulting multiplicities in the open channel are integers. Another way of getting to this formula directly in the unfolded theory with a charge conjugation invariant HIlbert space is to notice that, since $L$ commutes with all generators of $\cA$ it must be proportional to the identity in every subspace $\bH_{i}\otimes \bH_{i^+}$ of the Hilbert space \cite{Drukker:2010jp}. Therefore it can be written as a sum of projectors
\begin{equation}
    L_{\alpha} = \sum_{i}B_{\alpha}^{i}\sum_{\textbf{m}, \textbf{n}}\ket{i, \textbf{m}; i^+, \textbf{n}} \otimes \bra{i, \textbf{m}; i^+, \textbf{n}}
\end{equation}
where the sums over $\textbf{m}$ and $\textbf{n}$ run over descendants. Requiring that the Hilbert space twisted by $L$ has a consistent Hilbert space interpretation one again gets \eqref{eq:opaction}. Clearly the coefficients $B^{i}_{\alpha}$ tell us how the TDL acts on local operators. We also see that the number of independent TDLs matches that of physical permutation boundaries in the doubled theory. 

\paragraph{Untwisted case: minimal models}.

This analysis carries over directly to the minimal model case with charge conjugation invariant partition function using the half-family basis. With this choice of partition function we have a physical B-type boundary state for each half-family. In the doubled theory we have a permutation Ishibashi state for each half-family from which we construct a TDL for each half family. On the circle Hilbert space of the unfolded theory this acts as the sum of projectors
\begin{equation}
\begin{split}
     L_{a_1 a_2} &= \sum_{(a,c) \in Q_k}\frac{S_{a_1,a_2; ac}}{S_{00;ac}}\sum_{\textbf{m}, \textbf{n}}\ket{(a,c), \textbf{m}; (a^+,c^+), \textbf{n}} \otimes \bra{(a,c), \textbf{m}; (a^+,c^+), \textbf{n}}\\ 
     &= \sum_{(a,c) \in Q_k}\frac{S_{a_1,a_2; ac}}{S_{00;ac}}P_{ac} \, .
\end{split} 
\end{equation}
The primary states (of the bosonic subalgebra) appearing in the torus partition function can be labelled by a single half-family $\ket{(a,c)}\otimes\ket{(a^+,c^+)} = \ket{\Phi_{ac}}$ on which the TDLs act as
\begin{equation}
    L_{a_1,a_2}\ket{\Phi_{ac}} = \frac{S_{a_1,a_2; ac}}{S_{00;ac}}\ket{\Phi_{ac}}\, .
\end{equation}
We can work out the action on the full superconformal families simply changing basis. We have
\begin{equation}
     L_{a_1 a_2} = \sum_{(a,c) \in P'_k}\frac{S_{a_1,a_2; ac}}{S_{00;ac}} \left(P_{ac} + (-1)^{a_1+a_2}P_{k-a;c+k+2}\right)
\end{equation}
notice that the sum $P_{ac} + (-1)^{a_1+a_2}P_{k-a;c+k+2}$ projects on the full superconformal family labelled by $(a,c) \in P'_k$ if $[a_1+a_2]=0$, while if $[a_1 + a_2 ]=1$ it still projects on the same superconformal family but with a minus sign for the states with odd fermion number. Reintroducing the $(l,m,\lambda)$ parametrization we find the four types of TDLs
\begin{equation}\label{eq:TDLscfam}
\begin{split}
     & L_{lm; +, f} = \sum_{(l',m') \in P_k} \left(\frac{S^{\text{NSNS}}_{lm;l'm'}}{S^{\text{NSNS}}_{00;l'm'}}P^{\text{(NS)}}_{lm}+ f \frac{S^{\widetilde{\text{NS}} \text{R}}_{lm;l'm'}}{S^{\text{NSNS}}_{00;l'm'}}P^{\text{(R)}}_{lm}\right)\\
    & L_{lm;-,f} = \sum_{(l',m') \in P_k} \left(\frac{S^{\text{R}\widetilde{\text{NS}}}_{lm;l'm'}}{S^{\text{R}\widetilde{\text{NS}}}_{00;l'm'}}\widetilde{P}^{\text{(NS)}}_{lm}+ f\frac{S^{\widetilde{\text{R}}\widetilde{\text{R}}}_{lm;l'm'}}{S^{\text{R}\widetilde{\text{NS}}}_{00;l'm'}}\widetilde{P}^{\text{(R)}}_{lm}\right)
\end{split}
\end{equation}
where $f= \pm 1$ and $P^{(X)}_{lm}$ and $\widetilde{P}^{(X)}_{lm}$ project on the $(l,m)$ superconformal family in the $X$ sector without or with signs for the odd fermion number states.

\subsubsection{TDLs from Twisted Boundary Conditions}
In general we have also twisted boundary conditions for the CFT which also induce permutation boundary states for the doubled theory
\begin{equation}
\left(W_n^{i\,1}-(-1)^{h_i} \Omega\left(\overline{W}_{-n}^{i\,2}\right)\right)\ket{B}=\left(W_n^{i\, 2}-(-1)^{h_i} \Omega\left(\overline{W}_{-n}^{i\,1}\right)\right)\ket{B}=0\, .
\end{equation}
These corresponding TDLs in the unfolded theory commute with the generators of the chiral algebra only up to the automorphism $\Omega$
\begin{equation}
    L_{\Omega}W(z) = \Omega(W(z))L_{\Omega}
\end{equation}
and similarly for the anti-holomorphic side. The Ishibashi states $\ket{\cB_{ij}}_{\Omega}$ now have components in $\bH_{i}\otimes \bH_{\omega(i^+)} \otimes \bH_{j}\otimes \bH_{\omega(j^+)}$, and again only the diagonal ones with $i=j$ contribute to permutation boundaries. The analysis of the physical boundary states parallels that of the untwisted case as long as we use the twisted Ishibashi states and the solution
\begin{equation}
    B_\alpha^i = \frac{S_{i \alpha}}{S_{0 i}}
\end{equation}
is available as long as we are allowed to use all Ishibashi states, that is $M_{ij} = \delta_{j,\omega(i^+)}$. The expression for the TDL as an operator on the Hilbert space of the unfolded theory is now
\begin{equation}
     L_{\Omega, \alpha} = \sum_{i} \frac{S_{i \alpha}}{S_{0 i}}\sum_{\textbf{m}, \textbf{n}}\ket{i, \textbf{m}; \omega(i^+), \textbf{n}} \otimes \bra{i, \textbf{m}; \omega(i^+), \textbf{n}}
\end{equation}

\paragraph{Twisted case: minimal models}
We now pick a diagonal modular invariant partition function so that we have a physical A-type boundary state for each half-family, the corresponding topological lines are
\begin{equation}
\begin{split}
     L_{a_1 a_2} &= \sum_{(a,c) \in Q_k}\frac{S_{a_1,a_2; ac}}{S_{00;ac}}\sum_{\textbf{m}, \textbf{n}}\ket{(a,c), \textbf{m}; (a,c), \textbf{n}} \otimes \bra{(a,c), \textbf{m}; (a,c), \textbf{n}}\\ 
     &= \sum_{(a,c) \in Q_k}\frac{S_{a_1,a_2; ac}}{S_{00;ac}}P_{ac} \, .
\end{split} 
\end{equation}
In terms of the superconformal families we find again the expressions \eqref{eq:TDLscfam} where all projectors are now diagonal rather than charge conjugation invariant.

\newpage

\bibliographystyle{ytphys}
\baselineskip=0.85\baselineskip
\bibliography{bib}

\end{document}